\pdfoutput=1
\documentclass[iop]{emulateapj}

\usepackage{amsmath}
\usepackage{natbib}
\usepackage{color}
\usepackage{psfrag}
\usepackage{graphicx}
\usepackage{epsf}

\newcommand{\msun}{{\rm M}_{\sun}}

\newbox\grsign \setbox\grsign=\hbox{$>$}
\newdimen\grdimen \grdimen=\ht\grsign
\newbox\laxbox \newbox\gaxbox
\setbox\gaxbox=\hbox{\raise.5ex\hbox{$>$}\llap
     {\lower.5ex\hbox{$\sim$}}}\ht1=\grdimen\dp1=0pt
\setbox\laxbox=\hbox{\raise.5ex\hbox{$<$}\llap
     {\lower.5ex\hbox{$\sim$}}}\ht2=\grdimen\dp2=0pt

\shorttitle{Jet rotation from MHD shocks}
\shortauthors{Ch. Fendt}

\begin{document}

\title{Jet rotation driven by MHD shocks in helical magnetic fields}

\author{Christian Fendt}
\affil{Max Planck Institute for Astronomy, K\"onigstuhl 17,
D-69117 Heidelberg, Germany}
\email{fendt@mpia.de}

\begin{abstract}
In this paper we present a detailed numerical investigation of the hypothesis
that a rotation of astrophysical jets can be caused by magnetohydrodynamic 
shocks in a helical magnetic field.
Shock compression of the helical magnetic field results in a toroidal
Lorentz force component which will accelerate the jet material in toroidal
direction.
This process transforms magnetic angular momentum (magnetic stress) carried
along the jet into kinetic angular momentum (rotation).
The mechanism proposed here only works in a {\em helical} magnetic field
configuration.
We demonstrate the feasibility of this mechanism by axisymmetric MHD simulations
in 1.5D and 2.5D using the PLUTO code.
In our setup the jet is injected into the ambient gas with zero kinetic angular
momentum (no rotation).
Different dynamical parameters for jet propagation are applied such as the jet
internal Alfv\'en Mach number and fast magnetosonic Mach number, the density
contrast of jet to ambient medium, or the external sonic Mach number of the jet.
The mechanism we suggest should work for a variety of jet applications, e.g.
protostellar or extragalactic jets, and internal jet shocks (jet knots) 
or external shocks between the jet and ambient gas (entrainment).
For typical parameter values for protostellar jets, the numerically derived
rotation feature looks consistent with the observations, i.e. rotational
velocities of 0.1-1\% of the jet bulk velocity.
\end{abstract}

\section{Introduction}
Astrophysical jets are highly collimated beams of high velocity material,
observed in a variety of astronomical sources - among them young 
stellar objects (YSO), micro-quasars, or active galactic nuclei (AGN).
Somewhat less collimated beams of comparatively lower speed are
usually called outflows.
The current understanding of jet formation is that these outflows are launched
by {\em magnetohydrodynamic} (MHD) processes in the close vicinity of the
central object -- 
an accretion disk surrounding a protostar or a compact object
\citep{blan82,pudr83,came90,pudr07,cabr07}.

Over the past few years observational indication has been accumulated 
for axially symmetric differences in the radial velocity profile
across protostellar jets and outflows.
These radial velocity differences have repeatedly been interpreted as 
{\em outflow rotation}\footnote{In the following, for simplicity, we 
will omit such cautious wording regarding the existence of rotation in
jets/outflows - being aware, however, that outflow 
rotation has not yet been fully confirmed}.
For one of the best studied examples, the micro-jet from DG\,Tau, 
the suggested 
rotational velocity is about 6-15\,km/s at about $0.2\,$ arcsec or $40\,$AU
from the jet axis up to distances of 100\,AU from the star \citep{bacc02}.

As MHD outflows are launched from rotating sources by nature,
the rotation of the outflow itself seems to be a characteristic feature 
of MHD jet formation models -
from classic steady state models \citep{blan82}, 
non-relativistic \citep{ouye97,cass02} or 
recent relativistic simulations \citep{port10}.
The transverse stability of relativistic jets has been treated by
two-component simulations by \citet{meli07}.
When jets originate in the inner part of a rapidly rotating disk, the
outflow material is launched with a substantial amount of angular 
momentum.
Magnetohydrodynamic models further imply angular momentum conservation 
along the streamlines of the outflow, and, thus, jet material which
is persistently rotating.
Quantitative estimates of launching-induced jet rotation based on 
steady-state MHD models
seem to be in agreement with the observations \citep{ande03}.

There are several arguments why this picture may not hold in general.
The main argument is that the radial velocity profile is usually inferred
from emission lines from {\em shocked gas}. 
It is therefore questionable that classical steady-state MHD models can be
used to fit the velocity and angular momentum balance along the jet across 
these shocks.
Furthermore, traditional non self-similar steady-state models of MHD jets 
indicate on both large initial jet opening angles and large jet radii 
compared to the Alfv\'en radius $r_{\rm A}$.
In that case, as the rotational velocity decreases with $\sim 1/r$ for
$r > r_{\rm A}$, jet radii of $10^2 - 10^3 r_{\rm A}$ would result in a
negligible rotational velocity along the outer layers of a jet.
 
We like to note here that the outflow velocity profile itself might be 
explained without implementing jet rotation at all, and, thus, without 
assuming a MHD launching model for the outflow.
\citep{soke05} suggested a model in which a thick and twisted accretion disk
interacts with the outflow by entrainment. Inclination between the disk rotational
axis and the jet axis would lead to a different amount of deceleration on
both sides of the jet, thus visible in a velocity gradient across the jet.

In this paper, we propose a different explanation for the observationally indicated 
outflow rotation.
The model combines two basic features of MHD outflows - 
(i) the fact that the observed emission is from shocked gas, and 
(ii) the jet shock compressed helical field leads to a magnetic 
torque which accelerates the outflow material in toroidal direction.
Essentially, the jet kinetic angular momentum is retrieved from the 
magnetic angular momentum content of the jet across the shock.
First hints of such a mechanism have been observed in numerical simulations
by \citet{koes90} or \citet{uchi92,todo92}, but were not discussed in 
great detail, or even applied to observations\footnote{Observational 
indication of jet rotation had not been found by that time}.

In the following, we first summarize the observational findings
concerning outflow rotation in jets from young stars 
and then briefly mention the case of extragalactic jets.
We continue by discussing the governing MHD processes involved in 
our model, and present results of our MHD simulations.
These simulations are preliminary in the sense that our emphasis is on
the dynamical evolution of the outflow. 
This did not yet include further physical processes such as cooling 
(apart from one example simulation), magnetic diffusivity, or radiative
effects.

\section{Observational indication of outflow rotation in young stars}
Here we summarize the observational background discussing a few 
typical, but different example sources where outflow rotation is 
observationally indicated.

\subsection{Pc-scale jet propagation}
To our knowledge, first observational hints on jet rotation were
provided by Echelle spectroscopy of knots in the HH\,212 jet 
\citep{davi00}.
For one of the inner knots (SK1, 2000Au from the source), rotational 
velocities of 
$~1.5$\,km/s were derived, while other knots in the same 
jet/counter-jet did not follow the same sense of rotation (clockwise)
in spite of the almost perfect side-symmetry of HH\,212 
and the fact the sense of rotation of the central molecular disk 
is clockwise as well \citep{wise01}.

\citet{davi00} estimate the angular momentum of an inner accretion 
disk of 500\,AU.
Assuming that all that angular momentum will be removed by a jet with 
mass ejection rate of 10\% the accretion rate they conclude that
HH\,212 may rotate with about 2km/s.
If the disk angular momentum is removed by viscous transport
only, the jet rotational velocity would be lower.
We find this argument somewhat delusive as the high velocity jets
is in fact launched close to the inner edge of the disk.
Thus, the disk material which is ejected into the jet at these 
small radii has {\em low} specific angular momentum as it has lost
most of its initial angular momentum during the accretion process 
from larger to smaller radii. 

A corresponding search for outflow rotation in HH\,212 in SiO shocks
\citep{lee08} provided upper limits for rotational velocities only.
Applying angular momentum conservation for steady-state the inferred
jet launching radii can be constrained to $< 1$\,AU.
Most recent data of HH\,212 gave no clear signature of jet rotation,
but also cannot exclude rotation \citep{corr09}.

In the case of HH\,211, launching radii of $0.06 - 0.15\,$AU 
were derived from possible jet rotation of $1.5\,$km/s 
\citep{lee07,lee09}.

The NGC\,1333 IRAS\,4A2 system is another jet source for which
recent SiO observations indicate jet rotation by perpendicular
velocity gradients \citep{choi11}. These features extend as far as
almost 9000\,AU from the central protostar with derived rotational
speeds of about 3\,km/s and in corotation with the underlying
disk.

\subsection{Micro-jets from young stars}
The observed Doppler-shift of emission lines in the DG\,Tau micro-jet 
\citep{bacc02} give symmetric radial velocity differences of 6-15\,km/s 
for the low-velocity jet component.
It is difficult to imagine any other velocity pattern than {\em rotation}
to explain these radial velocity gradient across the jet.
The authors estimate the corresponding angular momentum flux in
the low velocity component to
$\dot{J} \simeq 3.8\times10^{-5}\msun{\rm yr^{-1} AU\,km\,s^{-1}}$.

More examples were detected by the same group \citep{coff04}, such as 
TH\,28 and RW\,Aur or LkH$\alpha$.
For RW\,Aur \citet{woit05} find toroidal velocities in the range 
of $5-30\,$km/s at distances 20 and 30\,AU away from the jet axis.
Furthermore, both bipolar lobes rotate in the same direction.

Another candidate source for jet rotation is HL\,Tau for which 
Fabry-Per\'ot interferometry has indicated a radial velocity gradient
across the jet with a slope of 6\,km/s/arcsec \citep{movs07}.
As for to DG\,Tau, it is the low velocity component which
shows a significant radial velocity slope while the radial velocity 
variation for the high velocity component is weak.

\subsection{Molecular outflows}
Indicate exists for rotation in molecular outflows as well.
From sub-mm data of the bipolar outflow Ori-S6 \citet{zapa09} find
bullets of in rotation of 5\,km/s, a CO shell around the jet rotating
with 2\,km/s (at 1000\,AU from the outflow axis), and a CO envelope
at distances of more than 2000\,AU from the source still rotating with
more then 0.5\,km/s, and rotation observed up to 25,000\,AU distance 
from the source.

Another example is CB\,26, a molecular outflow from a young T\,Tauri 
star \citep{laun09}.
Detailed kinematic modeling accompanied by synthetic CO maps suggest
rotational velocities of about 1\,km/s 100\,AU away from the outflow
axis and at 1000\,AU away from the source.
Alternative scenarios such as a precessing jet entraining 
material are discussed for this source and seem feasible as well.


\subsection{Magnetic field indication in protostellar jets}
In general, there is {\em indirect} indication of magnetic fields
being present in protostellar jets. 
One argument has always been the magnetic activity of protostars
showing large-scale flares on radio and X-rays and features of
dipolar accretion
indicating on a kG stellar surface field strength and also on a 
dipolar field distribution (see \citep{andr88, bouv07, cabr07}).

Concerning the large-scale field structure, there is the unique 
example of TTau where cyclotron radiation have been detected which 
is aligned with the jet outflow in this source \citep{ray97}.
The infered field strength is in the range of a G which is huge
compared to what is required by current modeling.

However, recent observations of massive star forming regions have 
detected maser emission in jet sources which allows for a direct 
measurement of the magnetic field strength. 
One example is the detection of the 3D field structure around the
protostar Ceph\,A\,HW\,2 with mG field strengths \citep{vlem10}.
Synchrotron emission indicating a 0.2 mG field has been found related to
the knots in the jet source HH80-81 \citep{carr10}.

\section{The case of extragalactic jets}
Observational indication of rotation is lacking so far for extragalactic jets. 
In fact, we do not even have direct measurements of propagation speed
for these sources.
On the other hand, clear evidence exists for a helical magnetic field structure 
in extragalactic jets indicated e.g. by gradients in the rotation measure 
across the jets (e.g. \citet{gabu04, lyut05, lain06}).

The existence of radio synchrotron emission along these jets on time scales
beyond the synchrotron cooling time requires re-heating of the jet material,
most probably by internal shocks.
Together, the helical field structure and the internal shocks, provide
the necessary prerequisites for the acceleration mechanism discussed 
below.
Therefore, signatures of jet rotation can also expected from these sources.

In fact, there is one jet source where azimuthal jet motion has been
detected. 
Imaging interferometry of emission lines have detected a {\em braided}
structure in the jet system of NGC\,4258, 
which could be modeled by a triple helix of jet motion on 5-10\,kpc 
scale \citep{ceci92}.
This kpc jet is aligned with a nuclear VLBA radio jet \citep{herr97}, 
thereby suggesting a physical connection. 
However, even if the helix model would explain the physical conditions
in the kpc jet, there seems to be no plausible explanation yet about
the intrinsic process causing such a complex outflow structure.

In this context it is interesting to note the relevance of jet rotation 
for modeling the heating of relativistic jet plasma by shear and centrifugal
acceleration for electrons \citep{rieg02, rieg04} or protons \citep{demp09}.

\section{MHD jets and rotation}
Magnetohydrodynamic jets are launched from rotating sources. 
The initial acceleration is by centrifugal forces along the magnetic field 
lines (magneto-centrifugal). 
Beyond the Alfv\'en point along each field line, the outflow material becomes
inertially so heavy that co-rotation of matter and magnetic field cannot longer
be maintained.
Thus, the field lines will be "bent" from a mainly poloidal into a 
helical geometry (a toroidal field component is induced).
This allows the matter to "slide" along the field lines in toroidal direction.
As a consequence, at large radii the jet material moves outwards in almost 
radial direction by conserving its total angular momentum. The angular velocity
decreases with radius.
 
Magnetohydrodynamic acceleration and collimation of outflows implies transfer 
of magnetic energy to kinetic energy, and, likewise, transfer of kinetic angular
momentum in magnetic angular momentum.
The exact balance between these quantities results from the local force-balance
throughout the outflow. 
The solution of the equations is complex and requires semi-analytical modeling
and/or numerical MHD simulations.

For understanding the considerations of this paper, it is helpful to re-call
some basic properties of axisymmetric, stationary-state theory of MHD winds 
and jets (see e.g. \citet{pudr07}).
In general, steady-state MHD implies the existence of conserved quantities
along the magnetic flux surfaces.
In the case of axisymmetry five such quantities can be defined.
The first one is the magnetic flux, 
$\Psi(r,z) \equiv  (1/2\pi )\int \vec{B}_{\rm p} d\vec{A}$,
indicating that magnetic field lines lie on magnetic flux surfaces.
The second one is the mass flux 
$\eta (\Psi) \equiv \int \rho \vec{v}_{\rm p} d\vec{A}$
per magnetic flux surface $\Psi$.
The third one is Ferraro's iso-rotation parameter 
$\Omega_{\rm F}(\Psi) \equiv ( v_{\rm p} B_{\phi}/B_{\rm p} - v_{\phi}) / r$,
often interpreted as angular velocity of the field lines.
The fourth one is the total (specific) angular momentum
\begin{equation}
\label{eqn-ang-mom}
L(\Psi) = r v_{\phi} - \frac{r B_{\phi}B_{\rm p}}{4 \pi \rho v_{\rm p}} 
\equiv \Omega_{\rm F} r_{\rm A}^2.
\end{equation}
The fifth one is the total (specific) energy
\begin{equation}
E(\Psi) = \frac{1}{2}v_{\rm p}^2 + \frac{1}{2}v_{\phi}^2 
        - \frac{r B_{\phi} B_{\rm p} \Omega_{\rm F}}{4 \pi \rho v_{\rm p}},
\end{equation}
where, for simplicity, we have omitted gravity and gas pressure (cold wind assumption).
In steady-state theory $E(\Psi)$, $L(\Psi)$, and $\eta(\Psi)$ are governed by 
the regularity condition across the fast magnetosonic point,
the Alfv\'en point at $r = r_{\rm A}$, and the
slow magnetosonic point, respectively.

Essentially, both the total energy and the total angular momentum 
(flux) consist of a magnetic and a kinetic contribution.
A classical Blandford-Payne-type outflow is launched with 
$v_{\phi}>> v_{\rm p}$ and $B_{\phi} < B_{\rm p}$.
Beyond the Alfv\'en point $v_{\phi}$ decreases and 
the magnetic field becomes dominated by $B_{\phi} > B_{\rm p}$.
Likewise, kinetic angular momentum of the ejected jet material, which 
is co-rotating with the disk, 
is increasingly transferred to the outflow magnetic angular momentum.

The existence of conserved quantities along the field lines can be used
to physically link the very inner region where the jet is launched
(not yet possible to resolve observationally)
to the asymptotic region of the outflow (accessible for observations).
\citet{ande03} have applied this formalism to the observed rotational
features in the DG\,Tau jet \citep{bacc02},
obtaining launching radii of 0.3-4.0\,AU.
While this approach could be applied to the low velocity jet component,
it does not fit, however, the high-velocity jet component of DG\,Tau.
Using a similar method for the bipolar jet from RW\,Aur,
\citet{woit05} find launching radii less than 0.5\,AU and a magnetic 
lever arm (Alfv\'en radius to launching radius) of about four.

In the following we will argue that the classical stationary MHD wind
approach discussed above is questionable as sole explanation of rotating
outflows. In the next section we will suggest and discuss an
alternative scenario.
Our main argument is that radial velocity measurements of stellar jets 
and outflows usually consider line emission of shocked gas or molecules. 
For example, \citet{lava00} prove shock excitation of forbidden emission 
lines in DG\,Tau and
\citet{woit05} explicitely state that rotation is more evident 
in the forbidden lines [OI] and [NII].
During shock transition, the jet material changes its dynamical state instantly.
Shock transitions are not included with the classical stationary state MHD wind
theory.
Instead, conservation laws must be applied in form of jump conditions,
and the direct connection of the observed rotating material with 
the disk by ideal MHD conservation laws is questionable.
This argument may hold in particular for the large scale jets and
outflows as e.g. for HH\,212 or molecular outflows.
The molecular material might also trace entrained interstellar
material. In this case, it has not been launched from the central 
region with intrinsic angular momentum.

Early work on steady-state MHD jets has suggested that jets do 
expand substantially from their launching area close to the star to 
the asymptotic (observed) regime. Expansion factors of 100-1000 
(from 0.1\,AU up to 100\,AU jet diameter) have been suggested.
These classical solutions predict Alfv\'en radii $R_{\rm A}$ of 
about ten foot point radii, $R_{\rm A} \simeq 10 R_{\rm F}$ and,
therefore, a substantial decrease of the toroidal outflow velocities 
for radii $r>>R_{\rm A}$ to $v_{\phi}/v_{\rm jet} < 0.001$.
This is due to the large initial opening angles which are a
result from non-self-similar MHD solutions 
(see \citet{fend95,fend96}, but also recent simulations of stellar
wind magnetospheres by \cite{matt08,matt10} resulting in a 
similar field geometry).
 
We conclude that while being consistent with stationary state MHD jet 
formation, 
the formalism suggested by Andersen et al. and subsequently applied 
by other authors is probably not a fully realistic picture to explain
the observed jet rotation.

\section{MHD shocks cranking jets}
In the following we discuss another framework of how jets and outflows
may be set in rotation.
The basic idea is that a compression of the toroidal magnetic field
component in MHD shocks results in a Lorentz torque in toroidal direction 
which accelerates the jet material.
The jet kinetic angular momentum is retrieved from the magnetic angular
momentum content of the jet by the shock.
This mechanism works only for {\bf helical} magnetic fields.
%
While in stationary state MHD the kinetic and magnetic angular momentum 
is gradually exchanged along the flux surfaces, the shock wave leads
to a sudden exchange.

\subsection{Lorentz torque}
The MHD Lorentz force is 
$\vec{F}_{\rm L} \sim \vec{j} \times \vec{B} \sim (\nabla\times\vec{B})\times\vec{B}$.
Jet rotation can be achieved by its toroidal component 
\begin{equation}
\vec{F}_{\phi} \sim (\nabla\times \vec{B}_{\phi})\times{B}_{\rm p}.
\end{equation}
Across the shock the toroidal field component is compressed, thus 
giving rise to a toroidal force component.
The compression ratio itself depends on the jet dynamics. 
For simple geometries the field compression can be estimated by the Rankine-Hugoniot 
jump conditions.
More complex configurations require numerical simulations to follow the shock evolution.

The magnetic flux is conserved across the shock\footnote{Index '1' denotes the 
upstream region and index '2' the downstream region}, 
$\left[ B_z \right]^2_1 = 0$, 
but the perpendicular magnetic field component typically undergoes a change.
In case of cylindrical coordinates (which we are using here), the $B_{\phi}$-component replaces the 
perpendicular component in Cartesian coordinates, resulting in an expression for the compression
\begin{equation}
\frac{\left[ B^2_{\phi} \right]^2_1}{B_z^2} = 
\frac{2 m_z^2 {\langle B_{\phi}\rangle}^2 / B_z^2}{m_z^2 - \left( B^2/4\pi {\langle\rho^{\star}\rangle}\right)}
\,\,\frac{\left[\rho^{\star}\right]^2_1}{{\langle\rho^{\star}\rangle}},
\end{equation}
\citep{uchi92},
where $m_z \equiv \rho v_z$ denotes the (conserved) axial momentum, 
$\langle\rho^{\star}\rangle \equiv 0.5\,(1/\rho_1 + 1/\rho_2)$
the averaged inverse density from upstream and downstream, 
$\left[ \rho^{\star} \right]^2_1$ the difference in the inverse density upstream 
and downstream, 
and ${\langle B_{\phi}\rangle}^2 \equiv {\langle B\rangle}^2 - B_z^2$ the average
toroidal field strength.
 
The dynamical structure of the propagating shock could be complex. 
As typically for this setup, four shocks are existent, accompanied by a contact
discontinuity, each satisfying the jump conditions (see e.g. \citet{uchi92, ryu95a}).
Considering the toroidal field strength, an increase within the first MHD (fast) 
shock is followed by a decrease across the second MHD (slow) shock, 
then again by an increase in the third (slow MHD) shock,
and another decrease in the fourth (fast MHD) shock
(going from downstream to upstream; see below and \citet{uchi92, ryu95a}). 

A vast number of multi-dimensional jet propagation simulations
have been published.
Only few of them apply a helical magnetic field configuration and some of those
have already provided numerical indication for toroidal velocity changes due 
to toroidal magnetic torques in jet shocks.
A connection to or predictions of the observable jet rotation has not
been made (such observations were not available in that time).
Early simulations of jet propagation do not show rotation \citep{clar86}
since treating a purely toroidal magnetic field only.
To our knowledge, the MHD simulations by \citet{koes90} were first indicating
rotational velocity changes caused by a toroidal magnetic torque 
across a shock.
Applying jet parameters such as Mach numbers of 4, plasma-$\beta = 1$, 
and a helical field configuration $B_{\phi} \simeq B_{\rm p}$ the rotational velocities
obtained were rather small $v_{\phi} < 0.1 c_{\rm s}$.

\citet{uchi92,todo92} present MHD simulations in 1.5D and 2.5D which in particular 
compare 1.5D simulations in cylindrical coordinates to analytical solutions of steady 
MHD shocks.
In their case, the rotational velocities obtained for $M_{\rm A} = 1.0, 2.5, 6.0$
simulations are comparatively high - in fact up to 10-40
speed, and, thus, not realistic compared to the observed values.
In fact, in a forthcoming paper on 3D simulations \citet{todo93} discuss such 
high azimuthal velocities as foremost indicator of a jet instability.

\citet{rose99} investigate jet stability under the influence of helical 
magnetic fields, but do not discuss jet rotation in particular.
Simulations by \citet{ston00} consider both steady and pulsed MHD jets 
with helical magnetic field, in particular their cooling and stability 
properties, but again do not stress jet rotation.

\subsection{Angular momentum budget}
A basic prerequisite for the applicability of the process discussed
in this paper is a sufficient amount of magnetic angular momentum 
available, which can be converted within the shock region.
We may estimate the angular momentum budget from numerical models of 
the jet formation region which provide solutions to the MHD equations
along the accelerating and expanding jet.

The amount of magnetic angular momentum flux $L_{\rm mag}$ in respect 
to the kinetic angular momentum flux $L_{\rm kin}$ depends crucially 
on the MHD state of the flow.
Analytical estimates \citep{blan82,pell92} show that for sub-Alfv\'enic
flows $L_{\rm mag} >> L_{\rm kin}$, while for marginally super-fast
flows $L_{\rm mag} \simeq 2 L_{\rm kin}$.
For a (poloidal) fast magnetosonic Mach number $M_{\rm FM} = 2$
the kinetic angular momentum flux dominates,
$L_{\rm mag}/L_{\rm kin} \simeq 2 /M_{\rm FM}^2 = 0.5$.

Numerical simulations of MHD jet formation from Keplerian disks show,
however, that typically only the inner jet 
(the part close to the jet axis) becomes super-fast magnetosonic
\citep{ouye97,kras99,fend02,fend06,port10}. 
This suggests an asymptotic jet structure consisting of a narrow core 
dominated by kinetic angular momentum, which is wrapped by an envelope 
dominated by magnetic angular momentum flux.
It is this envelope around the core jet, which i) (shock-) interacts 
with the ambient medium, and ii) is accessible to the observations.
Thus, for the purpose of this paper it is convenient to assume that a substantial
(i.e. sufficient) amount of magnetic angular momentum is present in the
asymptotic jet.

Numerical simulations of MHD jet formation from Keplerian disks also show
quite a variety in the asymptotic jet outflow they produce (and which is subject
to the present paper).
We refer to our past work \citep{fend06} which in very detail investigated
how the launching conditions as the disk magnetic field profile and strength
and also the mass flux distribution affects the asymtotic flow.
In particular the radial profile of the disk wind magnetization (steep or flat)
determines the collimation of the outflow (weakly, resp. strongly collimated).
We therefore decided to scan a rather large parameter range in the typical
jet characteristics as Mach number or plasma beta.

Recently \citet{rams11} have performed jet formation simulations spanning
from the disk surface well into the asymptotic regime, enclosing a huge, yet
unprecedented area of $256 \times 4096$\,AU size.
These simulations show 
i) that a well ordered, global field distribution
remains present also for the most distant outflow regime,
ii) that the asymptotic flows launched with different initial
  plasma-$\beta$ tend to approach asymptotic states with average
  plasma-$\beta$ close to unity (although there is still quite a
  spread during the simulation time scales considered), and 
iii) that flows of different initial plasma-$\beta$ may reach quite
  different asymptotic states concerning mass fluxes of velocities.

In summery, we decided to run our simulations for a variety of leading
jet parameters, considering also the fact that jets from protostars and
AGN could be intrinsically different.

\section{Model setup for the MHD simulations}
In order to quantify the toroidal torque in jets and the subsequent
toroidal velocity structure, we perform MHD simulations of jet
propagation applying different 
flow geometries, 
magnetic field distributions, and 
jet dynamical parameters.
In order to demonstrate the efficiency of the proposed mechanism, we inject
the jet material non-rotating, thus with vanishing kinetic angular momentum.

We use the PLUTO code \citep{mign07} to solve the standard set of ideal MHD equations.
We apply an ideal equation of state with polytropic index $\gamma =5/3$. 
We also discuss one example simulation for which cooling is considered.

In order to re-scale normalized code variables to astrophysical magnitudes,
one may apply typical numbers values for protostellar jets or AGN jets as 
shown in Tab.~\ref{tab:parajets}.

\begin{table}
\begin{center}
\caption{Typical astrophysical scaling parameters for jets from young stars and AGN}
\label{tab:parajets}
\begin{tabular}{ccc}
\noalign{\medskip} 
\tableline\tableline
\noalign{\smallskip} 
               & Stellar jets    & AGN jets      \\
\noalign{\smallskip}
\tableline
\noalign{\smallskip}
$v_{\rm jet}$ & $\simeq 300$\,km/s  & $\simeq 0.3 c $          \\
\noalign{\smallskip}
$\rho_{\rm jet}$ & $\simeq 10^4{\rm cm^{-3}}$  & $\simeq10^{-4}{\rm cm^{-3}}$          \\
\noalign{\smallskip}
$B_{\rm p,jet}$ & $10-100\mu$G  & 10-100\,G   \\
\noalign{\smallskip}
\tableline
\end{tabular}
\end{center}
\end{table}

\subsection{Simulations in 1.5D}
With "1.5D" we denote axisymmetric simulations taking into account 
all three spatial coordinates of the vector fields, 
however, putting one of the vector field component to zero
initially\footnote{That is the $r$-component in cylindrical coordinates.
We have also performed comparison simulations in Cartesian coordinates
with an initially vanishing $x$-component. 
For both cases we have approved that these components remain negligible 
during the simulation, i.e. $<10^{-14}$.}. 
We also limit the grid extension to only one cell in this direction.

The general setup is that of a typical MHD shock tube test simulation.
High-density/high pressure gas of high velocity is continuously injected 
from the left-hand side of the box (considered as the jet flow). 
Low-density/low pressure material is placed at the right-hand side of the 
box (considered as the ambient gas) and is allowed to move out of the box.
With this setup we extend previously existing simulations \citep{uchi92, ryu95a}
to the specific question of jet rotation and for parameter studies of that.
 
We apply different magnetic field configurations. 
For the simulations in cylindrical coordinates we assume an initial axial field 
$B_z(z)$ which is constant along the grid. 
The initial toroidal field $B_{\phi} (z)$ is either only present on the left 
side (upstream) or constant along the grid.
For comparison with the literature and in order to test the 1.5D approach in
cylindrical coordinates, we have computed the same problem also in Cartesian 
coordinates and have found perfect agreement (see appendix).

The numerical resolution varies from 5000 - 20000 grid cells for 20 physical 
scale lengths.

\subsection{Simulations in 2.5D}
With "2.5D" we denote axisymmetric simulations taking into account 
all three vector components.
By definition the simulations sustain axisymmetry as the $\phi$-derivatives
are not considered in this option of the code.
A jet of unity radius is injected into a box of ambient gas with
constant density, pressure, and longitudinal magnetic field.
We apply a non-equidistant grid with high resolution across the jet beam and the 
interaction region with the ambient gas ($r<1.5$, 100 cells), 
and a somewhat lower resolution in the ambient gas ($1.5<r<7.0$, 200 cells).

It would be interesting to extend the present investigation to the 3D case
in a future paper.
Although the principle mechanism proposed here will be the same in the 3D case,
the overall jet structure could be different and also the characteristic instabilities
affecting the jet flow.
In particular large-scale instabilities as the $m=1$ kink mode could appear
and disturb the jet flow. 
Such instabilities would develop beyond the Alfv\'en surface (thus in the super-Alf\'enic
flow), and may destroy the large-scale helical field structure \citep{moll09}.
This is a serious argument, however, it affects the standard model of jet rotation
(e.g. \citet{ande03}) even more.
On the other hand 3D simulations of supersonic  have revealed a quite similar 
general structure compared to the 2D case \citep{onei05}.
Also, 3D simulations of jet formation by \citet{ouye03} revealed a self-stabilizing
ability of the jet due to its axial spine.
We finally note the fact that we {\em do} observe large-scale collimated outflows 
which tells us that jet stability must be maintained by some (yet unknown) process.

\subsection{Main simulation parameters}
Our simulations are characterized by the following set of parameters.

%
\begin{table*}
\scriptsize
\begin{center}
\caption{Parameter of axisymmetric 1.5D simulations. 
 The external density is ${\rho_{\rm ext}}=1.0$ for all runs.
 The toroidal velocity is given before and after the shock front $v_{\phi,1},v_{\phi,2}$.
 Remarks: UT05 follows the same parameter setup as UT04, but considers
 cooling. UT06 follows the same parameter setup as UT04, except for the 
 negative rotation velocity.
 For comparison with Figs.~\ref{fig:sim_rj17},\ref{fig:sim_rj18} 
 note the PLUTO magnetic field normalization $B \rightarrow B/\sqrt{4\pi}$.
\label{tab:para-1da}
}
\begin{tabular}{ccccccccccccccc}
\tableline\tableline
\noalign{\smallskip} 
 ID     & 
 ${\rho_{\rm jet}}$      & 
 $v_{\rm p, jet}$        &
 $v_{\rm \phi, jet}$     &
 $P_{\rm jet}$           &
 $P_{\rm ext}$           &
 $B_{z}$                 &
 $B_{\phi, \rm jet}$     &
 $B_{\phi, \rm ext}$     &
 $M_{\rm Ap, jet}$       & 
 $M_{\rm A, jet}$        & 
 $M_{\rm F, jet}$        & 
 $M_{\rm S, ext}$        & 
 $v_{\phi,1},v_{\phi,2}$ \\
\noalign{\smallskip}
\tableline
\noalign{\medskip}
RJ03 & 2.5 & 1.0 & 0.0 & 1.53 & 0.27 & 0.11 & 1.70 & 0.0 & 50.0 & 3.5 & 0.88 & 1.5 & 0.0/-0.163 \\
RJ04 & 2.4 & 1.0 & 0.0 & 0.32 & 0.27 & 0.11 & 1.56 & 0.0 & 50.0 & 3.5 & 1.82 & 1.5 & 0.005/-0.235 \\
RJ06 & 2.4 & 1.0 & 0.0 &0.026 & 0.27 & 0.18 & 1.00 & 0.0 & 30.0 & 5.5 & 4.42 & 1.5 & 0.017/-0.22 \\

\noalign{\medskip}
RJ17 & 1.08 & 1.2 & 0.0 & 0.95 & 1.0 & 2.0 & 2.0 & 2.0 & 2.21 & 1.56 & 0.84 & 0.93 & +0.091/-0.088 \\
RJ18 & 1.08 & 1.2 & 0.0 & 0.95 & 1.0 & 2.0 & 2.0 & 0.0 & 2.21 & 1.56 & 0.84 & 0.93 & +0.084/-0.31 \\
RJ19 & 1.28 & 1.5 & 0.0 & 1.15 & 1.0 & 2.0 & 2.0 & 0.0 & 3.01 & 2.13 & 1.06 & 1.16 & +0.078/-0.29 \\

\noalign{\medskip}

RJ21 & 1.28 & 1.2 &-0.2 & 0.95 & 1.0 & 2.0 & 2.0 & 0.0 & 2.41 & 1.70 & 1.10 & 0.93 & -0.12/ -0.41 \\
RJ22 & 1.28 & 1.2 & 0.0 & 0.95 & 1.0 & 0.5 & 2.0 & 0.0 & 9.63 & 2.34 & 0.98 & 0.93 & +0.019/-0.33 \\
RJ23 & 1.28 & 1.2 & 0.0 & 0.95 & 1.0 & 2.0 & 0.5 & 0.0 & 2.41 & 2.34 & 0.98 & 0.93 & +0.029/-0.078 \\

\noalign{\medskip}
UT01 & 1.0 & 4.0 & 0.0 & 0.1 & 0.01 & 4.0 & 2.4 & 2.4 & 3.55 & 3.04 & 2.90 & 31.0 & +0.7/-0.7 \\
UT02 & 1.0 & 4.0 & 0.0 & 0.1 & 0.01 & 4.0 & 2.4 & 0.0 & 3.55 & 3.04 & 2.90 & 31.0 & +0.7/-0.4  \\
UT03 & 1.0 & 4.0 & 0.0 & 0.1 & 0.01 & 1.0 & 2.4 & 0.0 & 14.2 & 5.44 & 4.76 & 31.0 & +0.14/-0.54 \\
\noalign{\medskip}
UT04 & 1.0 & 4.0 & 0.1 & 0.1 & 0.01 & 1.0 & 2.4 & 0.0 & 14.2 & 5.44 & 4.76 & 31.0 & +0.24/-0.48 \\
UT05 & 1.0 & 4.0 & 0.1 & 0.1 & 0.01 & 1.0 & 2.4 & 0.0 & 14.2 & 5.44 & 4.76 & 31.0 & +0.25/-0.46 \\
UT06 & 1.0 & 4.0 &-0.1 & 0.1 & 0.01 & 1.0 & 2.4 & 0.0 & 14.2 & 5.44 & 4.76 & 31.0 & +0.043/-0.58 \\

\noalign{\medskip}
\tableline
\end{tabular}
\end{center}
\end{table*}

We apply different jet densities $\rho_{\rm jet}$, while the initial
ambient density is always the same, $\rho_{\rm ext} = 1.0$.
The jet inflow velocity is $v_{\rm jet}$. 
In the case of 2.5D simulations this is the maximum jet velocity as we
prescribe a cosine profile for the injection speed across the jet inlet.
The ambient gas is initially at rest.
The jet gas pressure is $P_{\rm jet}$ and the ambient gas pressure is $P_{\rm ext}$.
The strength of the magnetic field is parameterized by
the (squared) jet internal poloidal Alfv\'en Mach number, 
$$
M_{\rm Ap,jet}^2 \equiv \frac{4\pi \rho_{\rm jet} v_{\rm p,jet}^2}{B_{\rm p,jet}^2},
$$
%
%
and the (squared) jet internal total Alfv\'en Mach number,
$$
M_{\rm A,jet} \equiv
\frac{4\pi\rho_{\rm jet} v_{\rm p,jet}^2}{B_{\rm p,jet}^2 + B_{\phi,\rm jet}^2}.
$$
With the (squared) jet internal sound speed 
$c_{\rm s,jet}^2 = \gamma P_{\rm jet} / \rho_{\rm jet}$
and Mach number $M_{\rm S,jet} \equiv v_{\rm p,jet}/c_{\rm s,jet}$
the jet internal fast magnetosonic Mach number is
$$
M_{\rm F,jet} \equiv \left(M_{\rm A,jet}^{-2} + M_{\rm S,jet}^{-2}\right)^{-1/2}.
$$
The external Mach number relates the poloidal jet velocity to the sound 
speed in the external medium, 
$M_{\rm S,ext} \equiv v_{\rm p,jet}/c_{\rm s,ext}$.

Note that the Mach numbers are defined in respect to the poloidal velocity, 
considering MHD waves in the jet propagation direction.
Tables \ref{tab:para-1da},
       \ref{tab:para-1db} summarizes the parameter setup for the 1.5D simulations, 
whereas 
  Tab.~\ref{tab:para-2d} shows the parameter space for the 2.5D simulations.

The initial magnetic field $B_{\rm z}$-component is constant over the whole
computational domain.
We run simulations for which the toroidal field component $B_{\phi}$ is injected into
the ambient gas by the jet and others where a shock propagates along an already
existing  helical field structure.
For the jet toroidal field we have applied a linearly increasing radial profile
and a sine-profile. While the linear profile is force-free within the jet, we find 
the the sharp cut-off at the jet boundary leads to current sheets which disturb
the flow ab initio and could not be accepted. We therefore discuss the simulations
applying the (non force-free) sine profile only.

\section{Results and discussion}
%
We first discuss simulations resulting from a 1.5D setup.
These can be considered as toy simulations of axisymmetric shocks 
in a helical field and test cases of the model.
We then continue with 2.5D axisymmetric simulations of jet propagation 
with parameter ranges more comparable to observational values.

\subsection{Jet shock propagation in 1.5D}
Quite some literature exists on one-dimensional MHD shock simulations, 
as this setup provides one of the standard tests for MHD codes. 
In particular we refer to \citet{uchi92, ryu95a}.
We decided to repeat and extend a few of these 1.5D simulations for the
following reasons.
Firstly, they serve us as test case for our code for the 1.5D setup in 
cylindrical coordinates as we can directly compare them to simulations
in the literature in Cartesian coordinates which are widely used also
for analytic considerations.
Secondly, this simple setup will most clearly demonstrate the essential 
mechanism leading to a toroidal jet acceleration across a shock.
Thirdly, it allows us easily to explore the typical parameter space for
protostellar jets 
(see Tab.~\ref{tab:para-1da}).
In the majority of our simulations the jet is injected as non-rotating.
This simplification is made in order to demonstrate the acceleration
mechanism by the shock, and is in general only valid for a certain
iso-rotation parameter $\Omega_{\rm F} = (v_{\rm p} B_{\phi}) / (B_{\rm p} r) $.

We have performed several test simulations, of which two are shown in 
the appendix.  
That is simulation RJ11 repeating the simulation by \citet{ryu95a} (see their Fig.~2a),
done as well in Cartesian coordinates (Fig.~\ref{fig:ryu11}).
Simulation RJ09 is an extension of this parameter setup using however cylindrical
coordinates (Fig.~\ref{fig:ryu09}) and considering only one perpendicular 
vector component (no radial magnetic field, no radial motion).

\begin{figure*}
\centering
\includegraphics[width=6cm]{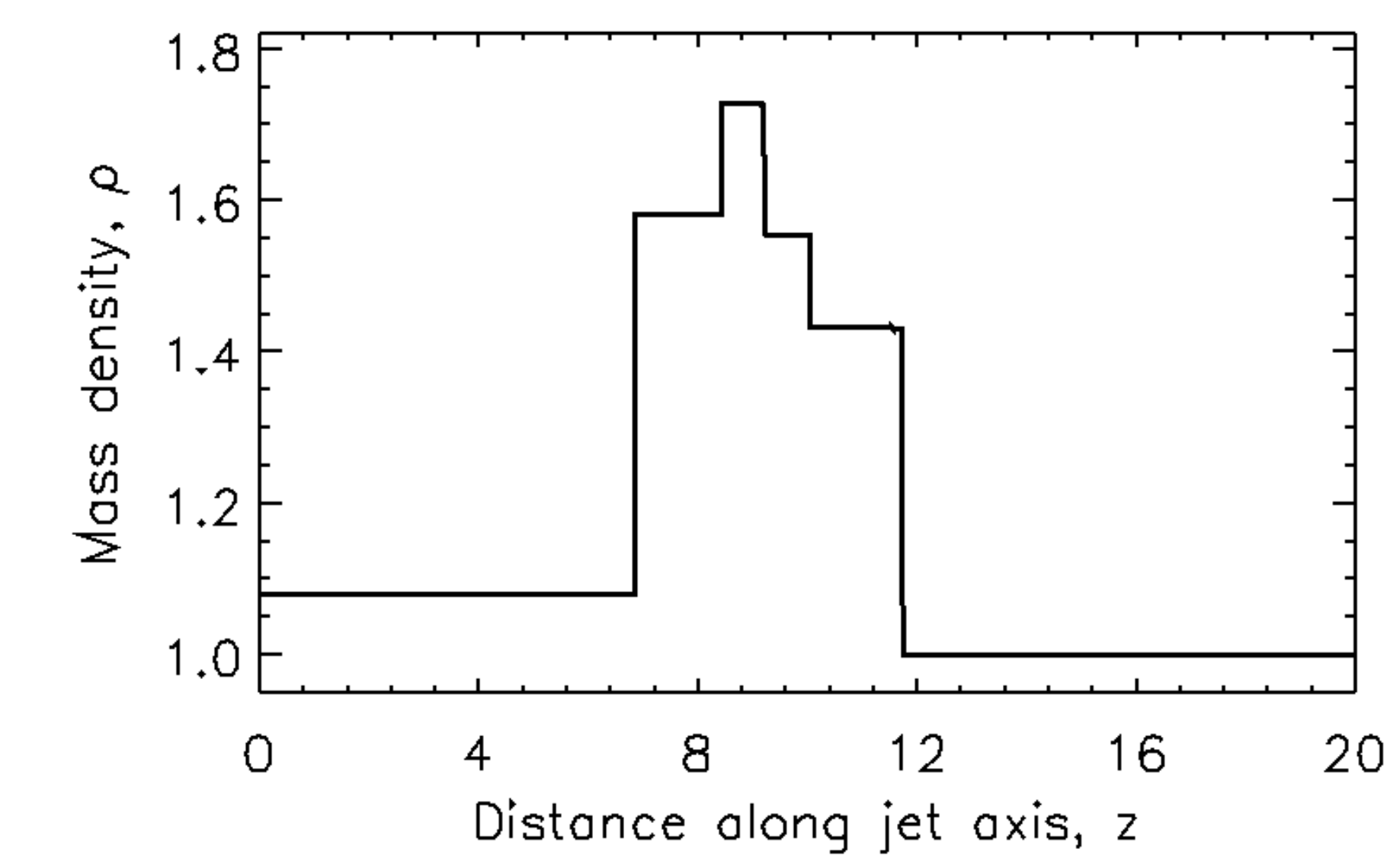}    
\includegraphics[width=6cm]{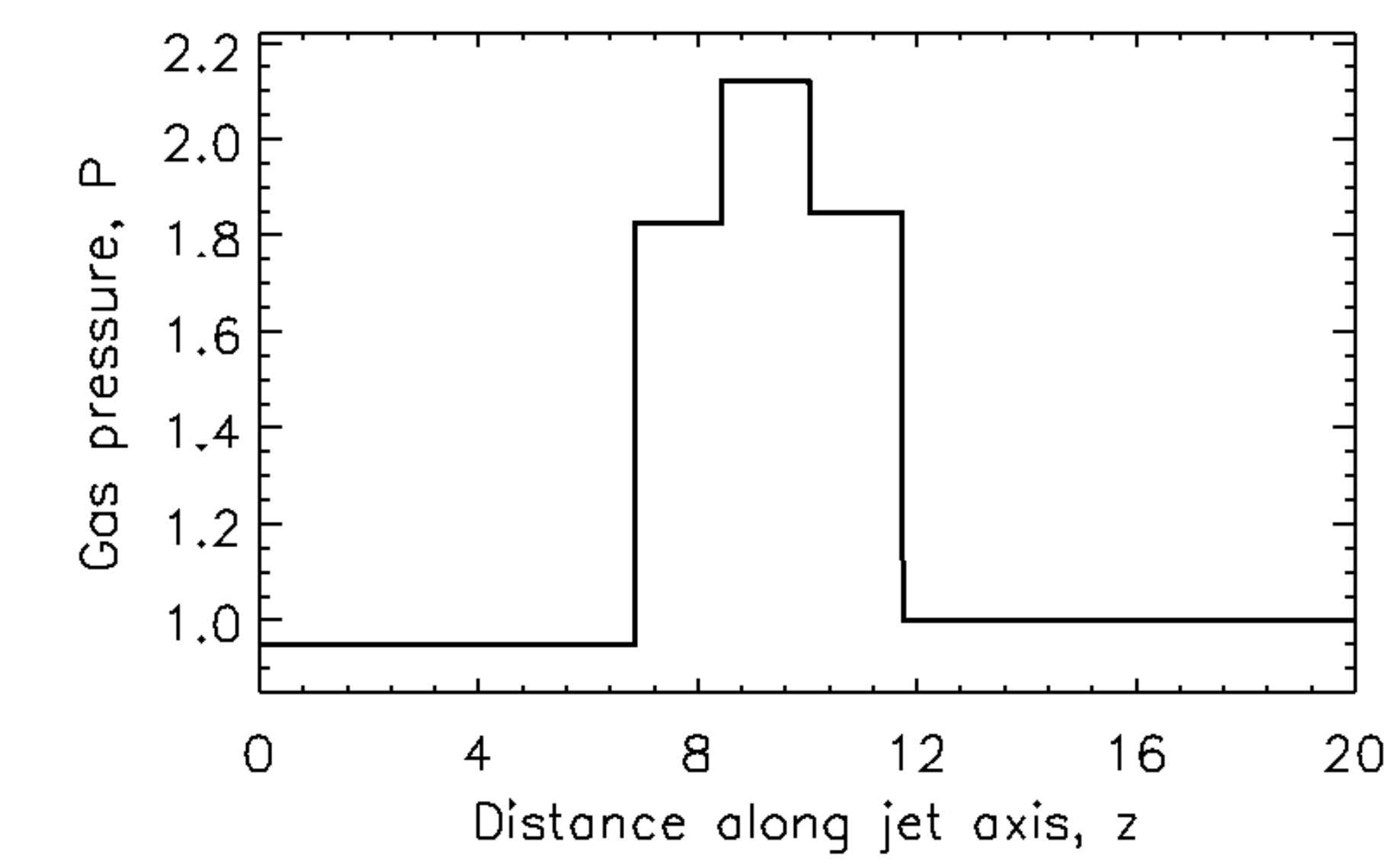}\\  
\includegraphics[width=6cm]{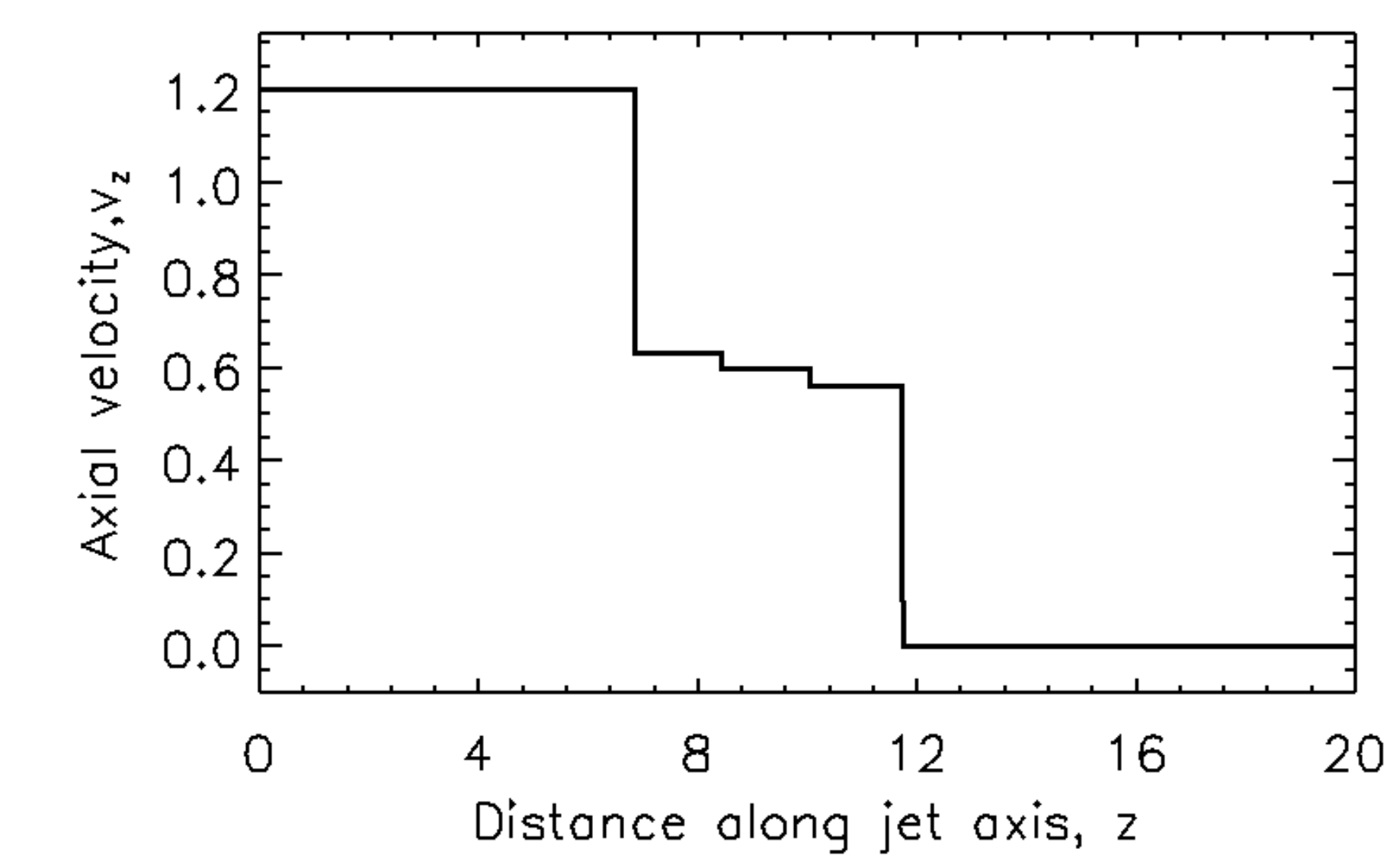}    
\includegraphics[width=6cm]{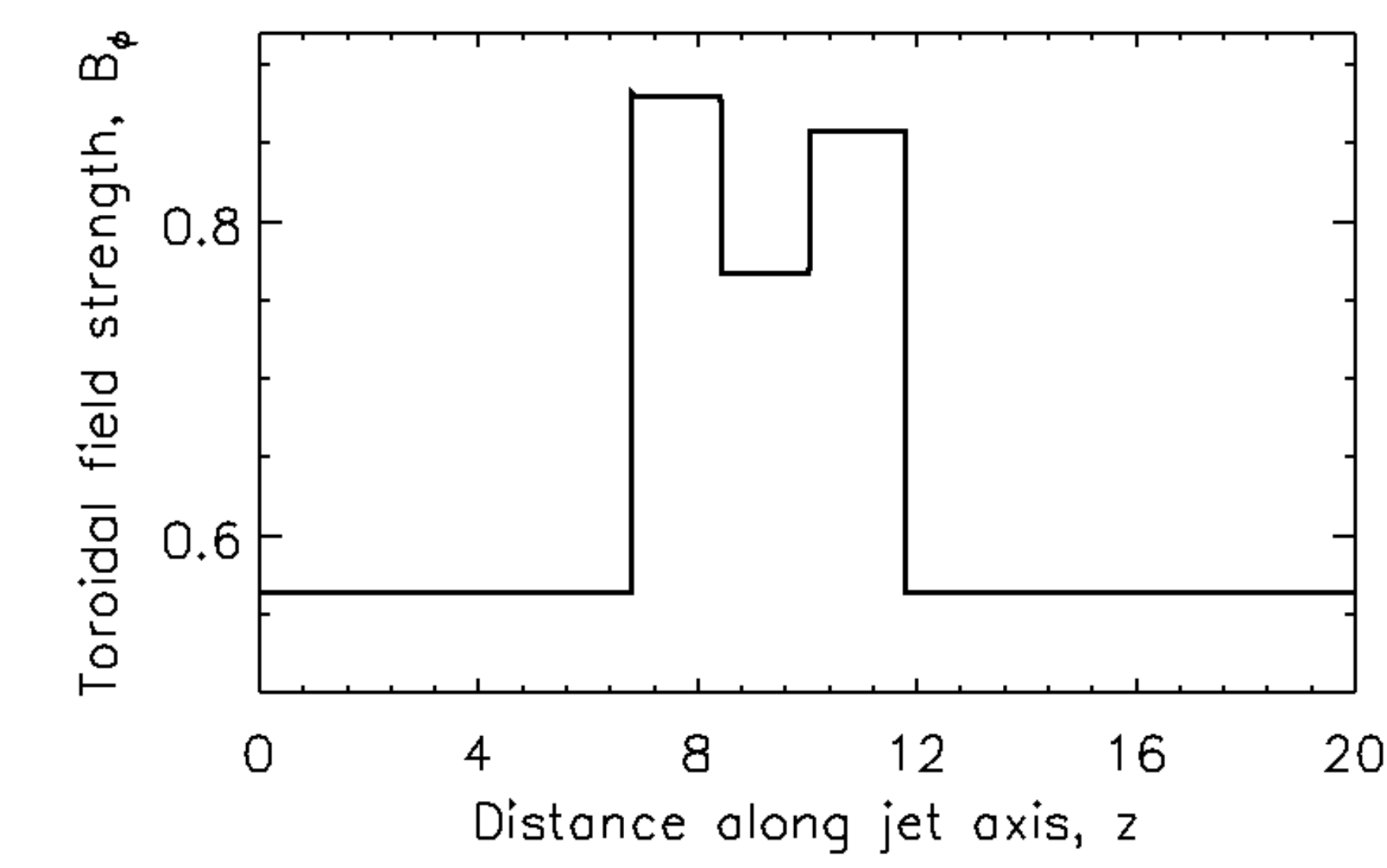}\\  
\includegraphics[width=6cm]{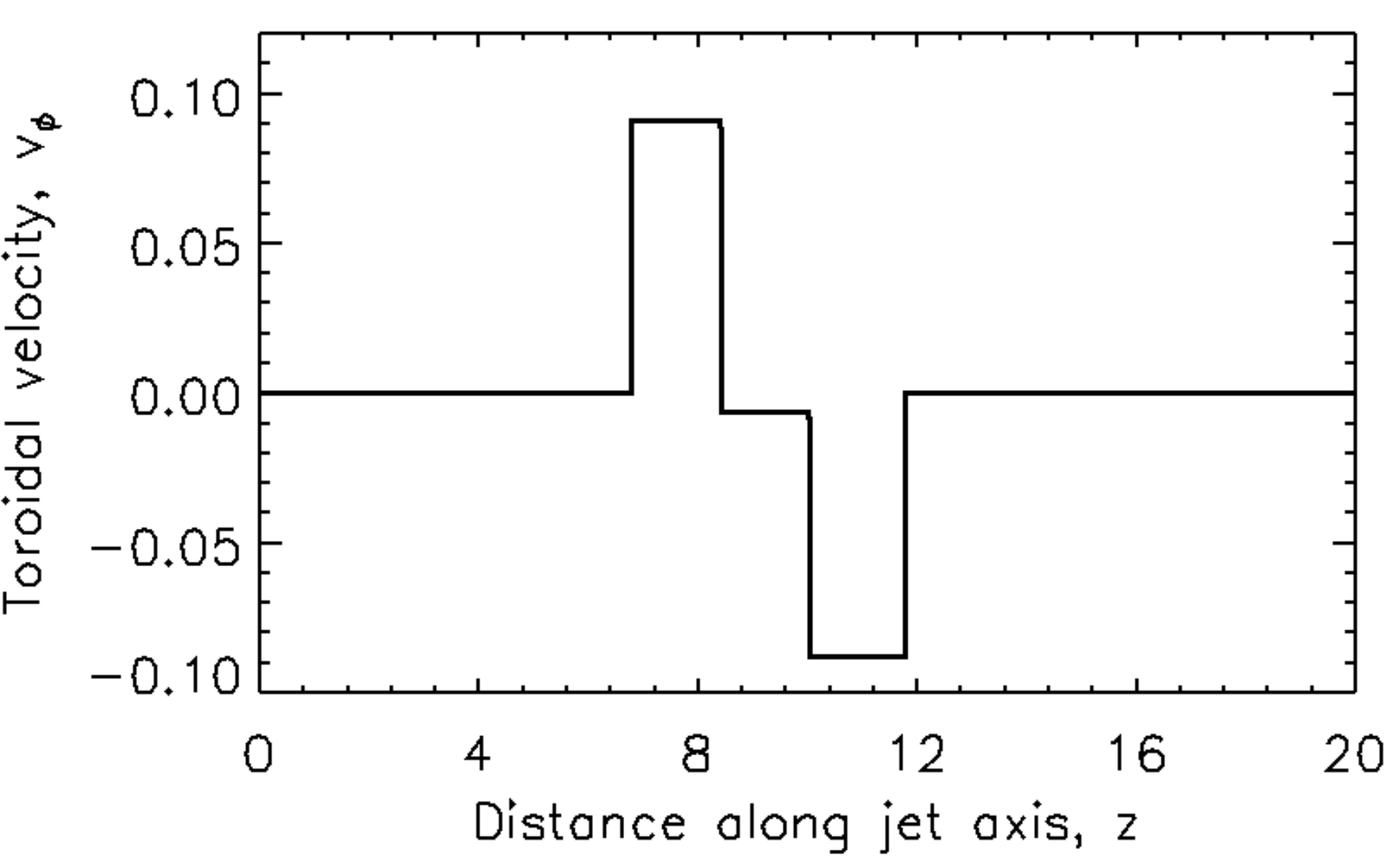}    
\includegraphics[width=6cm]{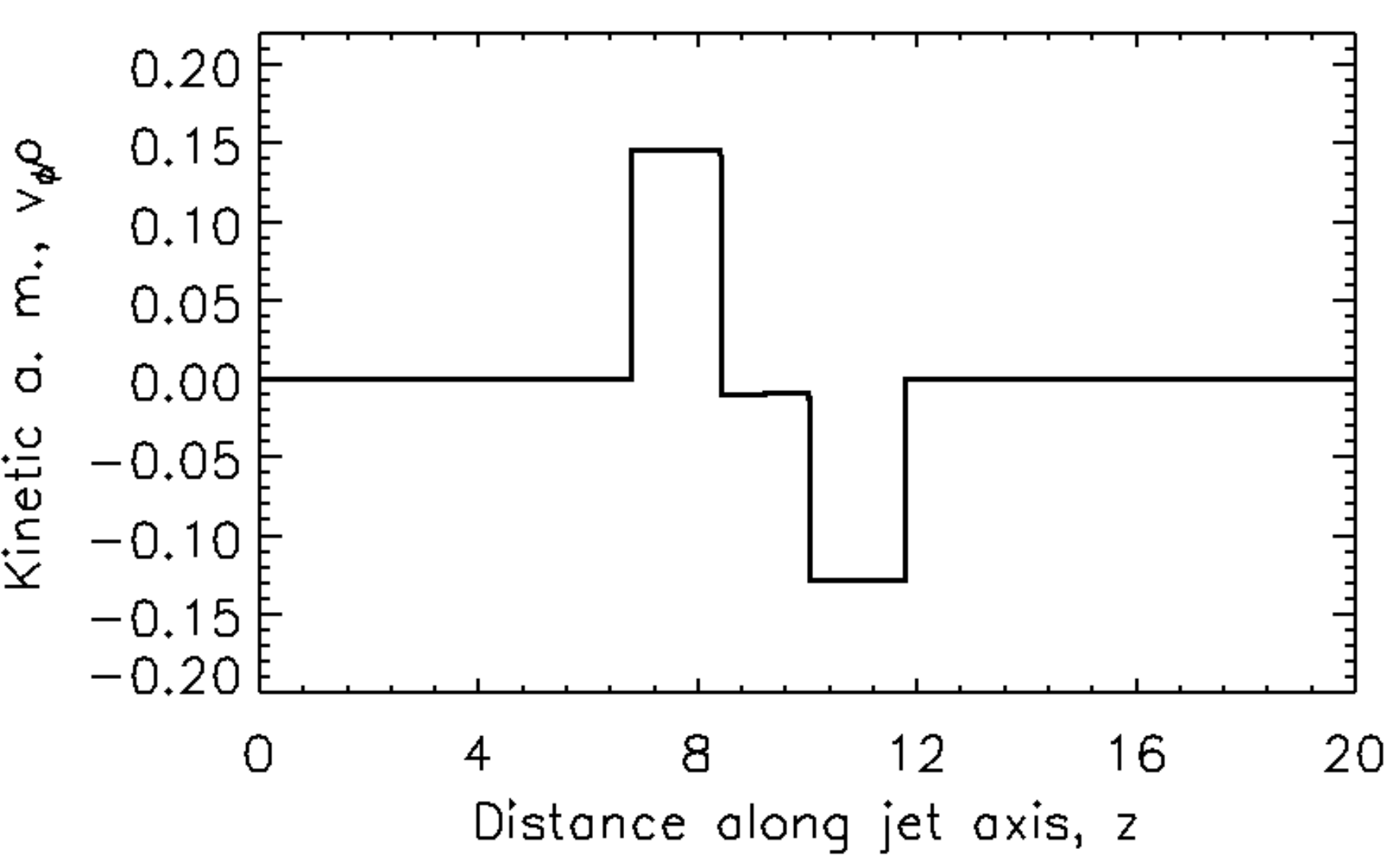}    
\caption{Axisymmetric 1.5D shock simulation RJ17 in cylindrical coordinates.
 Shock evolution at dynamical time $t=2$ with a grid resolution 
 of 1.000 cells / unit length.
 Shown is density; gas pressure; axial, and toroidal velocities;
 axial, and toroidal magnetic field strengths, and the kinetic angular 
 momentum (at $r=1$) (from top left to bottom right). 
\label{fig:sim_rj17}
}
\end{figure*}

Our first science example is a shock moving along a helical magnetic field (run RJ17), 
similar to the setup of \citet{ryu95a}.
This is comparable to a jet internal shock propagating within the jet channel, 
caused e.g. by a sudden increase of jet injection velocity.
Figure \ref{fig:sim_rj17} shows the dynamical state after two dynamical time scales.
The typical shock structure is visible, consisting of four shocks which propagate back
resp. forth in respect to the contact discontinuity.
The hydrodynamical state of the different shock regions varies,
each satisfying the Rankine-Hugoniot jump conditions.
The toroidal velocity panel in Fig.~\ref{fig:sim_rj17} clearly shows how the 
material is set into rotation when crossing the shock fronts.

Across the shock the toroidal field becomes compressed,
and the steeper field gradient implies a toroidal torque on the shocked 
material.
The material downstream and upstream the shock rotates in opposite
direction, seen also in axial profile of kinetic angular momentum.
This is a consequence of a change of sign in the toroidal Lorentz 
force 
$F_{\phi} \sim B_z \partial B_{\phi}/\partial z$ across the shock.
With increasing time, the shock evolution demonstrates the 
continuous conversion of magnetic angular momentum flux into kinetic 
angular momentum flux 
(not shown here, but see 
Figs.~\ref{fig:sim_am_evol_R01},\ref{fig:sim_am_evol_R12}  
below).
The the domain-integrated kinetic angular momentum increases in time.
Due to the downstream boundary condition for the toroidal magnetic field, 
the net amount of kinetic angular momentum along the flow remains zero:
Implicitly, the outflow conditions for angular momentum is kept the same 
as the inflow condition.

\begin{figure*}
\centering
\includegraphics[width=6cm]{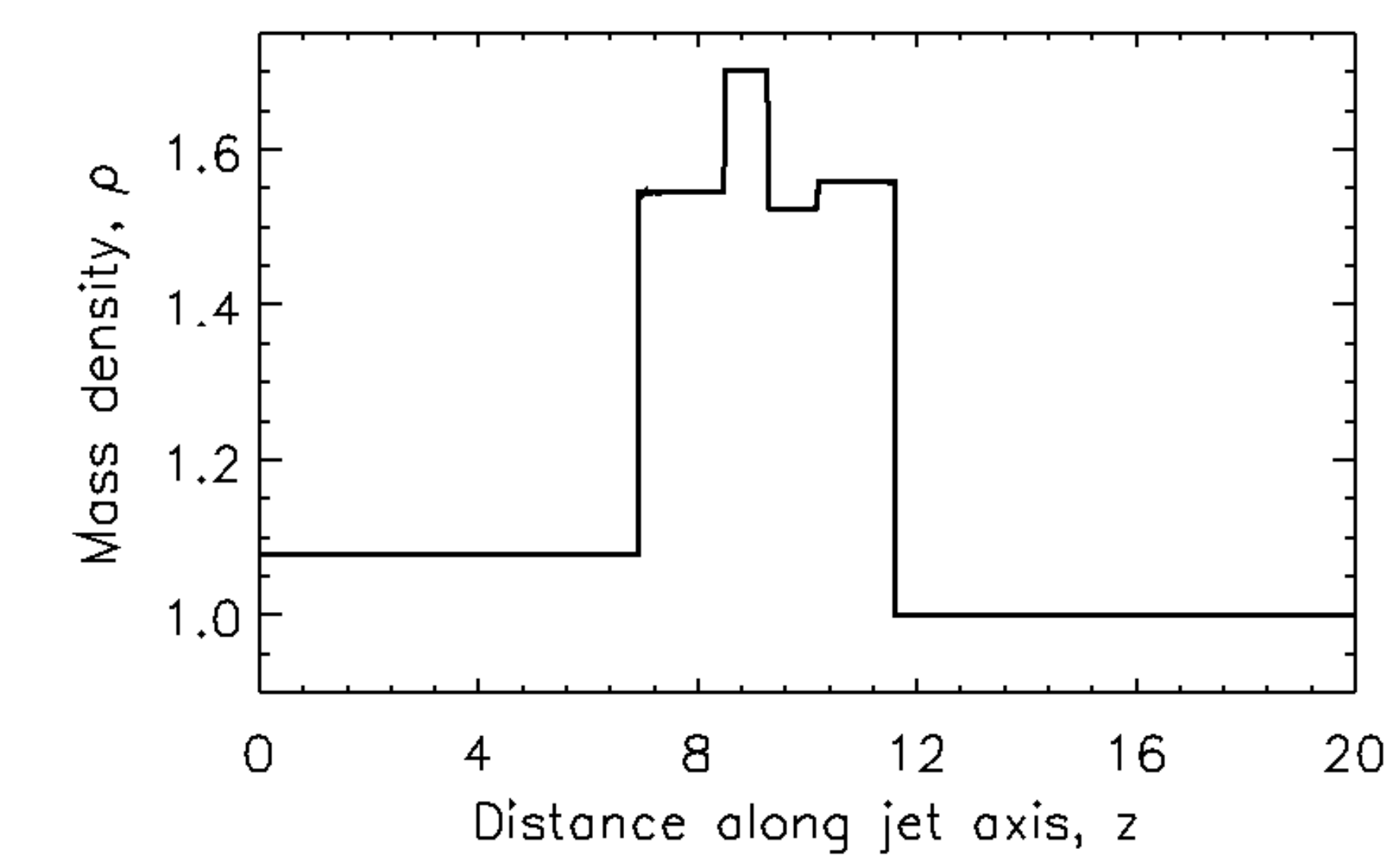}   
\includegraphics[width=6cm]{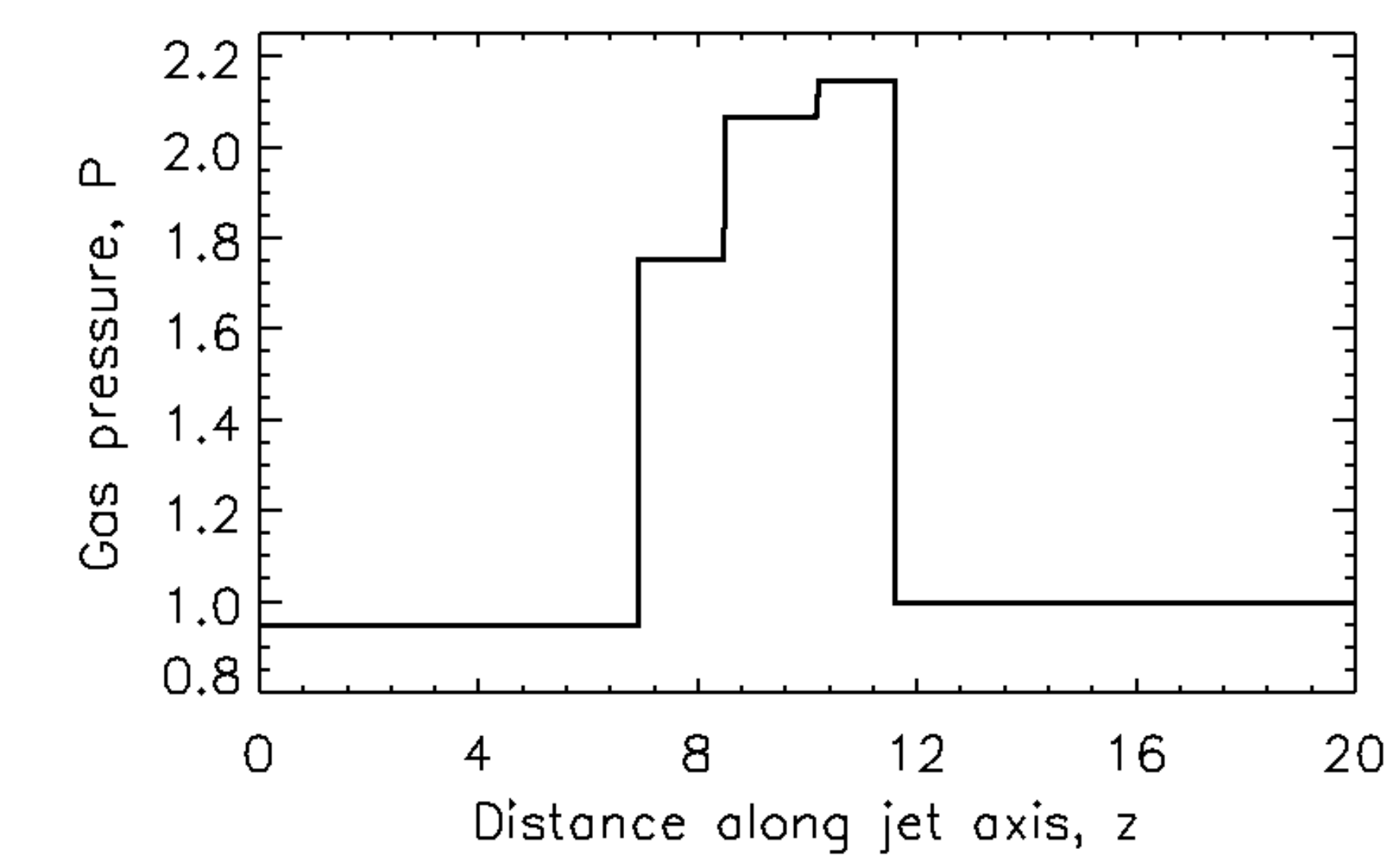}\\ 
\includegraphics[width=6cm]{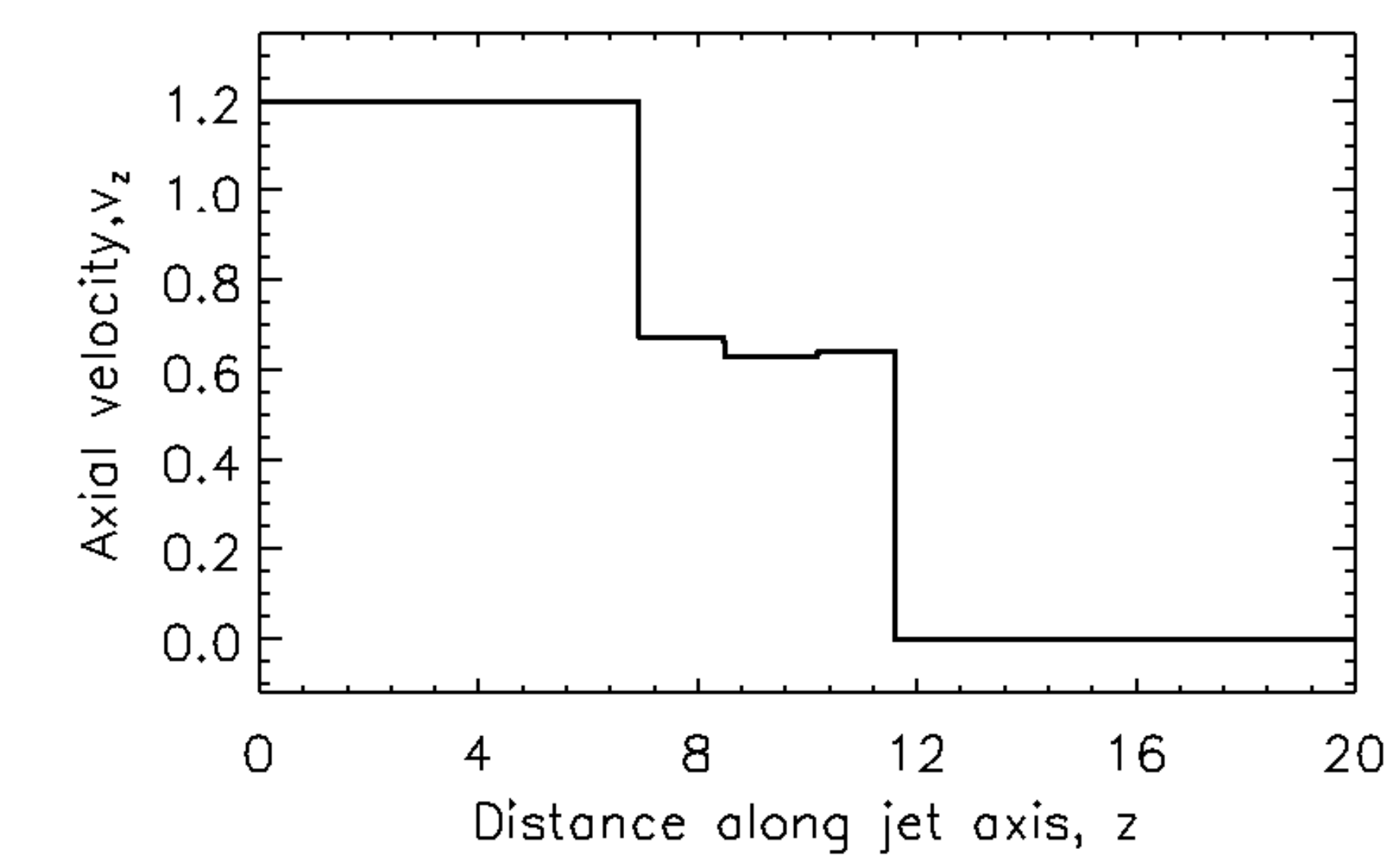}   
\includegraphics[width=6cm]{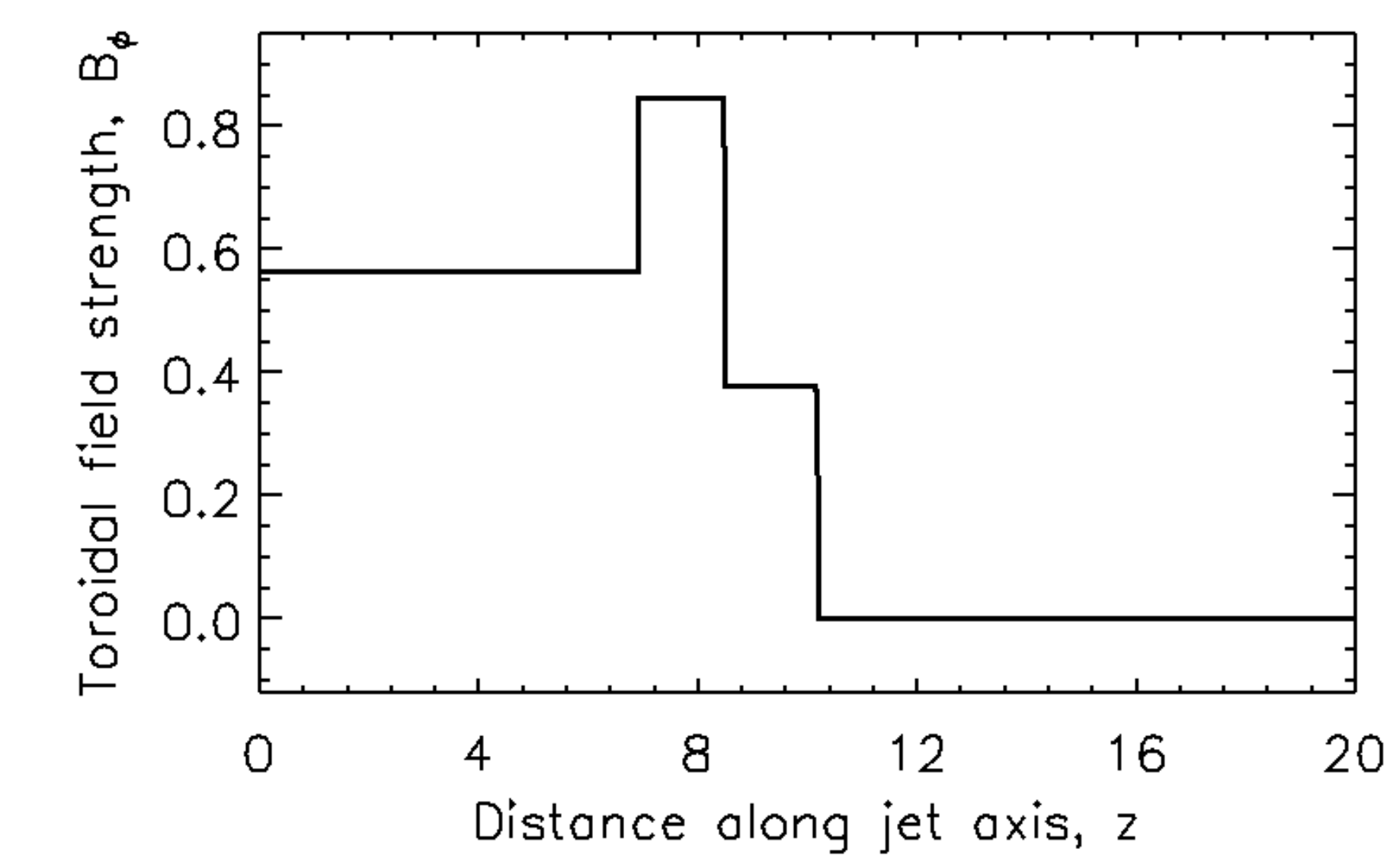}\\ 
\includegraphics[width=6cm]{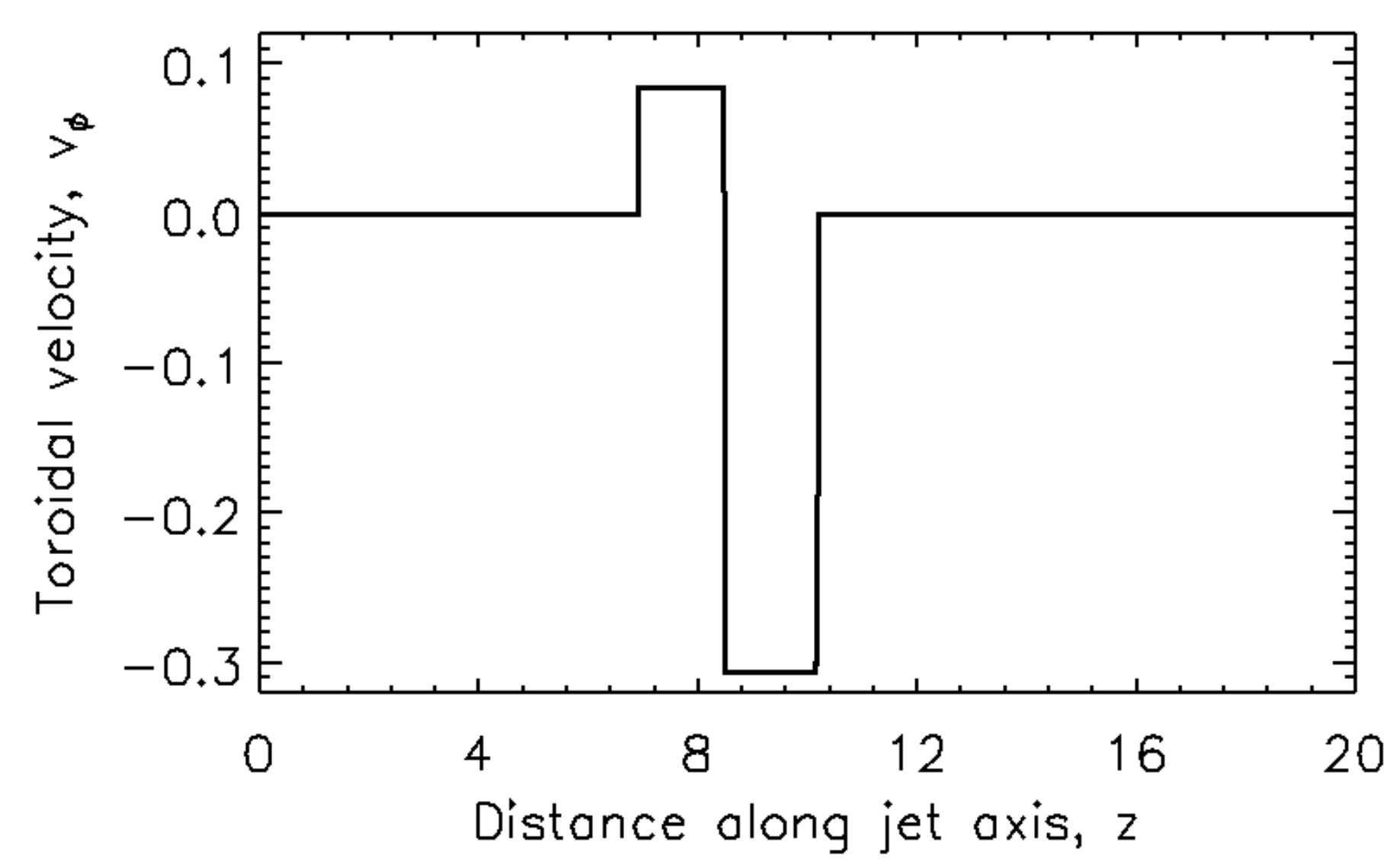}   
\includegraphics[width=6cm]{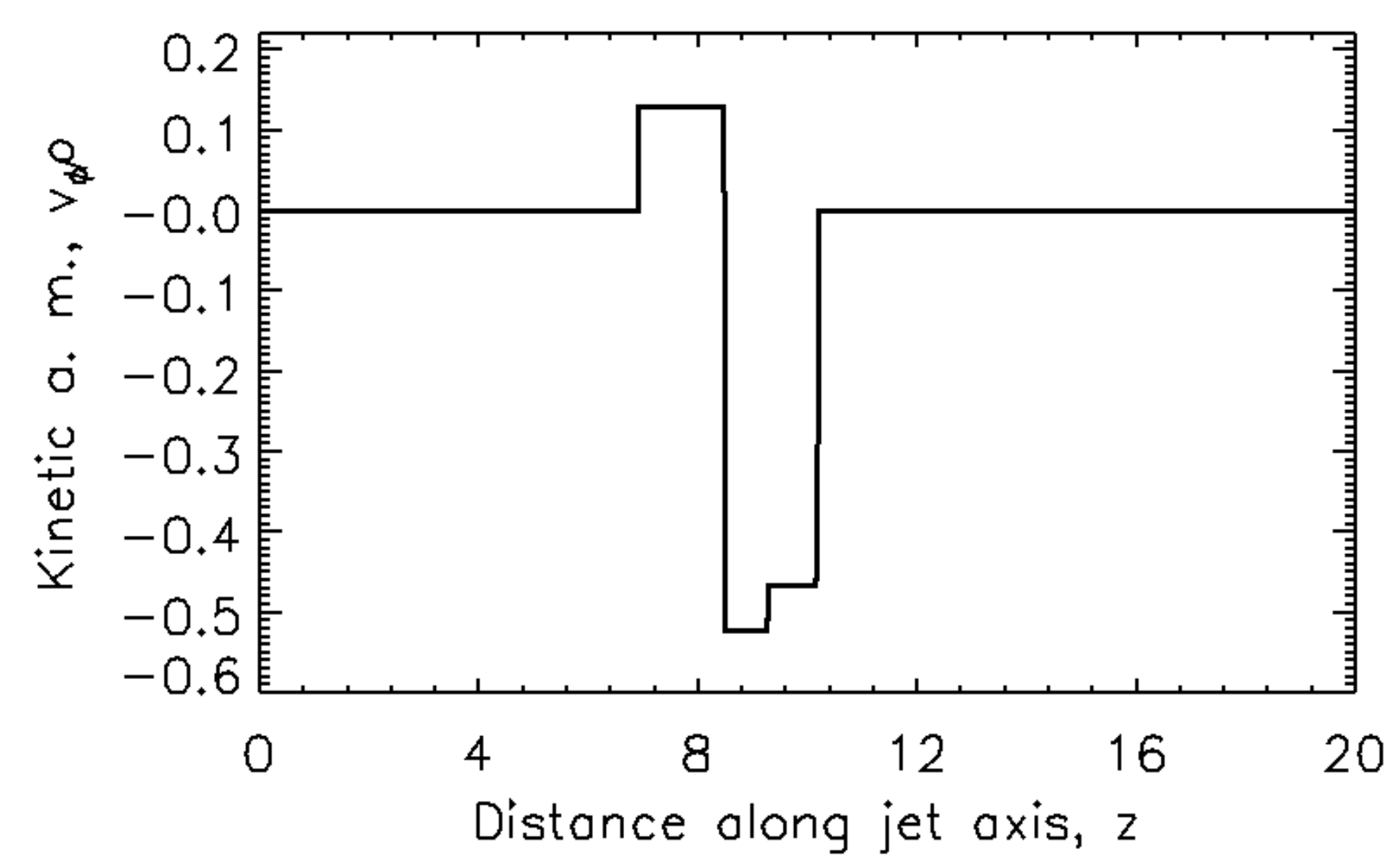}   
\caption{Axisymmetric 1.5D simulation RJ18 in cylindrical coordinates.
 Shock evolution at dynamical time $t=2$ with a grid resolution 
 of 1.000 cells / unit length.
 Shown is density; gas pressure; axial, and toroidal velocities;
 axial, and toroidal magnetic field strengths, and the kinetic angular 
 momentum (at $r=1$) (from top left to bottom right). 
 The toroidal magnetic field in the ambient medium is vanishing (compare to
 simulation RJ17 in Fig.~\ref{fig:sim_rj17}).
\label{fig:sim_rj18}
}
\end{figure*}

Our next example RJ18 applies similar flow parameters, but corresponds 
to a model setup where a jet is injected into an ambient medium threaded by an axial
magnetic field only (the ambient toroidal magnetic field is set to zero).
Figure \ref{fig:sim_rj18} shows the dynamical state at the same time as for RJ17.
The difference comes from the magnetic angular momentum flux which now decreases 
along the flow, while being transferred to kinetic angular momentum.
For the parameters applied 
the toroidal velocities obtained are about 
25\% of the jet velocity, and much higher than the observed values.

Simulations RJ17 and RJ18 exhibit a similar hydrodynamic shock 
structure (considering $v_{\rm jet}, \rho, P$). 
Because of the stronger toroidal field coiling, the MHD torque and, 
thus, the resulting toroidal velocities are increases by a factor 
of three.


\begin{figure}
\centering
\includegraphics[width=7cm]{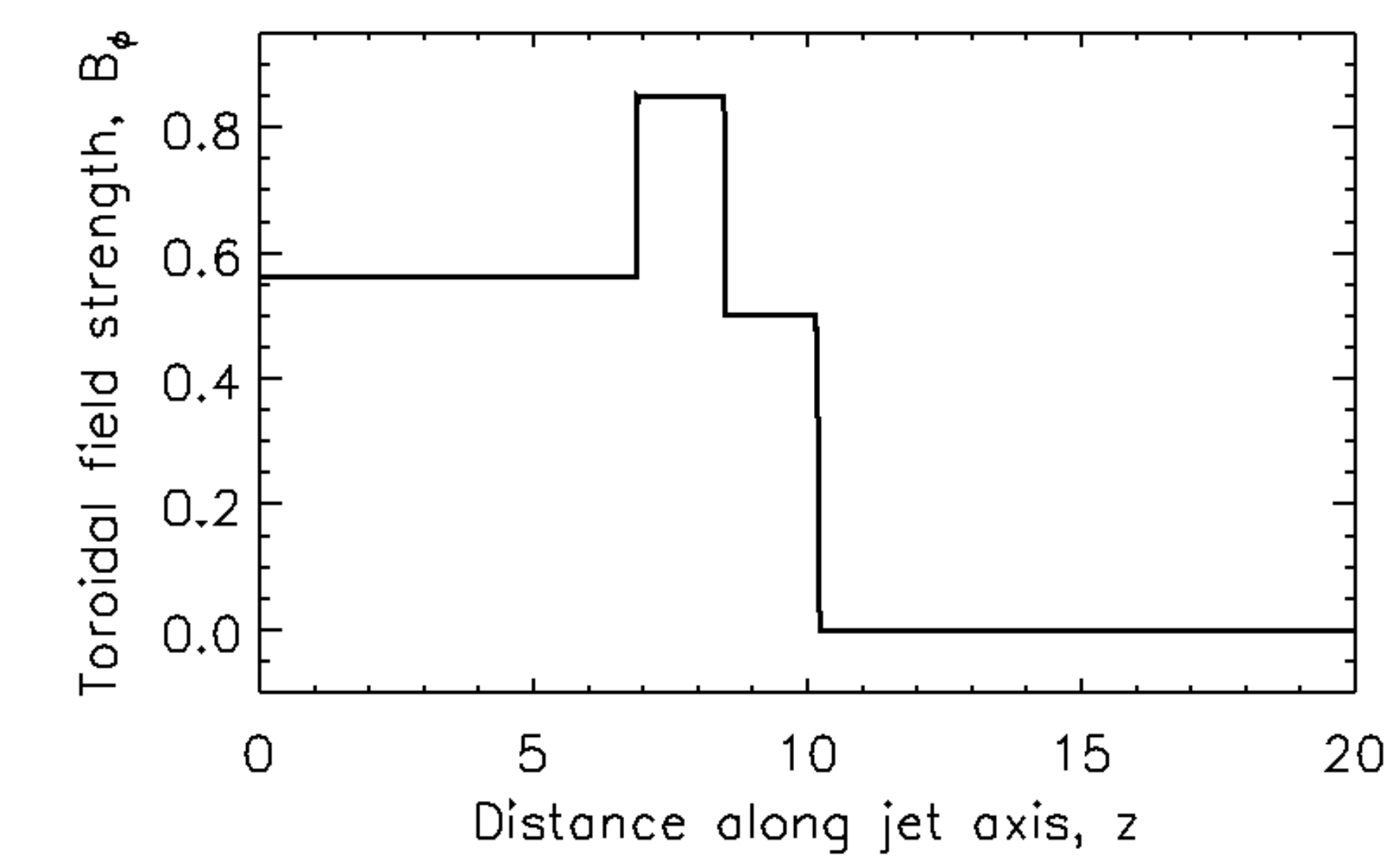}  
\includegraphics[width=7cm]{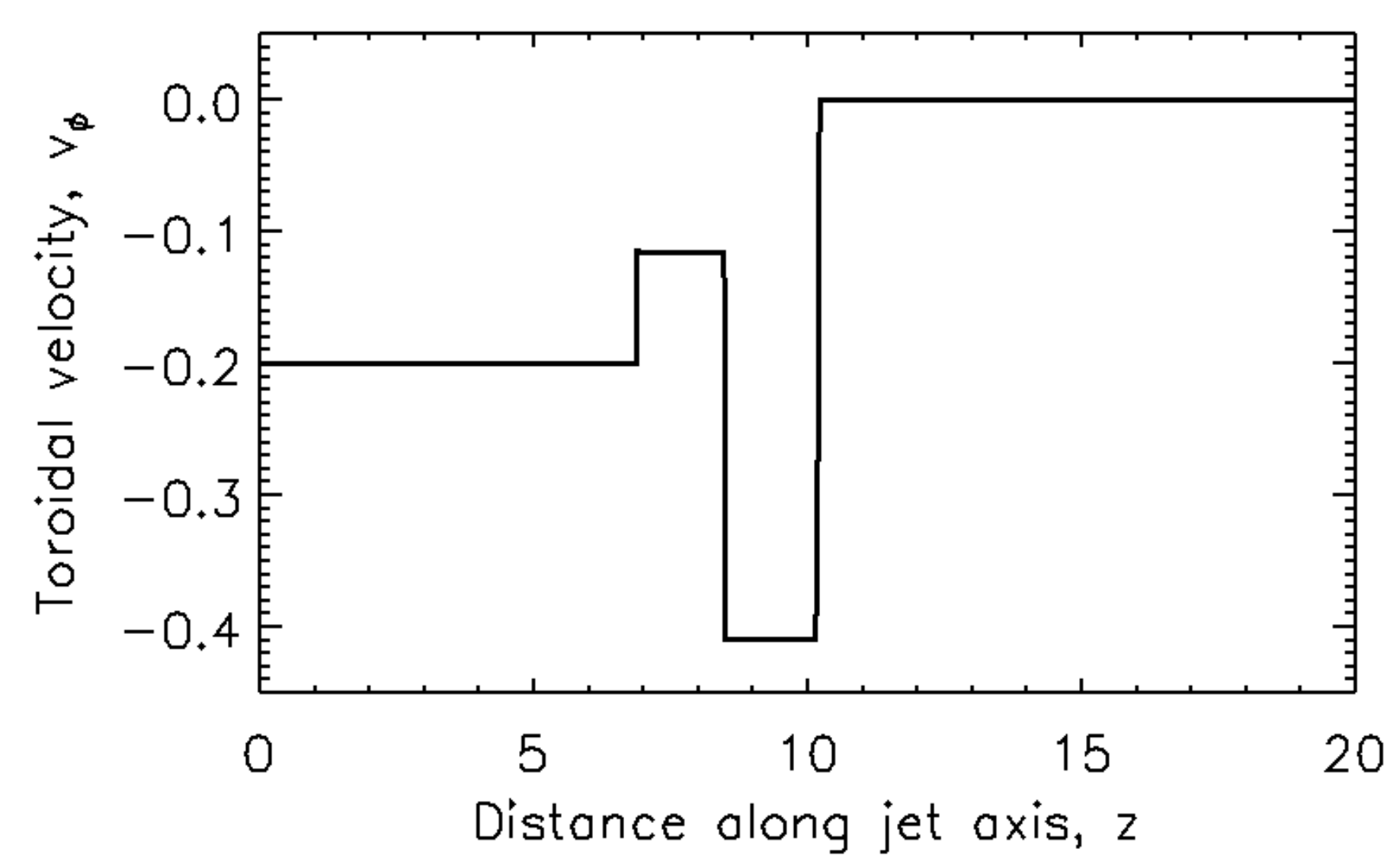}  
\includegraphics[width=7cm]{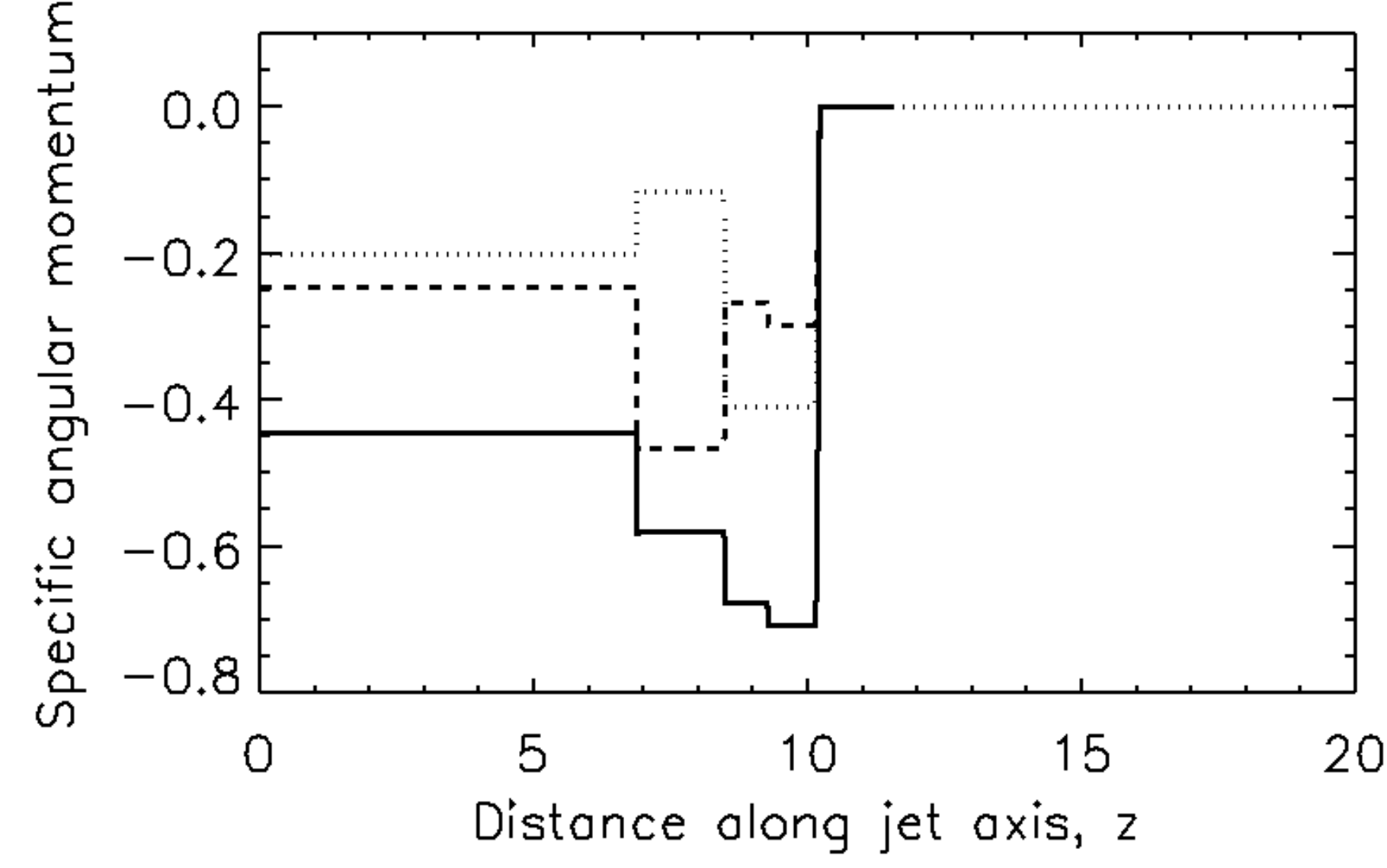}  
\caption{Axisymmetric 1.5D simulation of a propagating MHD shock.
 Simulation RJ21 considers a rotating jet injected into an ambient medium
 with $v_{\phi, \rm jet} = -0.2$, but otherwise the same parameters as 
 RJ18 (see Fig.\,\ref{fig:sim_rj18}).
 Shown is the shock evolution at dynamical time $t=2$ with a grid resolution
 of 1.000 cells / unit length.
 The bottom panel shows the
 specific total angular momentum $\rho v_{\phi}$ ({\it solid}), 
 with the magnetic ({\it dashed}), and
 hydrodynamic ({\it dotted}) contribution.
\label{fig:sim_rj21}
}
\end{figure}

Figure \ref{fig:sim_rj21} shows simulation RJ21 in which a jet {\rm rotating} with 
$v_{\phi, \rm jet} = -0.2$ is injected into the ambient medium. 
By choice, the injected toroidal field direction is anti-aligned with the toroidal 
velocity.
The shock structure of RJ21 is very similar to RJ18 which has otherwise the same
hydrodynamic and magnetic parameters. 
Comparing RJ18 to RJ21, the resulting maximum jet rotational velocity is 
enhanced from $v_{\phi} = -0.31$ to $v_{\phi} = -0.41$,
while the rotational velocity jump across the receding shock 
is the same in both cases, $\Delta v_{\phi} = 0.74$.
$v_{\phi} = -0.41$, but also from $v_{\phi} = -0.31$ to $v_{\phi} = -0.41$.
Figure \ref{fig:sim_rj21} also shows the hydrodynamic and the magnetic angular
momentum contribution to the total angular momentum budget 
(see Eq.~\ref{eqn-ang-mom}, but note that we measure the a.m. at unity radius $r=1$).
For each of the four shocks the total specific angular momentum is conserved 
in the shock frame satisfying the Rankine-Hugoniot conditions.
It can be clearly seen that the magnetic angular momentum decreases across
the shock as the kinetic angular momentum contribution increases.

\begin{figure*}
\centering
\includegraphics[width=7cm]{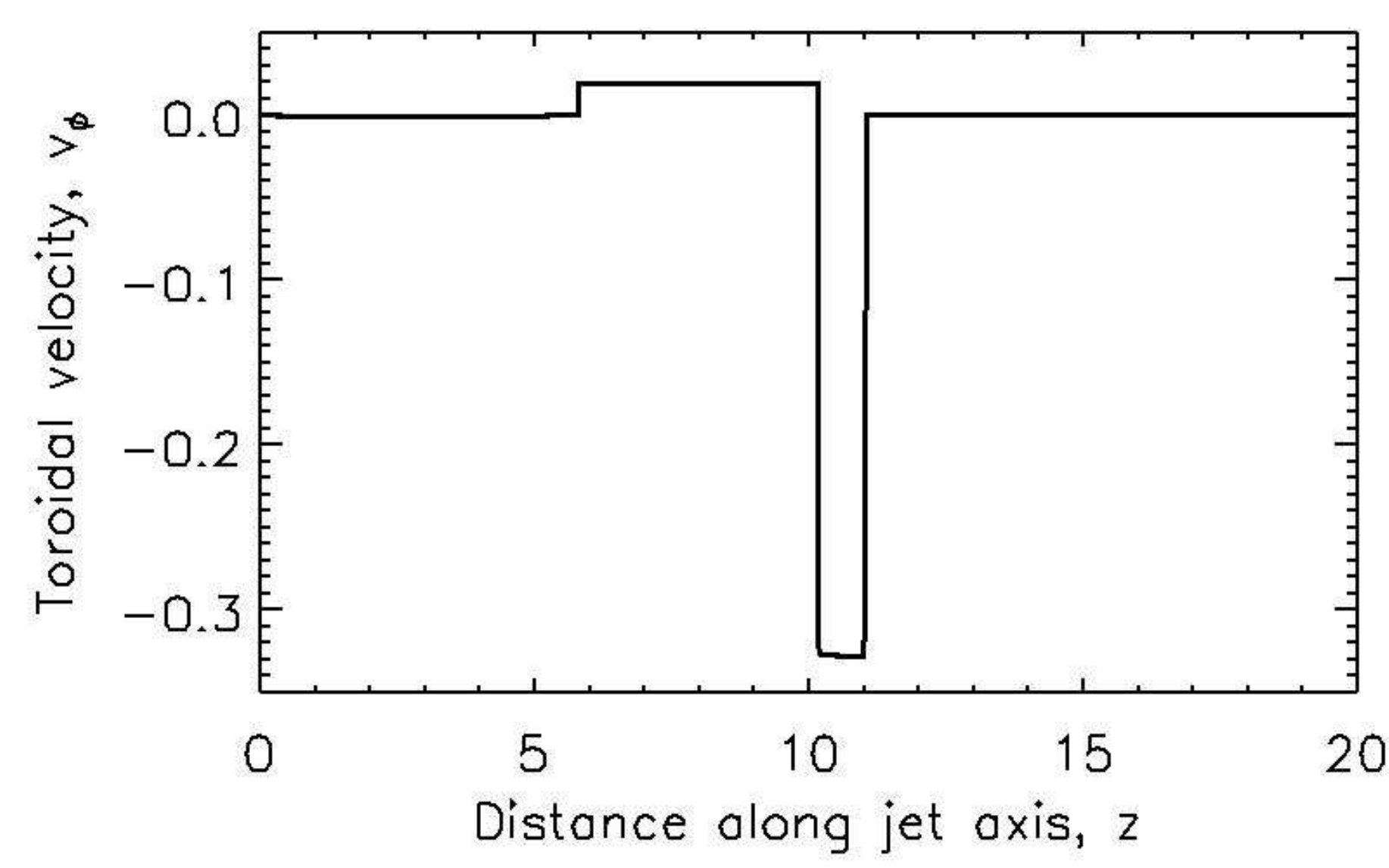}    
\includegraphics[width=7cm]{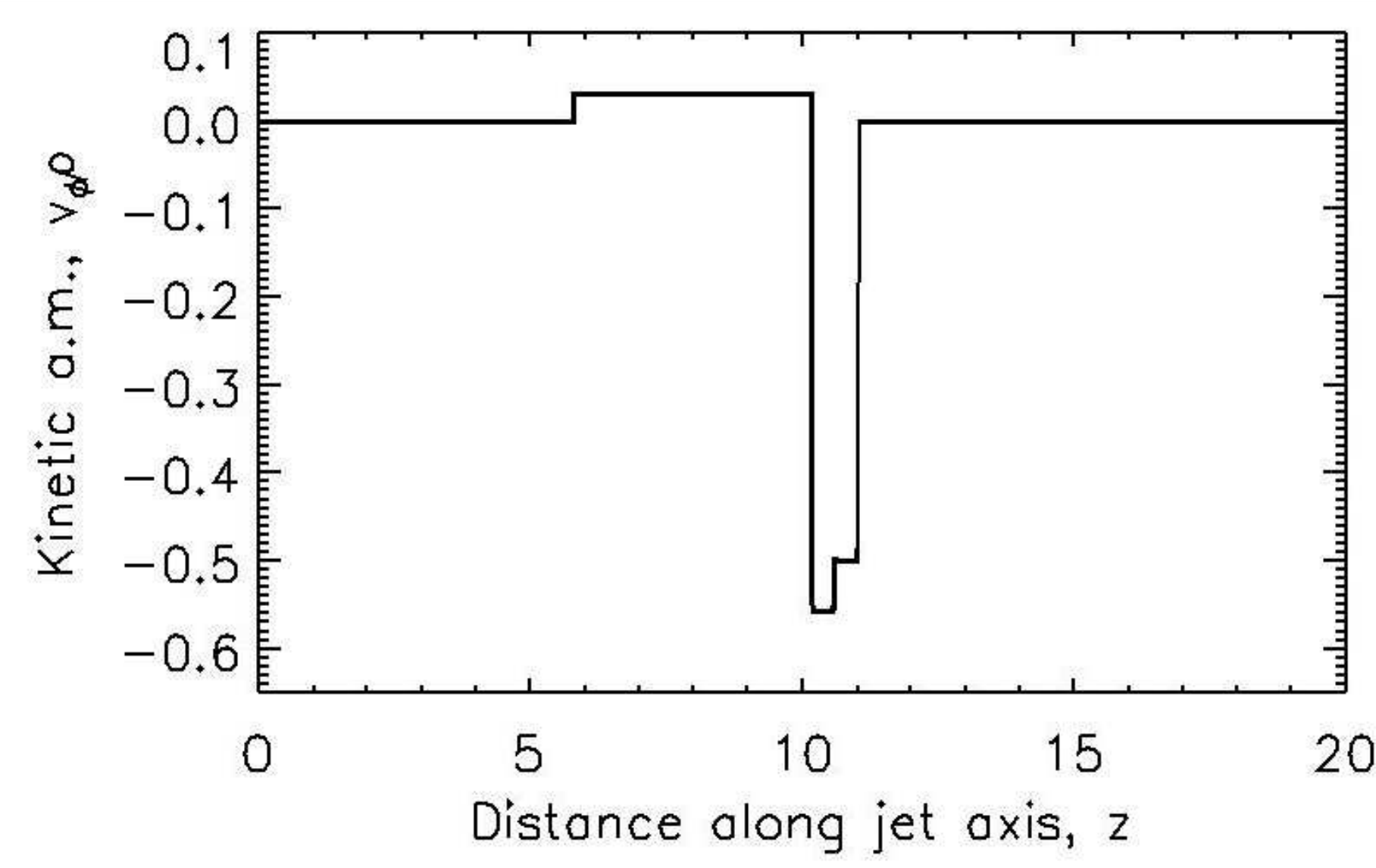}\\  
\includegraphics[width=7cm]{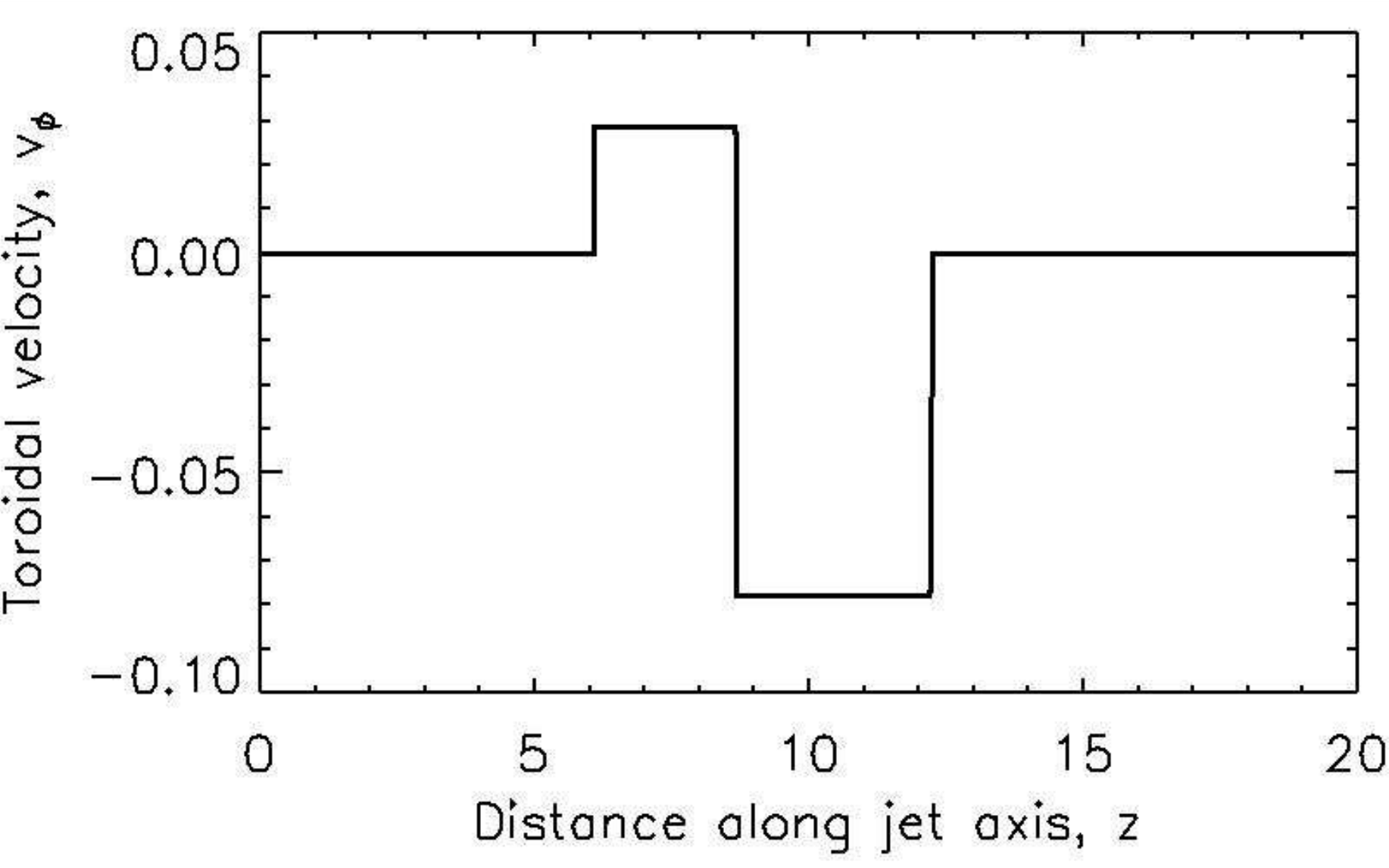}    
\includegraphics[width=7cm]{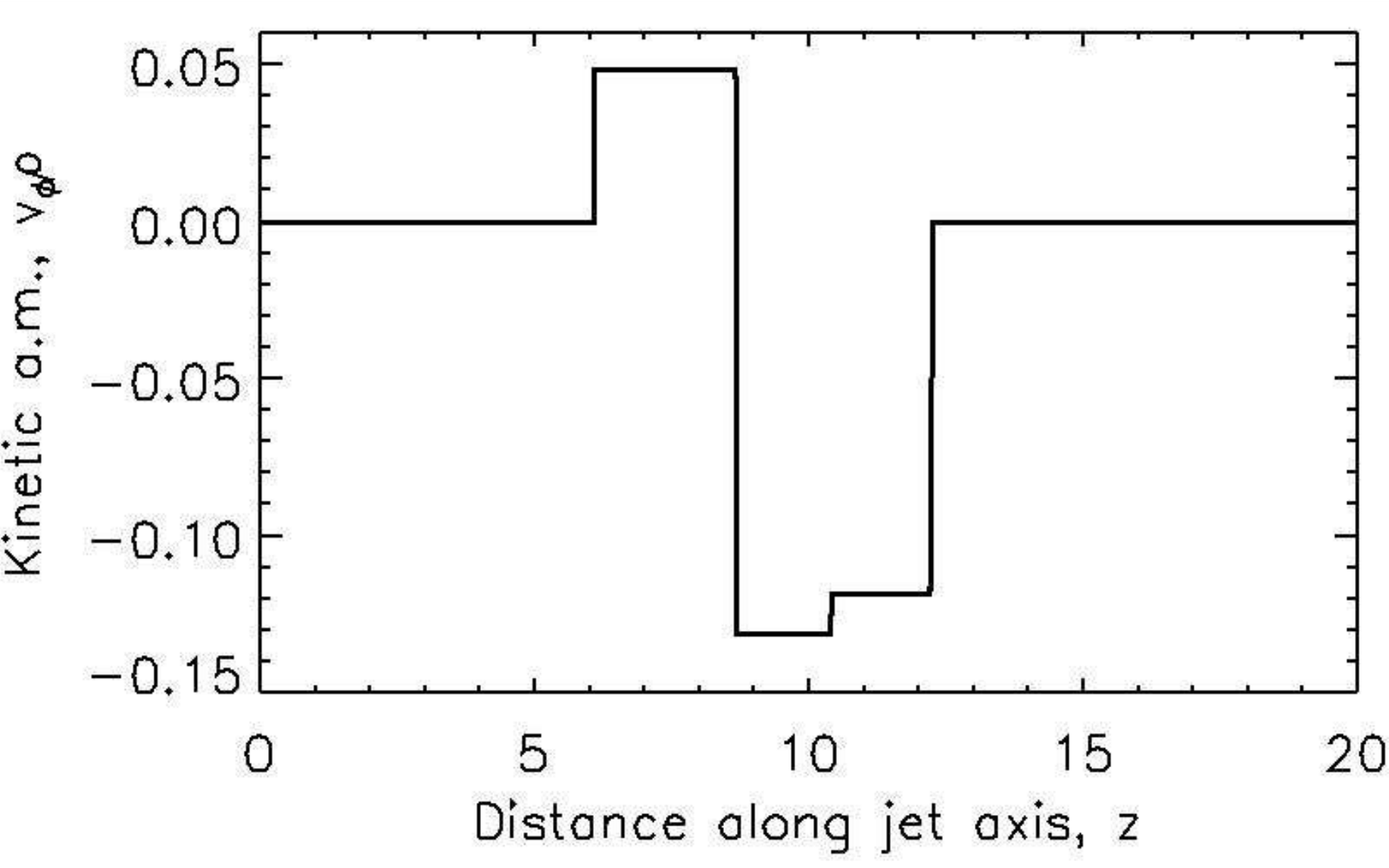}    
\caption{Axisymmetric 1.5D simulations RJ22 (top) and RJ23 (bottom) 
 considering jets with same magnetic angular momentum 
 $\sim B_{\phi}B_z$ (and Poynting flux), 
 but different strength of the field components, 
 $B_{\phi} = 4 B_z$ (RJ22), $B_z = 4 B_{\phi}$ (RJ23), respectively
 (with otherwise the same parameters).
 Shown is the toroidal velocity ({\it left}), and kinetic angular 
 momentum ({\it right}) at dynamical time $t=4$.
\label{fig:sim_rj2223a}
}
\end{figure*}


\begin{figure}
\centering
\includegraphics[width=7cm]{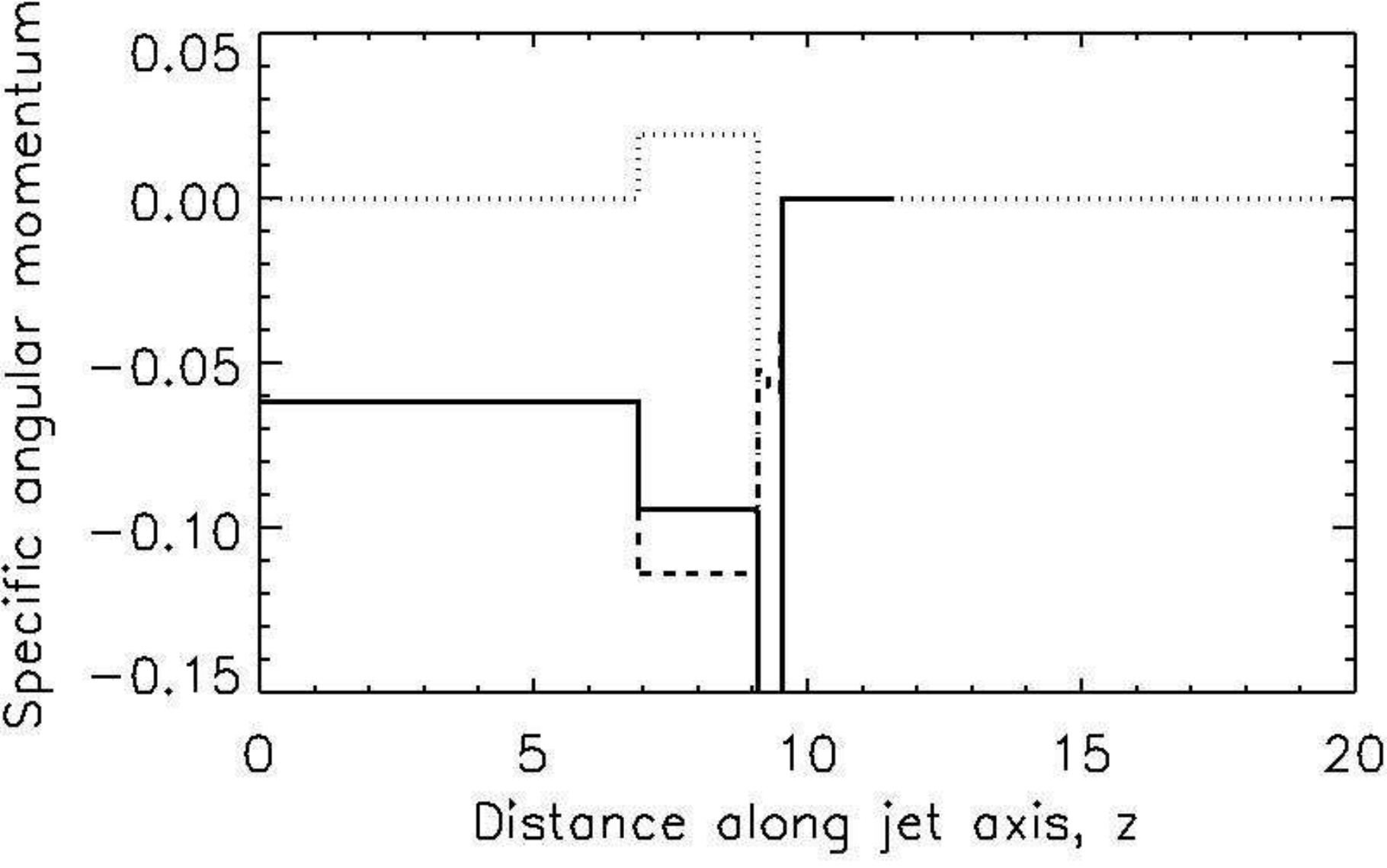}  
\includegraphics[width=7cm]{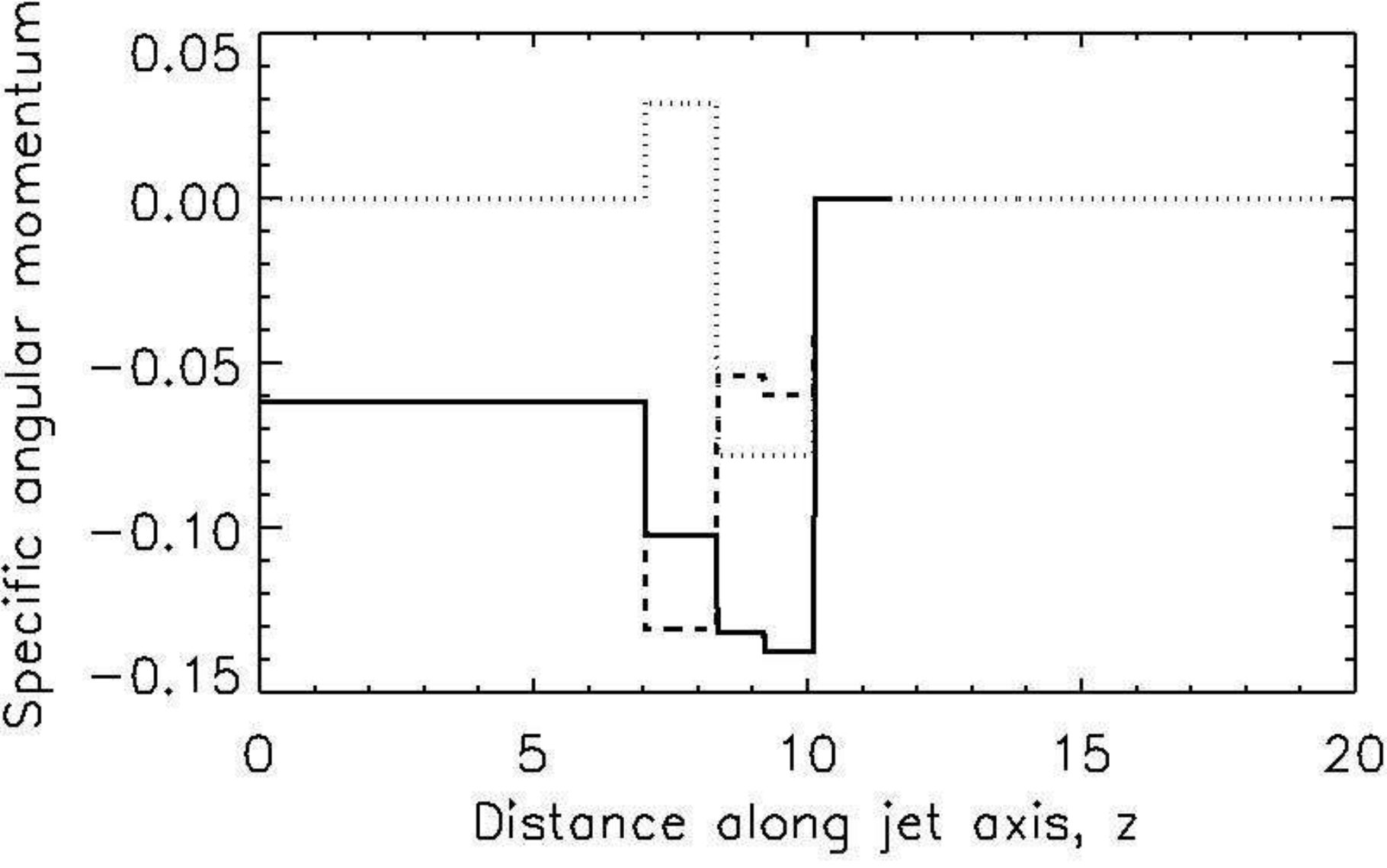}  
\caption{Axisymmetric 1.5D simulations RJ22 (top) and RJ23 (bottom) 
 considering jets with the same magnetic angular momentum 
 $\sim B_{\phi}B_z$ (and Poynting flux), 
 but different strength of the field components, 
 $B_{\phi} = 4 B_z$ (RJ22), $B_z = 4 B_{\phi}$ (RJ23), respectively
 (with otherwise the same parameters).
 Shown is the specific total angular momentum ({\it solid}), with the 
 magnetic ({\it dashed}), and hydrodynamic ({\it dotted}) contribution.
\label{fig:sim_rj2223b}
}
\end{figure}

It is also interesting to compare simulations RJ22 and RJ23 
(see Figs.~\ref{fig:sim_rj2223a}, \ref{fig:sim_rj2223b}).
Both start off with the same initial magnetic angular momentum budget 
and Poynting flux $\sim B_{\rm p} B_{\phi}$, 
however, RJ22 has $B_{\phi} = 4 B_{\rm p}$,
while for RJ23 it is $B_{\rm p} = 4 B_{\phi}$. 
From this, one may expect also a similar toroidal magnetic torque
and, thus, a similar toroidal velocity, 
in particular since a similar compression rate in the hydrodynamic
parameters and also in $B_{\phi}$ applies.
However, we find that the resulting maximum toroidal velocities 
substantially change from 0.19 to 0.29  for the receding shock, 
or from -0.33 to -0.078 for the preceding shock and similarly for
the kinetic angular momentum (see Fig.~\ref{fig:sim_rj2223a}). 
The reason is that the stronger poloidal field in RJ22 broadens the 
shock structure and thus distributes the kinetic angular momentum over
a wider spatial area. This is a typical property of C-shocks.
Note, however, that the total, volume-integrated kinetic angular momentum 
in the shocked region is the same since the volume of the shocked gas 
differs (the shock widens from $\Delta r = 1.5$ to $3.5$ for the preceding 
shock).

We have also performed a comparison simulation with applying a 
simple power-law cooling function. 
Our simulations indicate a minor effect of cooling on jet rotation.
The shock width decreases slightly, as the gas pressure decreases
by cooling (this is a $10\%$-effect for the chosen cooling function).
The gas density and toroidal magnetic field strength increase
resulting in a slightly increased maximum toroidal velocity.
These preliminary results need to be further investigated in a future
paper.

In summary, our 1.5D simulations clearly show how a jet flow in a
helical magnetic field is forced into rotation, when the toroidal field
component is compressed across the shock, thus exerting a Lorentz force\
on the jet material.
Typically, jet rotation velocities are about $3-10\%$ of the jet axial 
velocity for Alfv\'en Mach numbers $M_{\rm Ap,jet} \simeq 2 - 3$
(see Tab.~\ref{tab:para-1da}).

\subsection{Jet shock propagation in 2.5D}
We now discuss axisymmetric simulations taking into account 
all 3 vector components.
This approach allows to investigate also the lateral expansion
of the jet and its interaction with the ambient medium.
We have run a parameter study, mainly considering different values
for the jet magnetosonic Mach numbers and plasma beta 
(see Tab.~\ref{tab:para-2d}).

\begin{table*}
\scriptsize
\begin{center}
\caption{Parameter study of axisymmetric 2.5D jet simulations. 
The ambient density is ${\rho_{\rm ext}}=1.0$ for all simulations.
The minimum and maximum values for the toroidal velocity 
$<v_{\phi, \rm jet, (min,max)}>$ are derived from a histogram statistics,
i.e. indicating the maxima for the {\em bulk} toroidal motion (there
is always a tail in the toroidal velocity distribution of even higher
velocities). The toroidal velocity maxima are estimates when the jet flow
has penetrated most of the computational domain, but its bow shock has not
yet left the grid.
\label{tab:para-2d}
}
\begin{tabular}{ccccccccccc}
\noalign{\smallskip} 
\tableline\tableline
\noalign{\smallskip} 
 ID                      & 
 ${\rho_{\rm jet}}$      & 
 $v_{\rm jet}$           &
 $P_{\rm jet}$           &
 $P_{\rm ext}$           &
 $M_{\rm Ap, jet}$       & 
 $M_{\rm A, jet}$        & 
 $M_{\rm S, ext}$        & 
 $\beta$                 &
 $<v_{\phi, \rm jet, (min,max)}>$ &   
 Remarks        \\
\noalign{\smallskip}
\tableline
\noalign{\medskip}

R01 & 5.5 & 1.0 & 0.006 & 0.0015 & 2000.0 & 300.0 & 10.0 & 20.1 & -0.00014, 0.000065 &  \\ 

\noalign{\medskip}

R02 & 0.5 & 1.0  & 0.002 & 0.006 & 200.0 & 30.0 & 10.0 & 10.1 & -0.0023, 0.04 & \\ 

\noalign{\medskip}

R03 & 1.0 & 1.0  & 0.015 & 0.006 & 200.0 & 30.0 & 10.0 & 10.1 & -0.028, 0.023 & \\ 

\noalign{\medskip}

R04 & 1.0 & 1.0  & 0.012 & 0.024 & 200.0 & 30.0 & 10.0 & 5.0 & -0.022, 0.025 & \\ 

\noalign{\medskip}

R05 & 1.0 & 1.0  & 0.9 & 0.15 & 20.0 & 3.0 & 10.0 & 2.1 & -0.42, 0.25 & \\ 

\noalign{\medskip}

R06 & 1.0 & 1.0  & 0.9 & 0.15 &  50.0 & 3.0 & 10.0 & 1.0 & -0.67, 0.40 &\\ 

\noalign{\medskip}

R07 & 1.0 & 1.0  & 0.9 & 0.15 &  50.0 & 3.0 & 10.0 & 2.0 & -0.55, 0.40 &\\ 

\noalign{\medskip}

R08 & 1.0 & 1.0  & 0.5 & 0.15 & 50.0 & 3.0 & 0.5 & 2.0 & -0.22, 0.15 & \\ 

\noalign{\medskip}

R09 & 1.0 & 1.0  & 0.5 & 0.15 & 50.0 & 3.0 & 0.5 & 2.0 & -0.15, 0.07 & \\ 

\noalign{\medskip}

R11 & 1.0 & 1.0 & 0.1 & 0.15 & 20.0 & 3.0 & 2.0 & 1.1 & -0.18, 0.05  \\ 

\noalign{\medskip}

R12 & 1.0 & 1.0  & 0.001 & 0.024 & 200.0 & 30.0 & 5.0 & 1.1 & -0.0023, 0.015 &  \\ 

\noalign{\medskip}

R13 & 1.0 & 1.0  &0.0001 & 0.024 & 200.0 & 30.0 & 5.0 & 0.1 & -0.0031, 0.010 &   \\ 

\noalign{\medskip}

R15 & 1.0 & 1.0  & 0.01 & 0.067 & 50.0 & 10.0 & 3.0 & 1.1 & -0.03, 0.05 & $\Delta v_{\rm inj}\sim\frac{1}{2}\sin(0.1 t)$  \\ 

\noalign{\medskip}

R16 & 1.0 & 1.0  & 0.001 & 0.024 & 50.0 & 10.0 & 3.0 & 1.1 & -0.007, 0.007 & $\Delta v_{\rm inj}\sim\frac{1}{2}\sin(0.3 t)$ \\ 

\noalign{\medskip}

R17 & 1.0 & 3.0  & 0.01 & 0.22 & 200.0 & 30.0 & 5.0 & 1.1 & -0.0035, 0.013 & $\Delta v_{\rm inj}\sim\frac{1}{2}(\sin 0.5 t)^3$ \\ 

\noalign{\medskip}

R19 & 1.0 & 1.0  & 0.05 & 0.15 & 50.0 & 5.0 & 2.0 & 1.1 & -0.035, 0.05 &  $\Delta v_{\rm inj}\sim\frac{1}{2}(\sin 0.2 t)^3$ \\ 
    \\
\noalign{\medskip}
\tableline
\end{tabular}
\end{center}
\end{table*}

The general boundary condition applied for the jet toroidal magnetic 
field follows a profile 
$B_{\phi} = B_{\phi,\rm max} ( \sin((2.0 r - 0.5) ) \pi + 1.0)$ for $r<1.0$,
and vanishes for larger radii. 
This is not a force-free configuration, however, it avoids the strong current
sheet usually launched by sawtooth-like profiles.
The initial poloidal field distribution is constant and purely longitudinal.
The radial profile of the jet injection velocity follows a cosine profile,
$v_{\rm inj}(r) = v_{\rm jet} \cos(r \pi/2)$ for $r<1$.
We have also run simulations with a time-dependent injection velocity,
$v_{\rm jet}(t) = 1.0 + 0.5*\sin(0.3 t)$, thus 
$v_{\rm inj}(t,r) = v_{\rm inj}(r) * (1.0 + 0.5*\sin(0.3 t)) $,
or, alternatively with a longer period $\sim (1.0 + 0.5*\sin(0.1 t))$.

\begin{figure*}
\centering
 \includegraphics[width=5.cm,height=9cm]{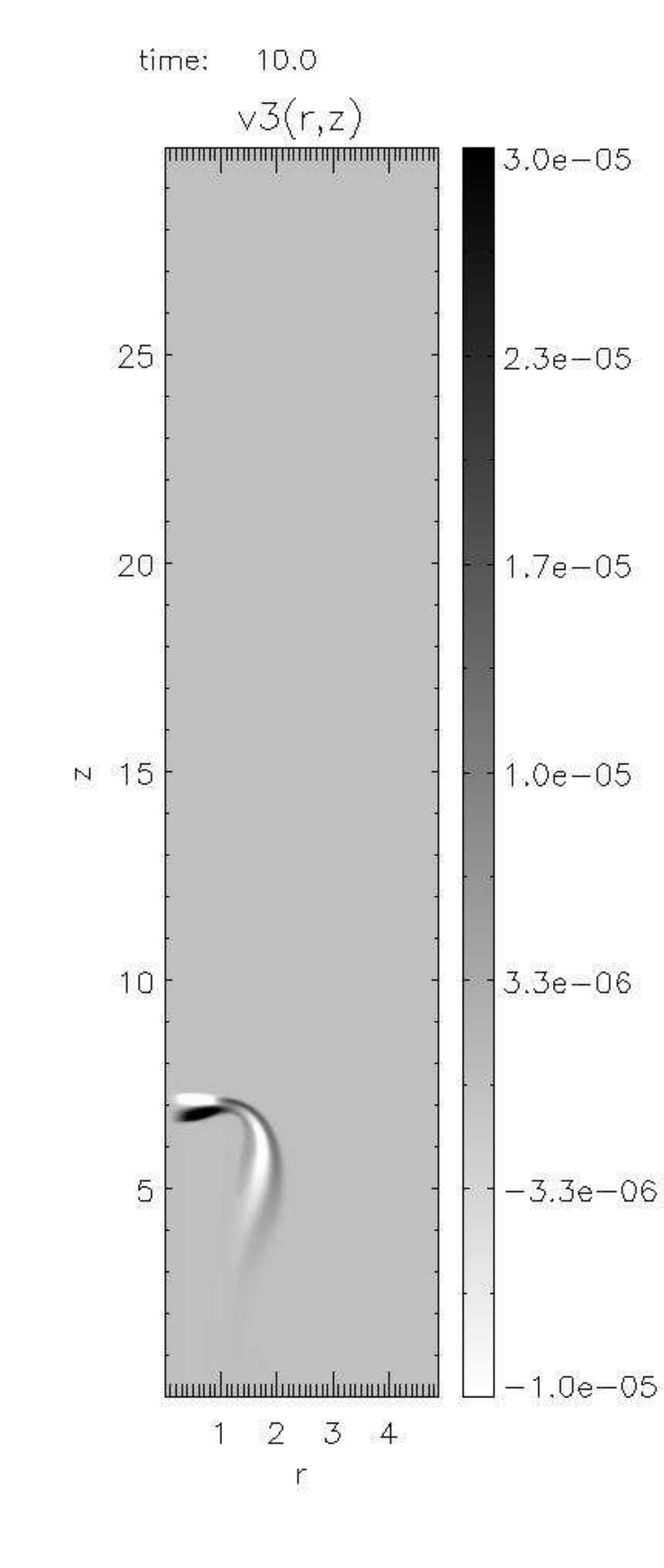}  
 \includegraphics[width=5.cm,height=9cm]{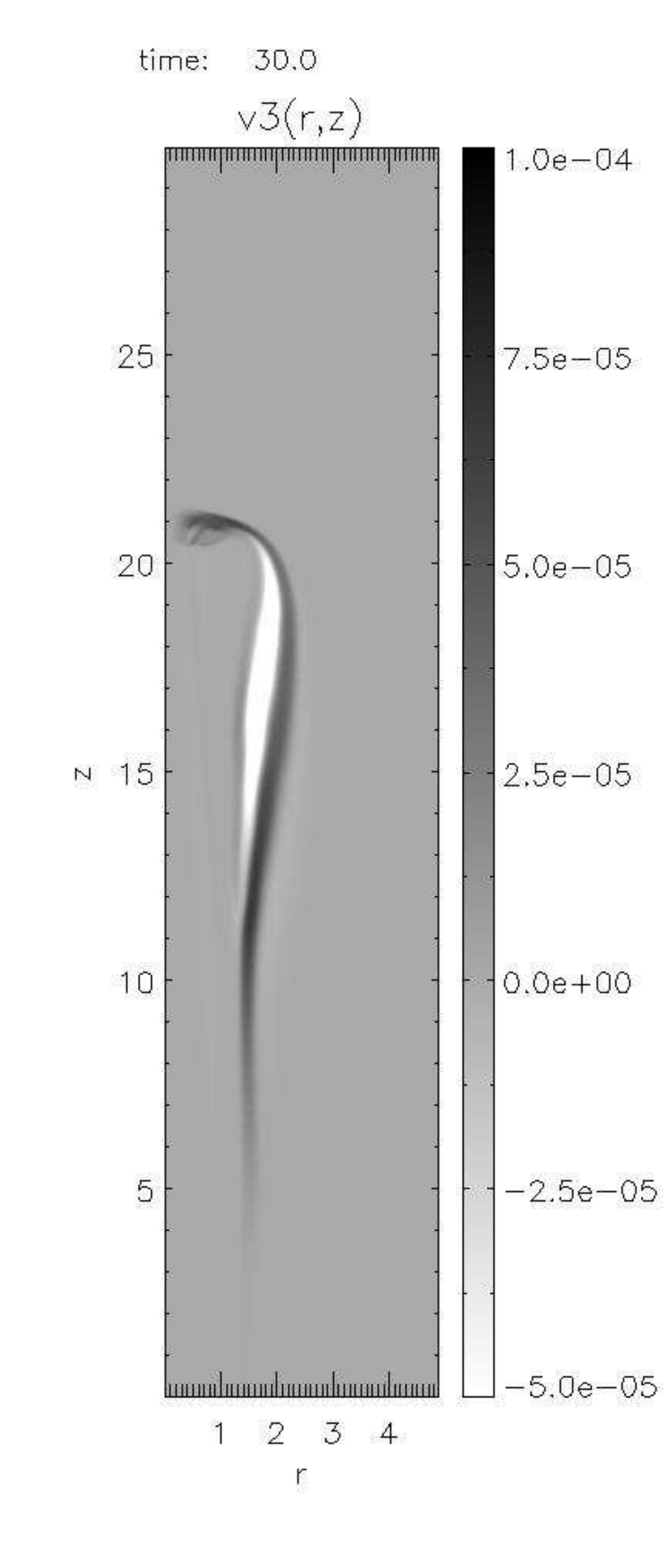}  
 \includegraphics[width=5.cm,height=9cm]{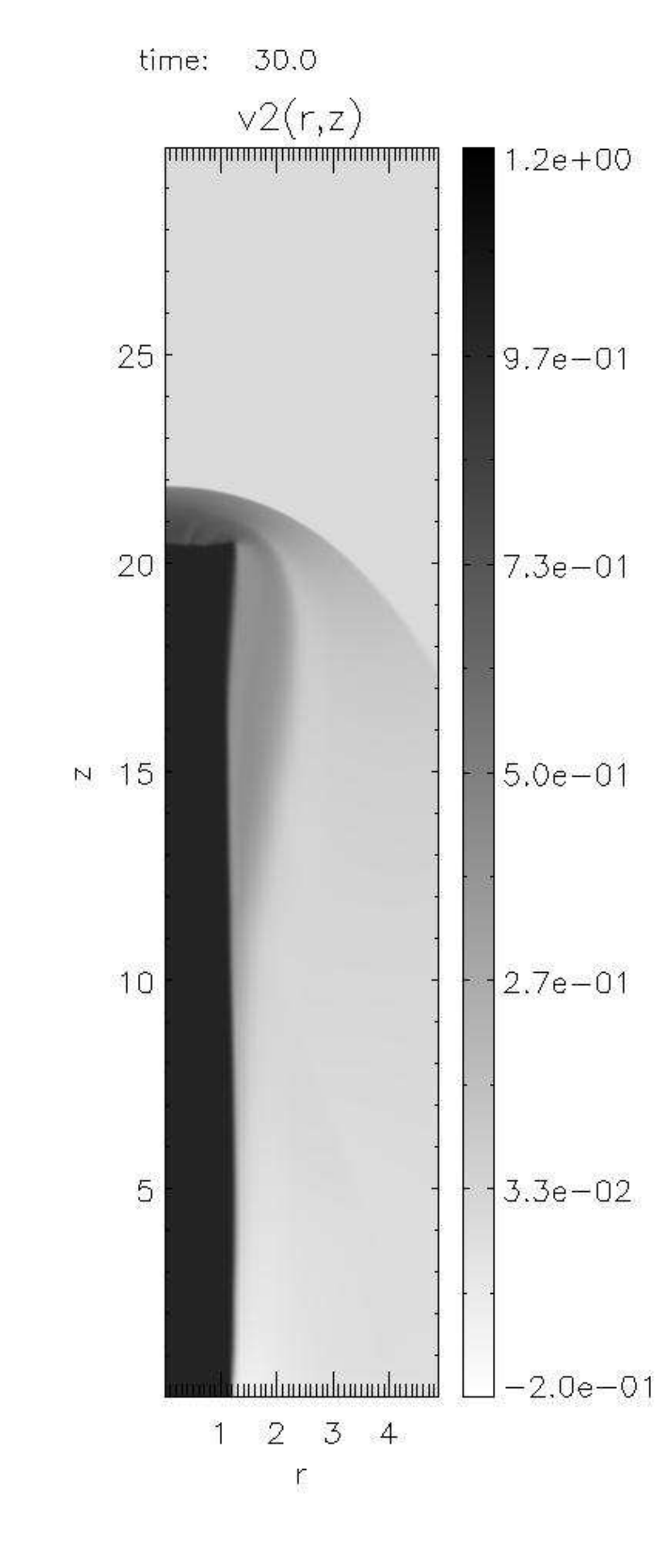}\\ 
 \includegraphics[width=5.cm,height=9cm]{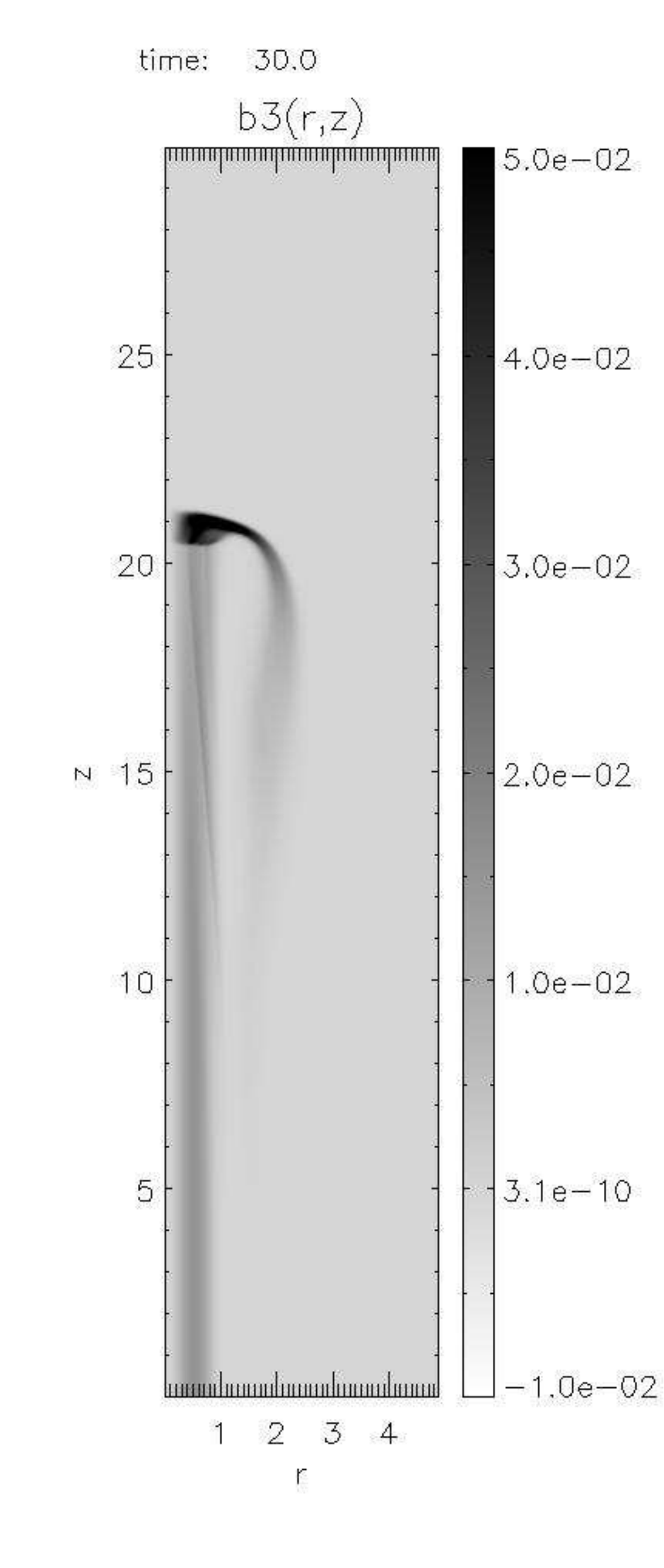}   
 \includegraphics[width=5.cm,height=9cm]{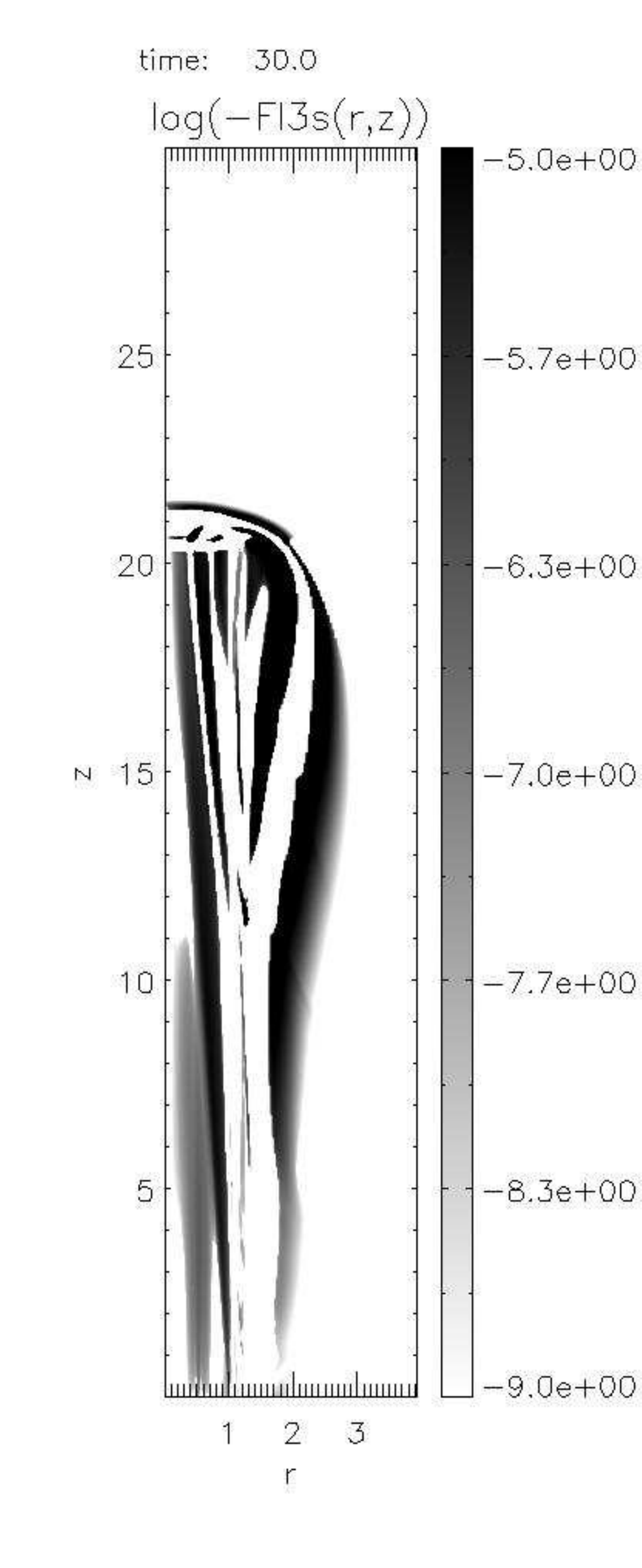}   
 \includegraphics[width=5.cm,height=9cm]{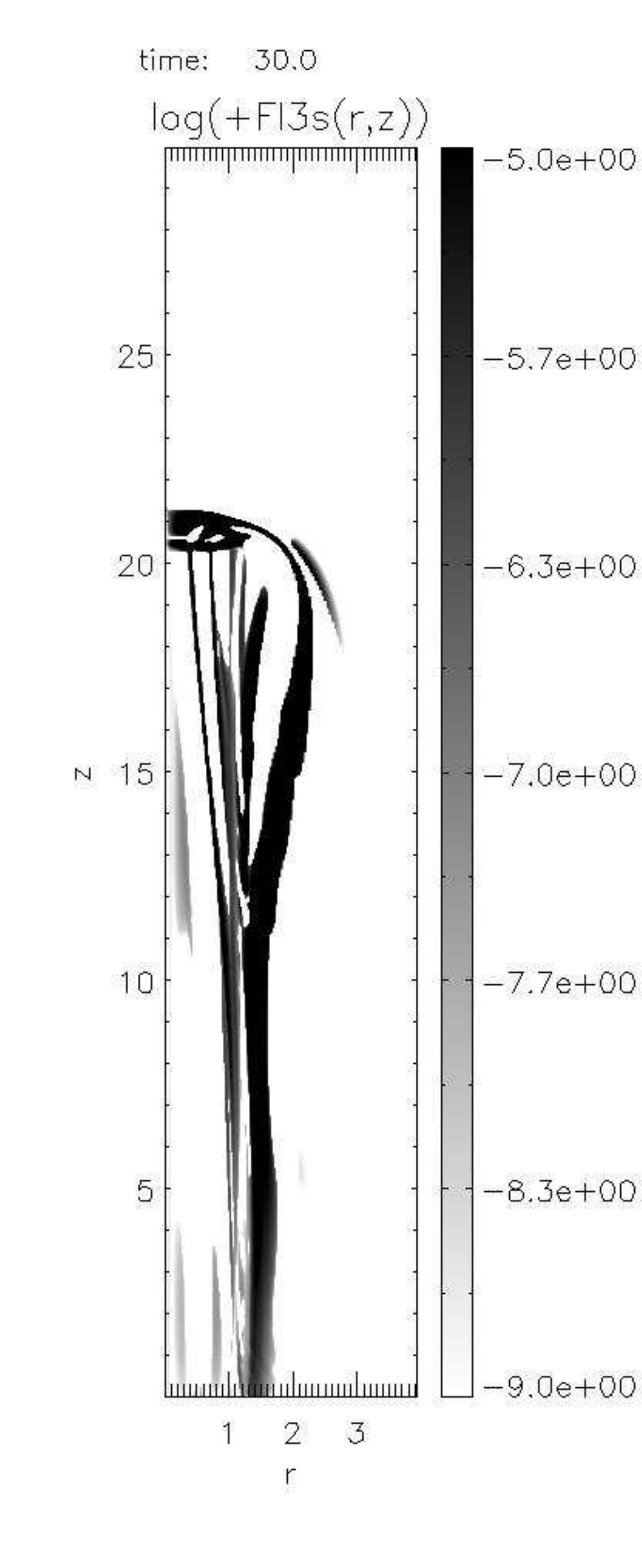}   
\caption{Axisymmetric 2.5D simulation run R01 considers a high (external) 
 Mach number jet with weak magnetic field.
 Shown is (from {\it top left} to {\it bottom right}) the toroidal velocity 
 at dynamical time $t=10, 30$, and for the latter also the axial velocity 
 $v_z$, the toroidal field strength $B_{\phi}$, 
 and the negative and positive toroidal Lorentz force $r F_{\rm L,\phi}$
 (on a log scale).
 Shown is the inner part $r < 5.0$ of the whole grid of $(r<7.0, z<30.0)$.
 The inner, equidistant grid consists of $(100 \times 810)$ cells for 
 $(r<1.5, z<30.0)$. Attached to that is a logarithmically scaled grid 
 for $r>1.5$.
 The jet nozzle is located between $0 < r < 1.0$.
\label{fig:sim_run01}
}
\end{figure*}

Simulation run R01 considers a high external Mach number $M_{\rm s,ext} = 10$,
a comparatively weak jet magnetic field corresponding to high magnetosonic
Mach numbers, and a high plasma beta - thus, a jet rather close to the 
hydrodynamical limit.
Figure \ref{fig:sim_run01} shows the time evolution of the toroidal 
velocity of this jet for the dynamical time steps $t=10, 30, 40$,
and the corresponding toroidal magnetic field strength, axial velocity,
and toroidal torque for time $t=30$. 
This images show the inner domain $r<5.0$ of the simulation box $r<7.0$.

The simulation shows the rotating material surrounding the inner jet 
beam in a cocoon-like structure.
The injected jet material is non-rotating (by definition of the boundary
condition).
For run R01 the obtained rotational velocities are comparatively low 
$v_{\phi} < 10^{-4} v_{\rm jet}$,
resulting from the low magnetic field strength in spite of the high 
shock compression ratio in the near-hydrodynamical limit.

During the initial time steps the termination shock is rotating with positive sense, 
the bow shock with negative sense.
The toroidal magnetic field compression is clearly seen in the termination shock,
but also in surrounding cocoon. The cocoon toroidal field is enhanced
partly by the radial expansion of the shocked gas, partly
by winding up the poloidal field by the rotating jet material.

The map of the toroidal torque demonstrates the interrelation between
Lorentz force and rotation.
The jet head is rotating in positive direction ($t=40$, grey blackish colors)
with the toroidal torque in the same direction (white colors).
The rotation velocity measured at a certain time is, however,
resulting from the time-integrated torque on a parcel of material along 
its path.

\begin{figure*}
\centering
\includegraphics[width=5.cm,height=9cm]{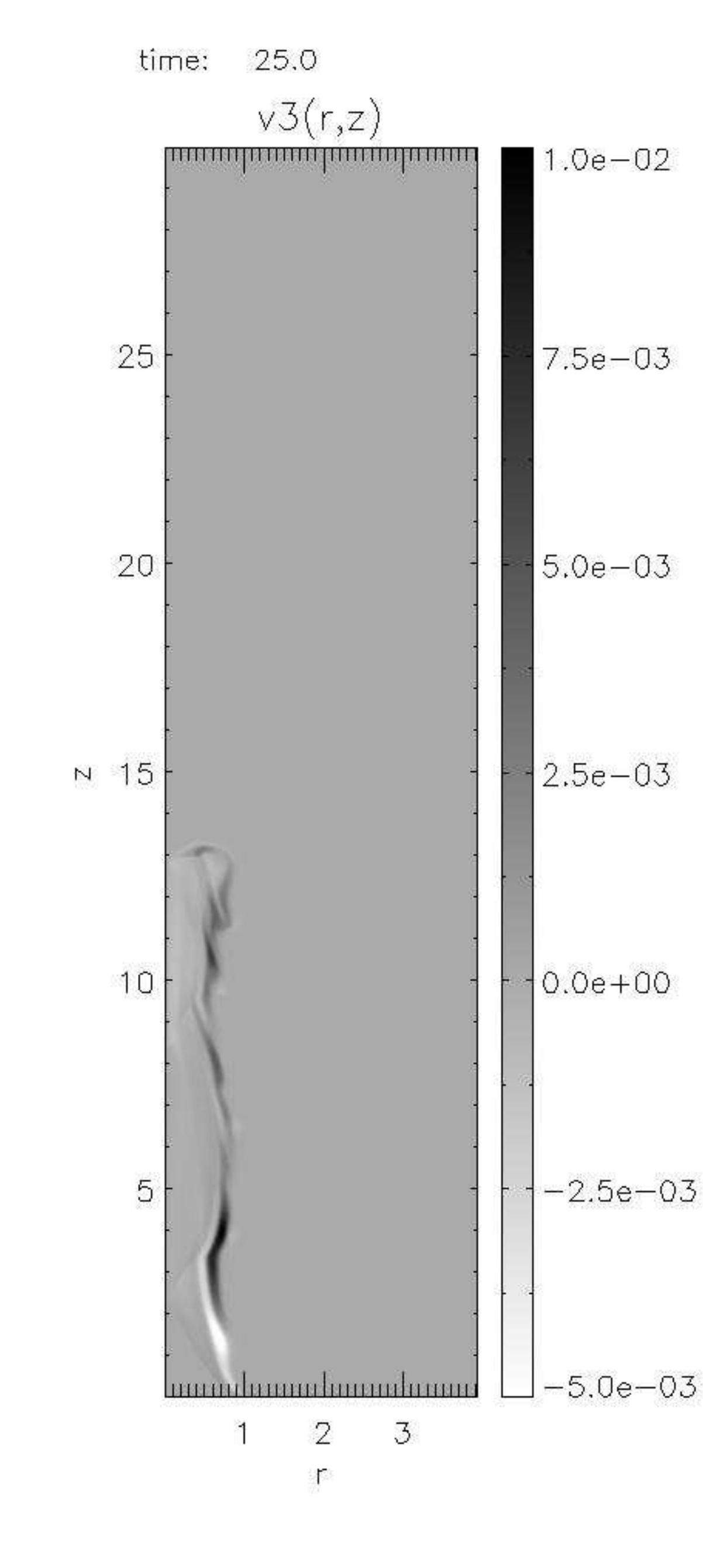}  
\includegraphics[width=5.cm,height=9cm]{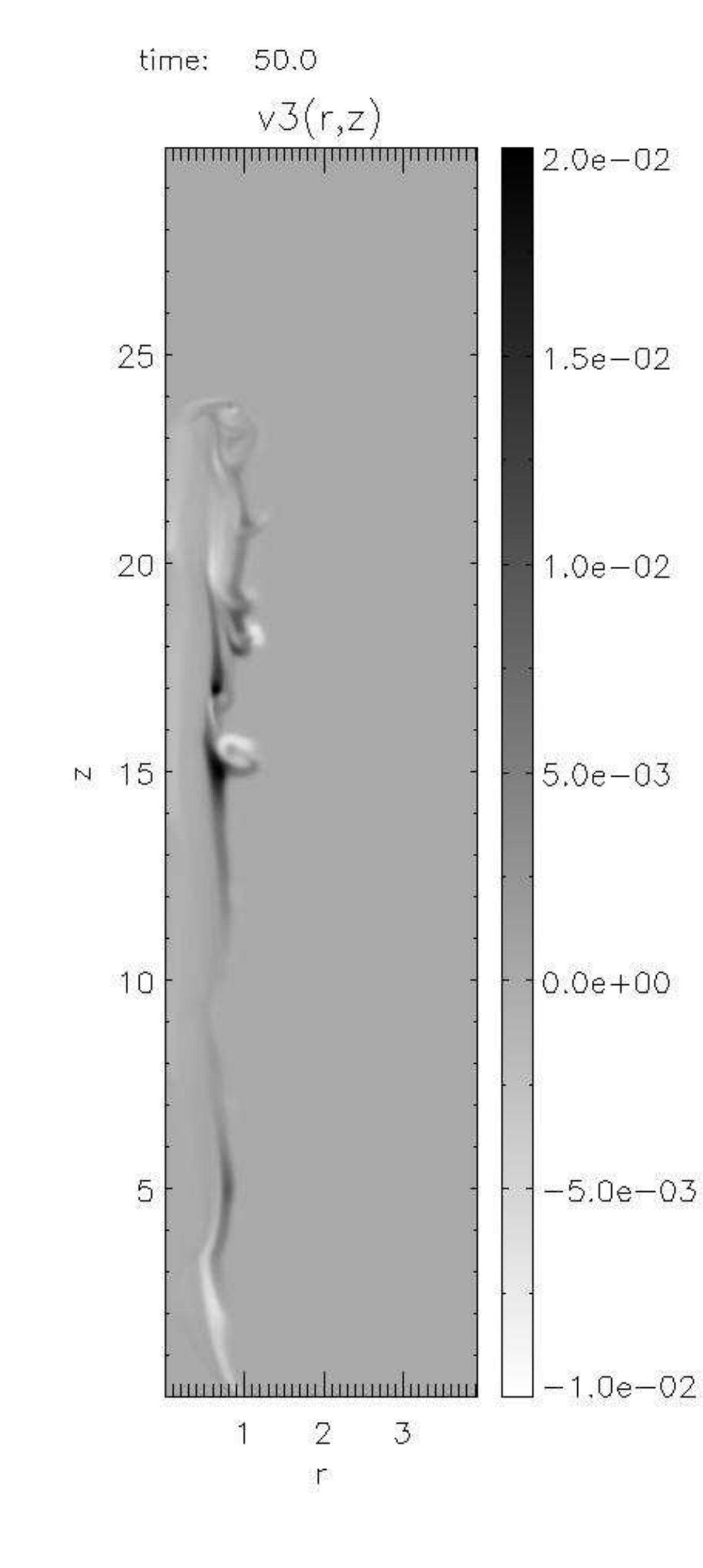}  
\includegraphics[width=5.cm,height=9cm]{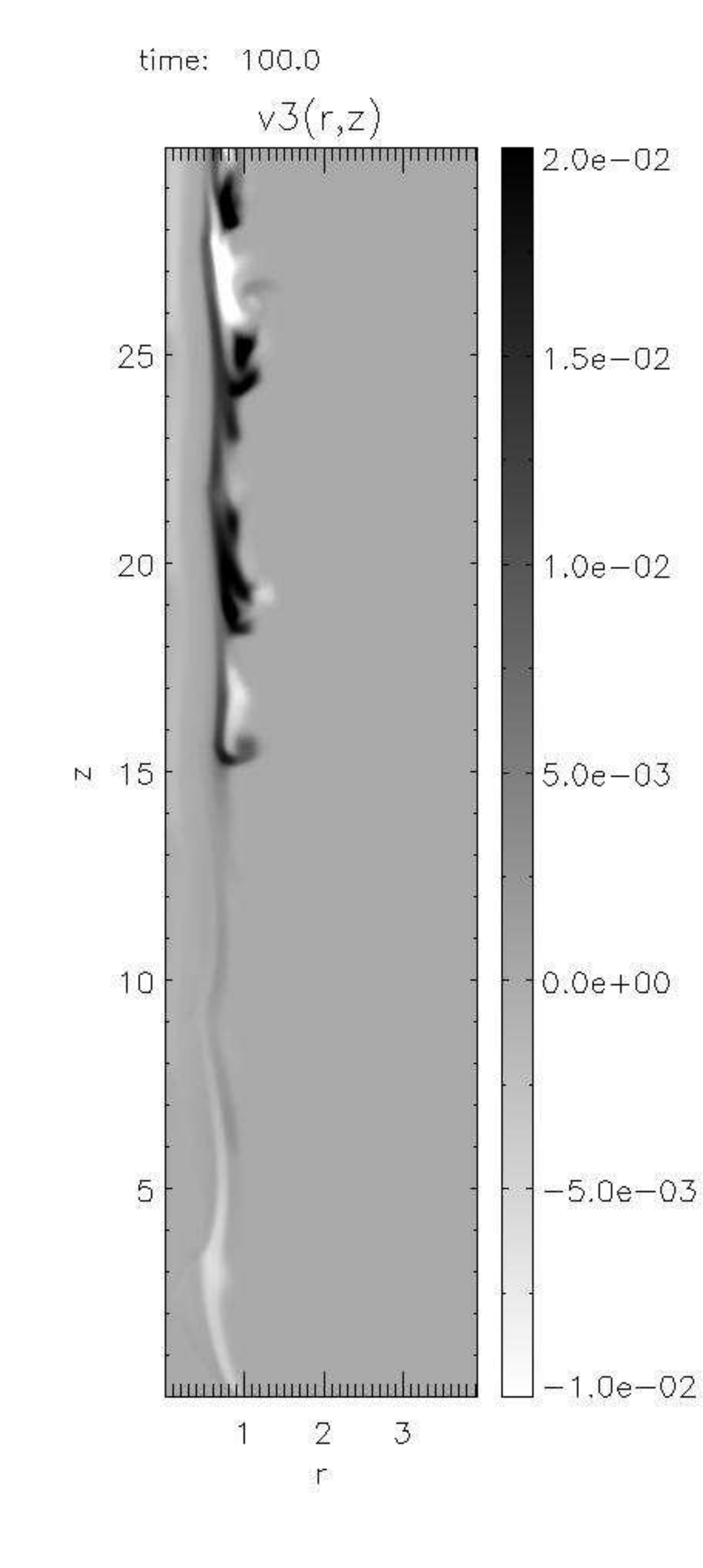}  
\caption{Axisymmetric simulation run R12 considers a high (external) 
 Mach number jet with weak magnetic field, however, 
 representing a less extreme case compared to run R01.
 Shown is toroidal velocity at time steps $t=25, 50, 100$ 
 (from {\it left} to {\it right}) for a sub-grid of $r<4.0$ of the
 whole computational domain $(0.0<r<10.0, 0.0<z<30.0)$.
\label{fig:sim_run12a}
}
\end{figure*}

Simulation run R12 considers a high (external) Mach number jet with weak 
magnetic field, similar to R01, but representing a less extreme case
(see Fig.~\ref{fig:sim_run12a}).
The toroidal velocities induced range from -0.07 to 0.033 corresponding to
1-3 \%. This is in the range of the observationally indicated velocities.
Also the jet dynamical parameters, 
$M_{\rm A, jet} =30$, $M_{\rm S, ext} = 5$, are observationally indicated.
Simulation run R13 considers the same parameters except a lower plasma-$\beta$,
thus a stronger poloidal magnetic field.

For jets with lower Mach numbers (e.g. R09, R10) even higher 
toroidal velocities can be produced.
The main contribution to the rotating material is from entrained 
(and shocked) ambient gas.

\begin{figure*}
\centering
 \includegraphics[width=4.0cm,height=6cm]{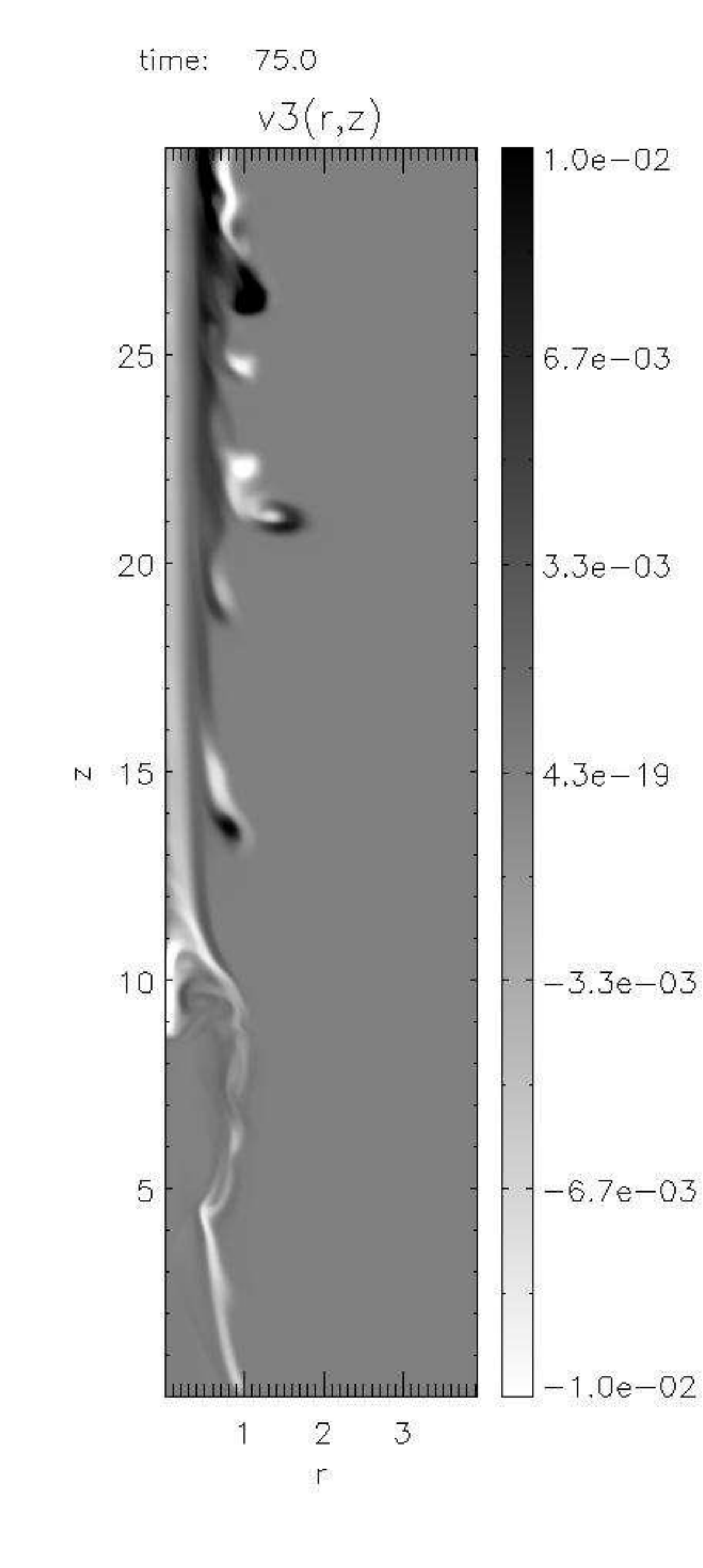}   
 \includegraphics[width=4.0cm,height=6cm]{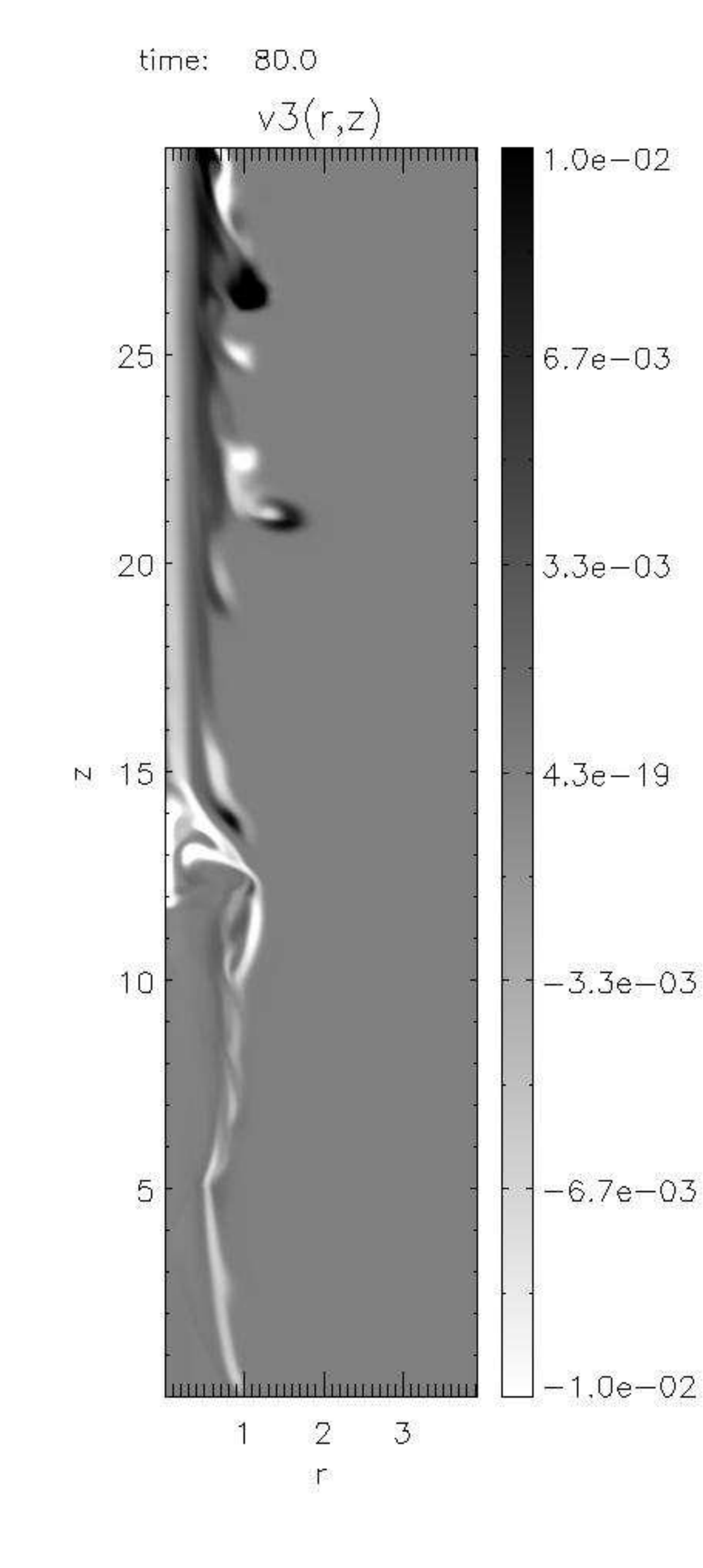}   
 \includegraphics[width=4.0cm,height=6cm]{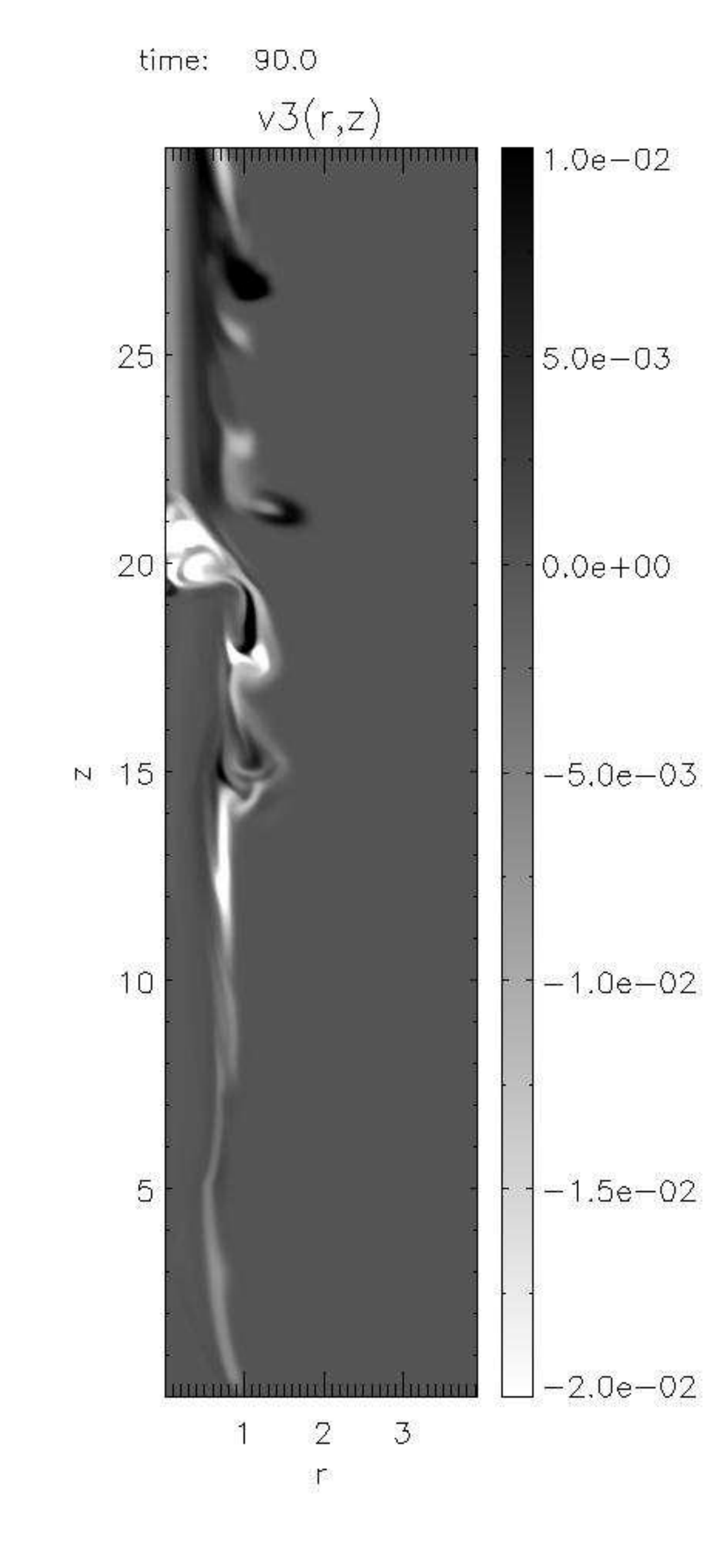}   
 \includegraphics[width=4.0cm,height=6cm]{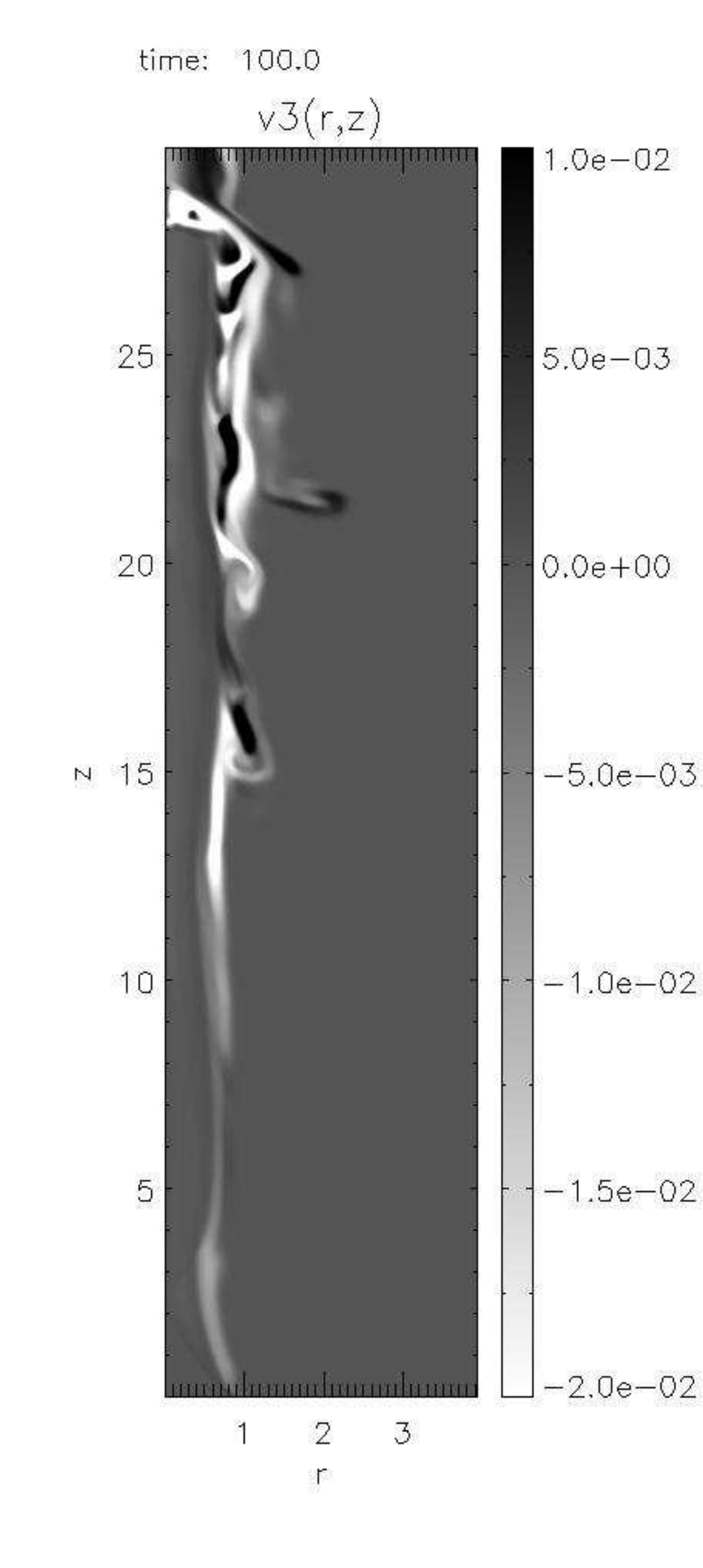}\\ 
 \includegraphics[width=4.0cm,height=6cm]{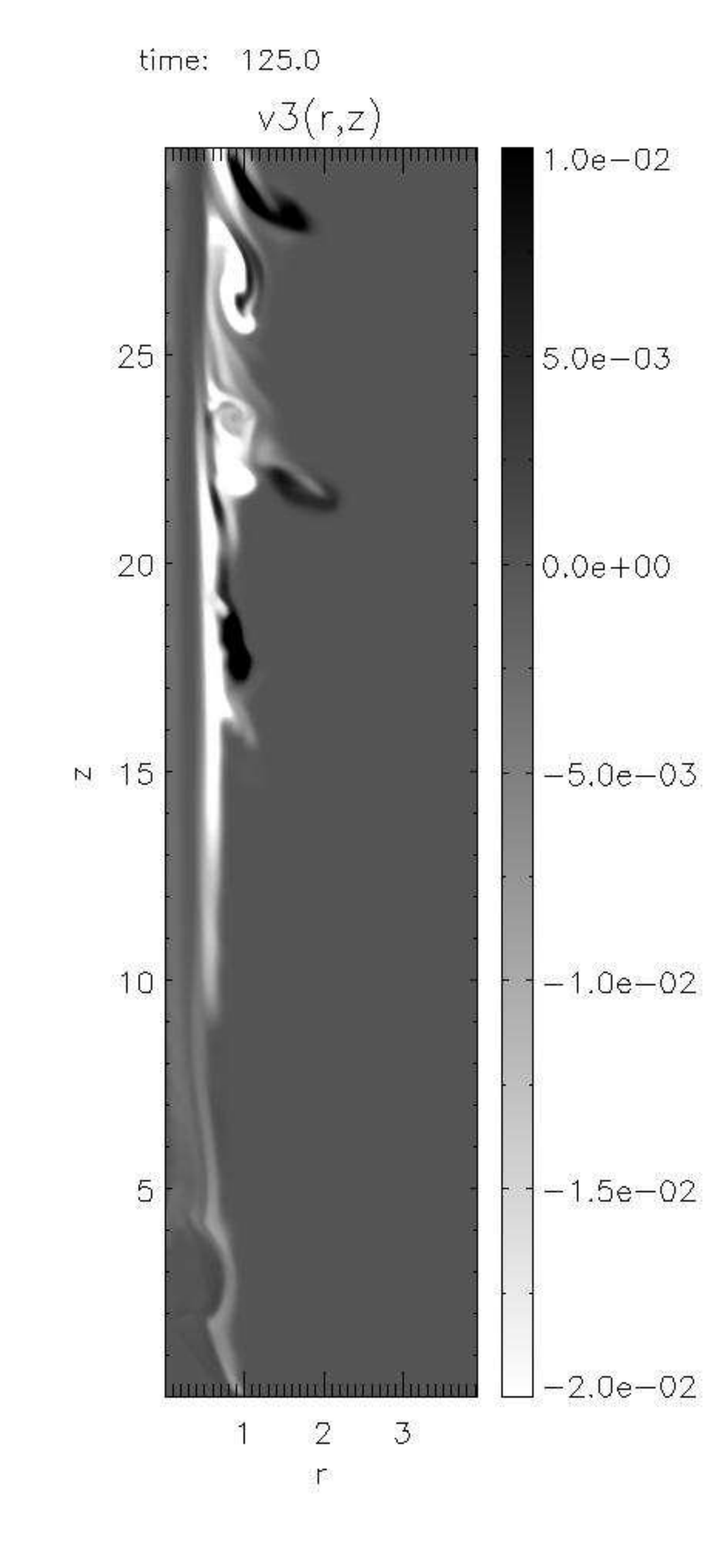}   
 \includegraphics[width=4.0cm,height=6cm]{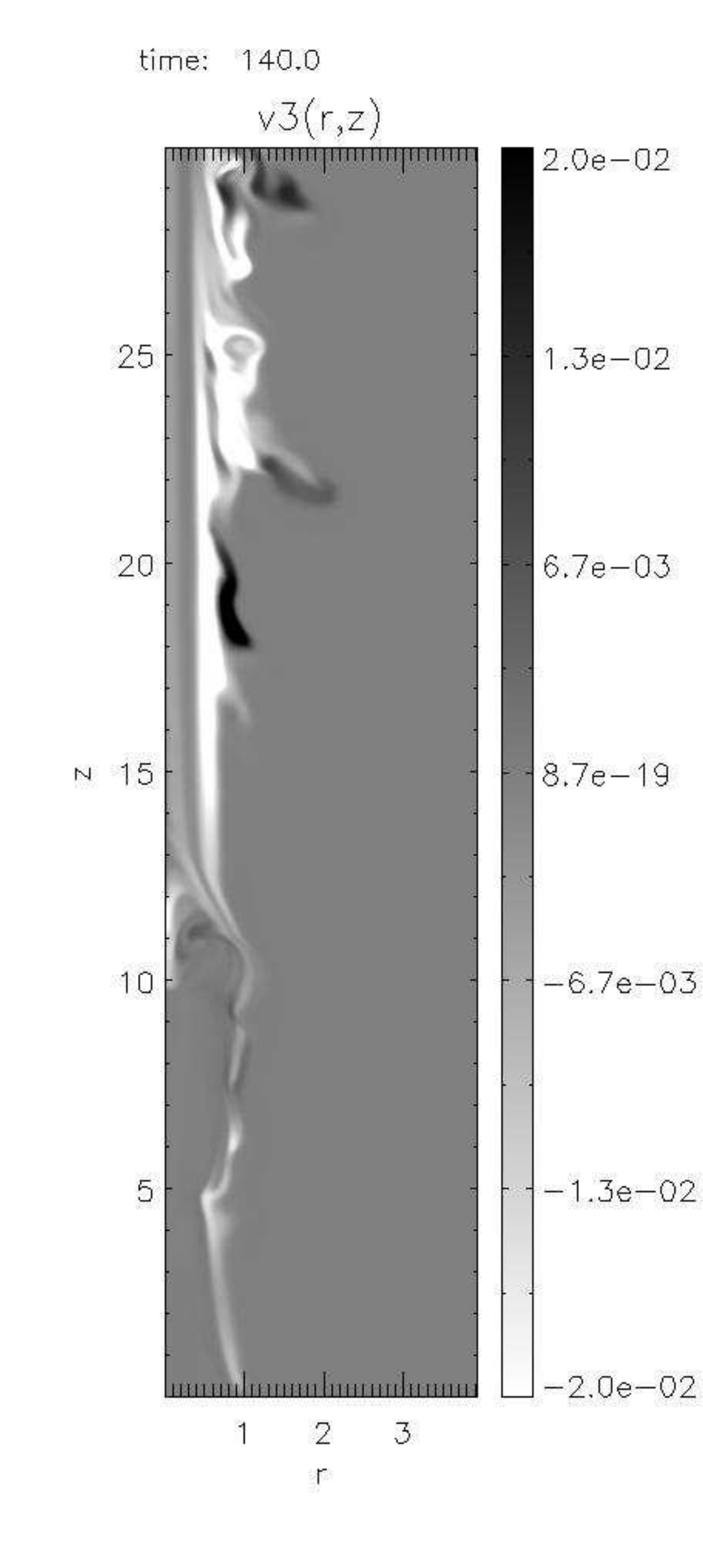}   
 \includegraphics[width=4.0cm,height=6cm]{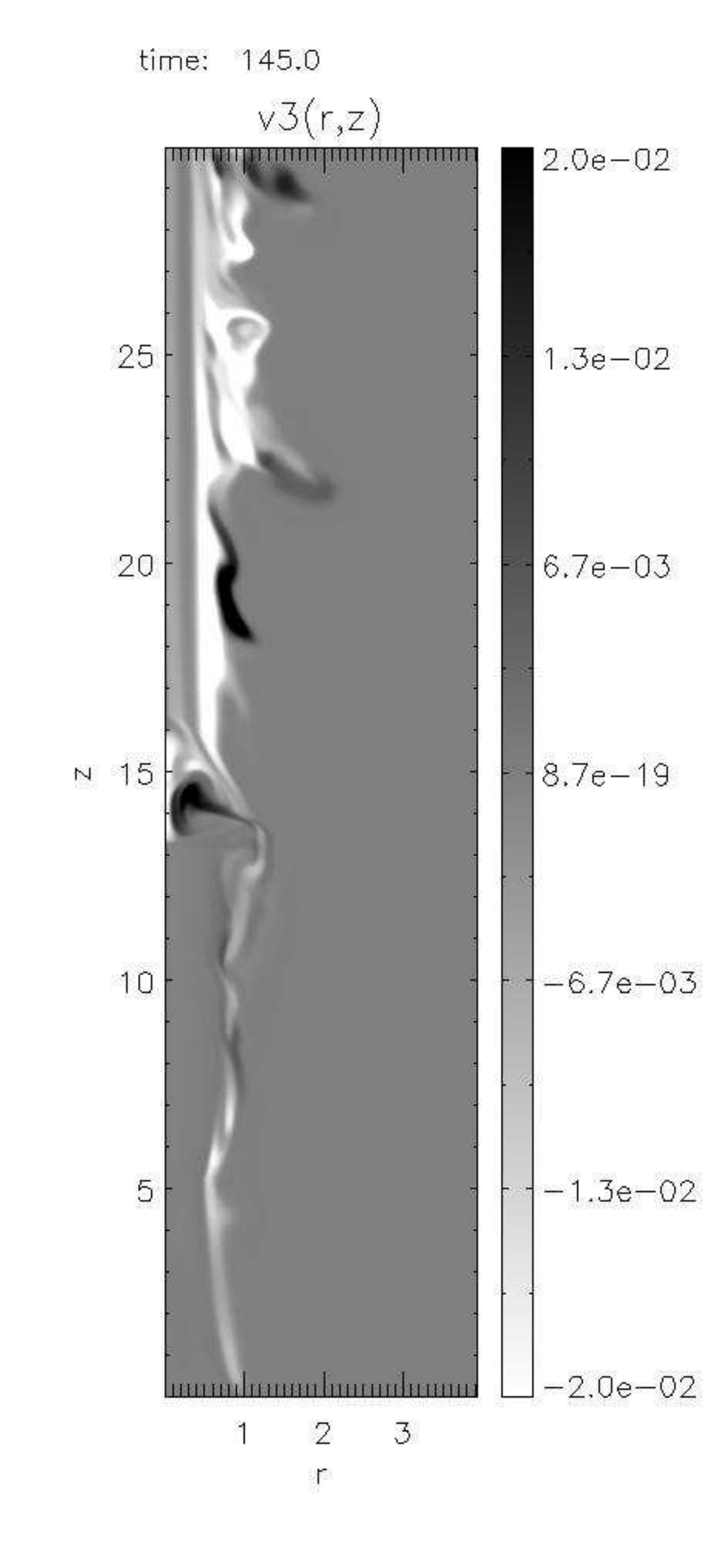}   
 \includegraphics[width=4.0cm,height=6cm]{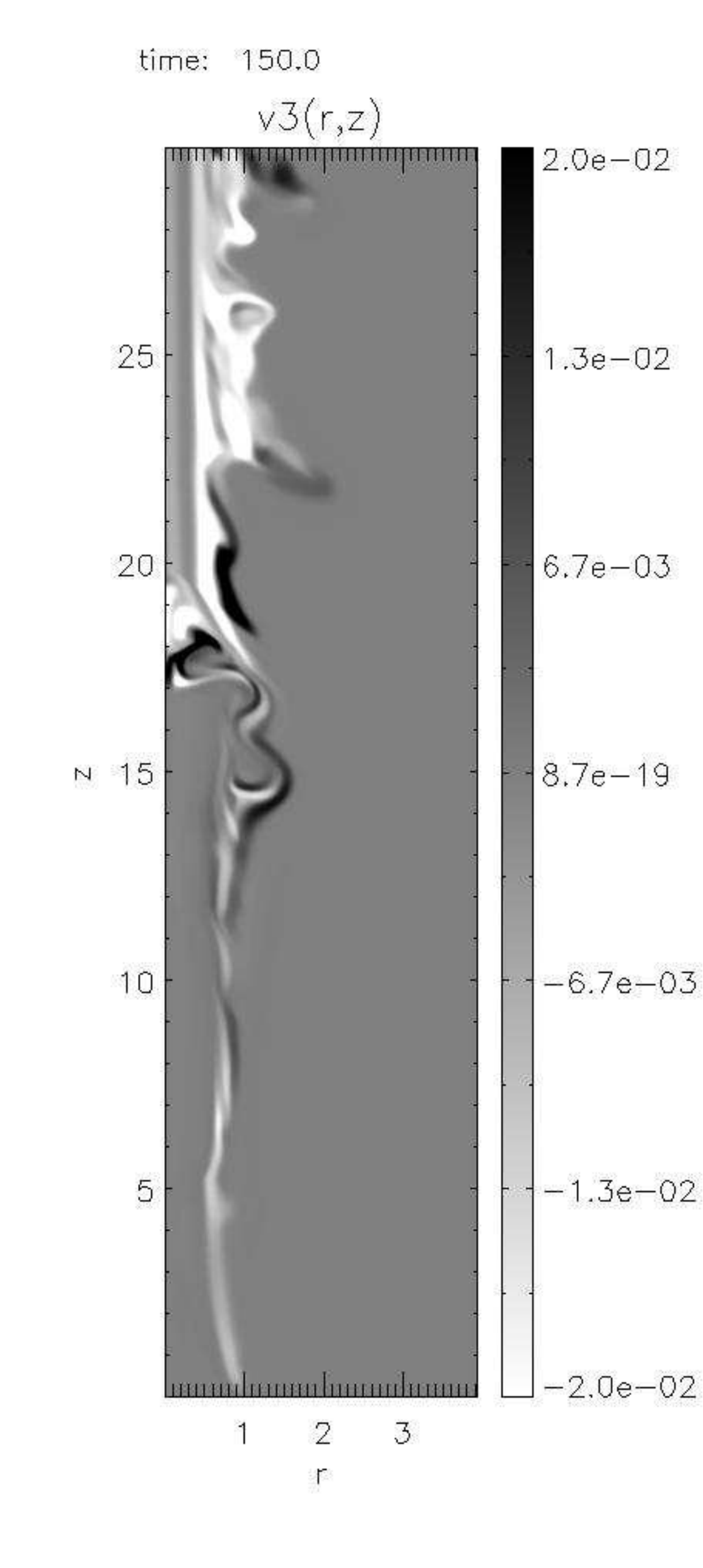}\\ 
 \includegraphics[width=4.0cm,height=6cm]{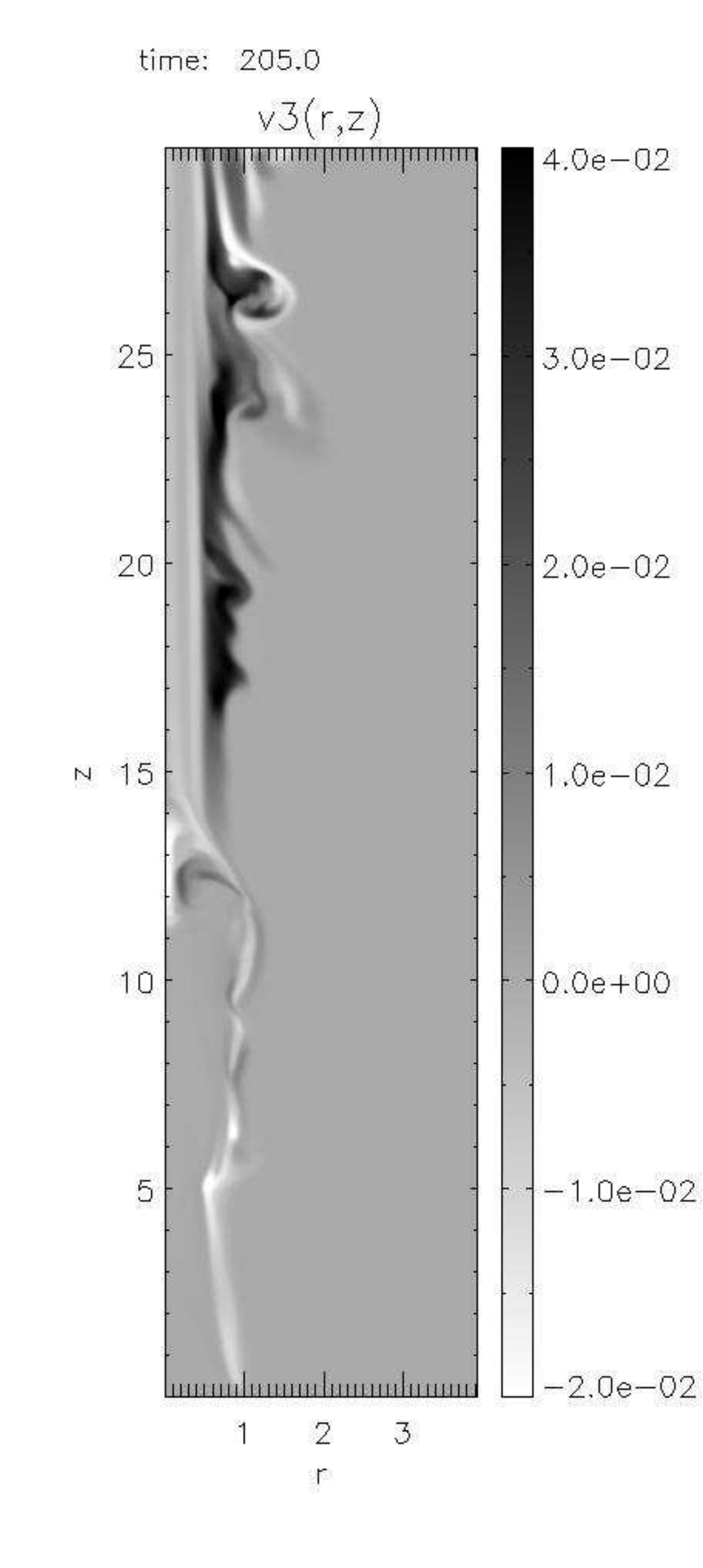}   
 \includegraphics[width=4.0cm,height=6cm]{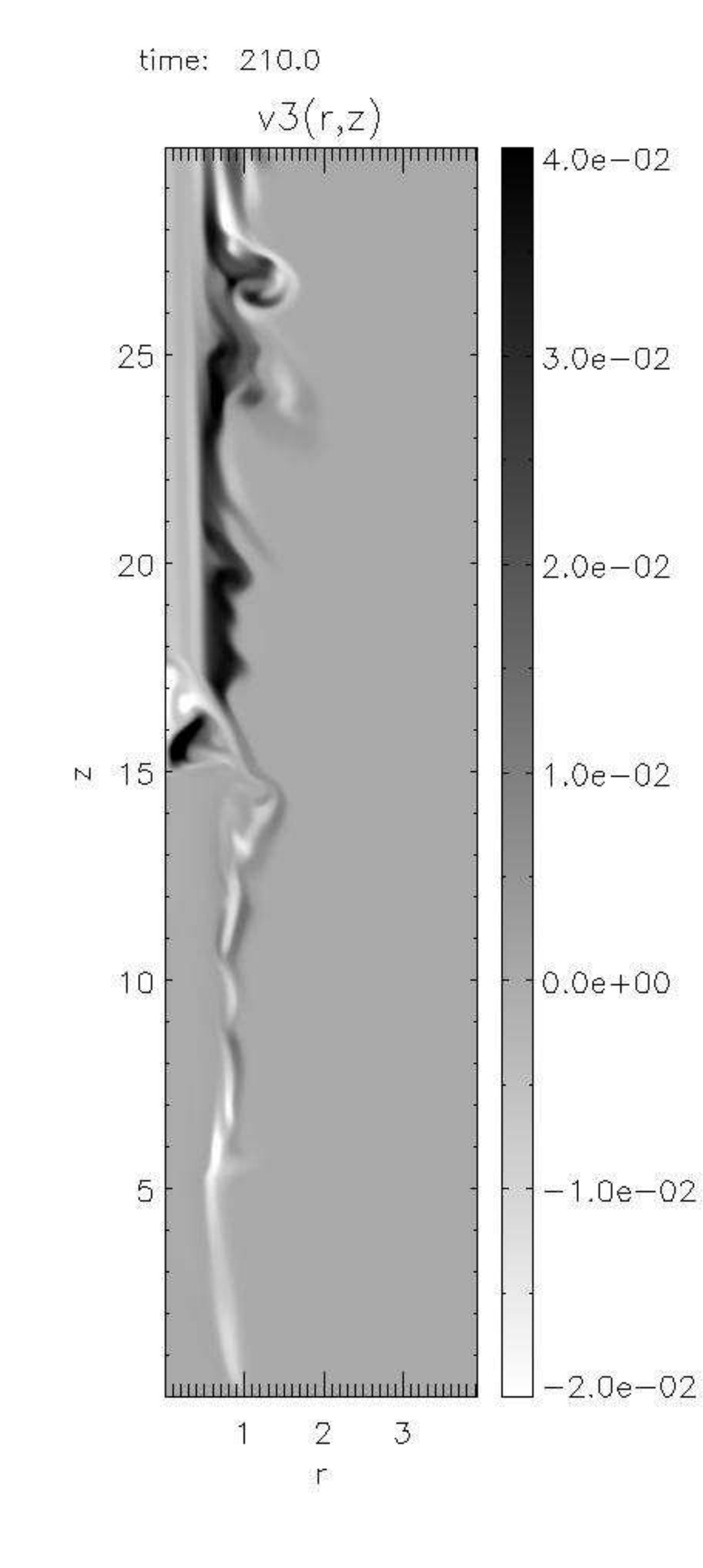}   
 \includegraphics[width=4.0cm,height=6cm]{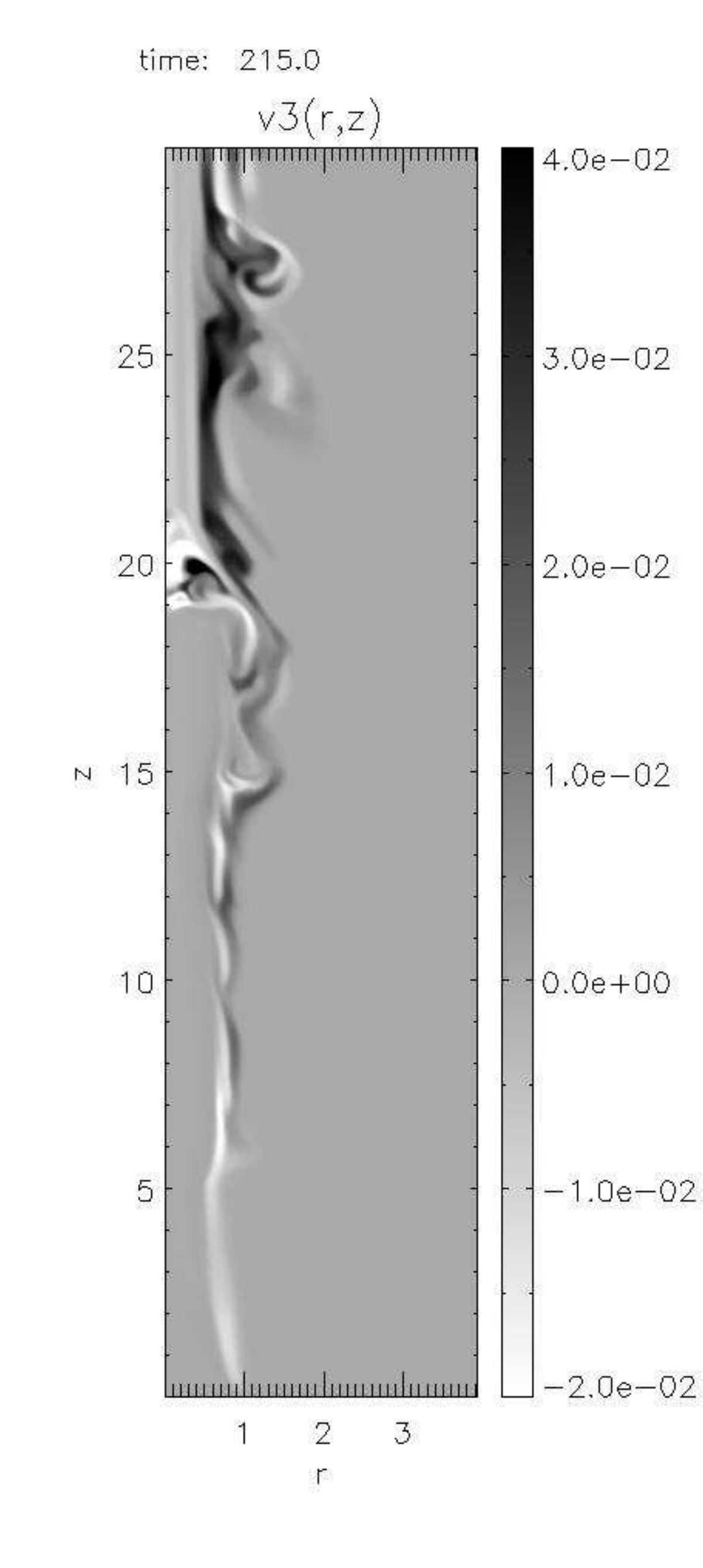}   
 \includegraphics[width=4.0cm,height=6cm]{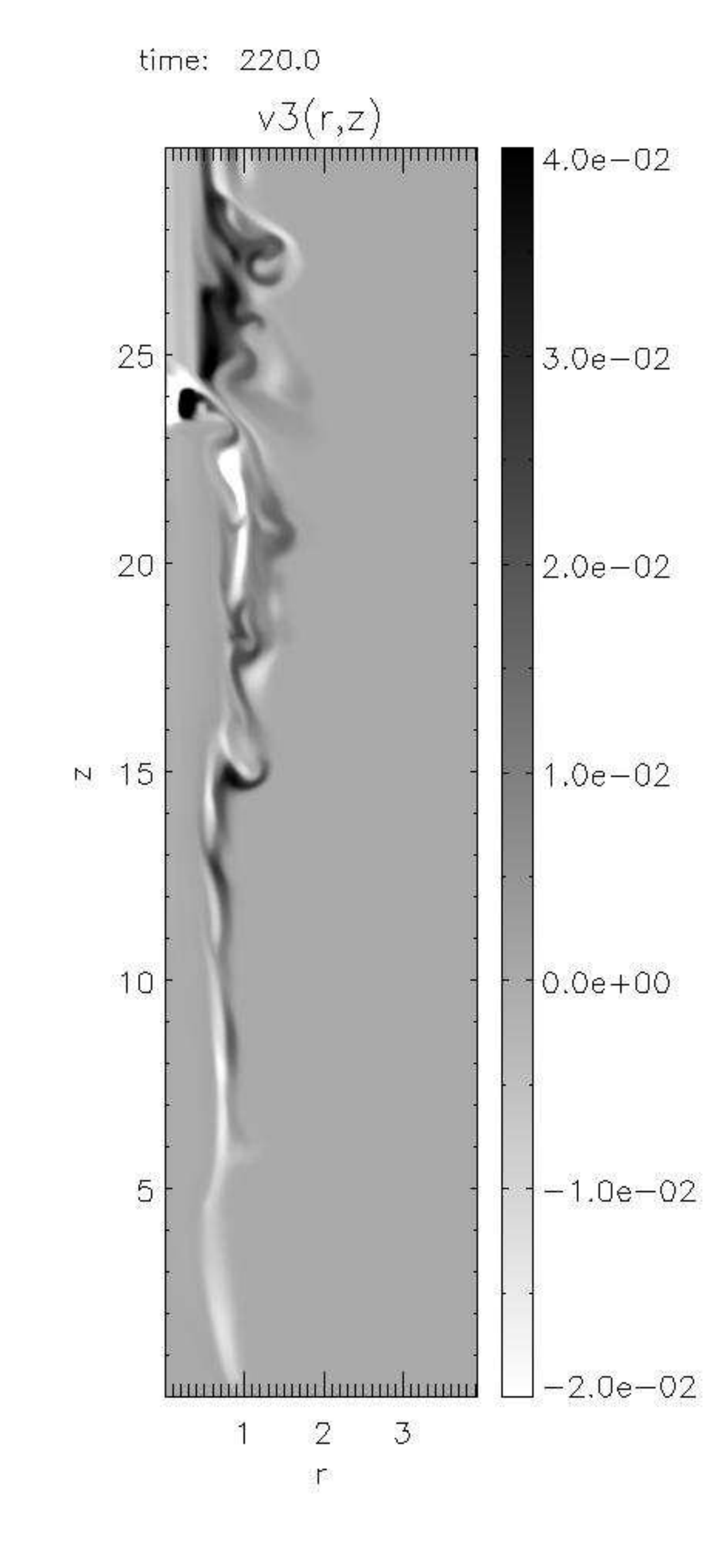}   
\caption{Axisymmetric simulation run R16 considers a high Mach number jet 
 similar to R12, however with a time-dependent injection speed, 
 thus generating internal shock waves moving along the jet.
 Shown is the toroidal velocity at dynamical time steps
 $t=75, 80, 90, 100, 125, 140, 145, 150, 205, 210, 215, 220$.
 (from {\it top left} to {\it bottom right}) for
 a sub-grid of $r<4.0$ of the whole computational domain 
 $(0.0<r<10.0, 0.0<z<30.0)$.
 The jet nozzle is between $0 < r < 1.0$.
\label{fig:sim_run16a}
}
\end{figure*}

\begin{figure*}
\centering
 \includegraphics[width=4.0cm,height=6cm]{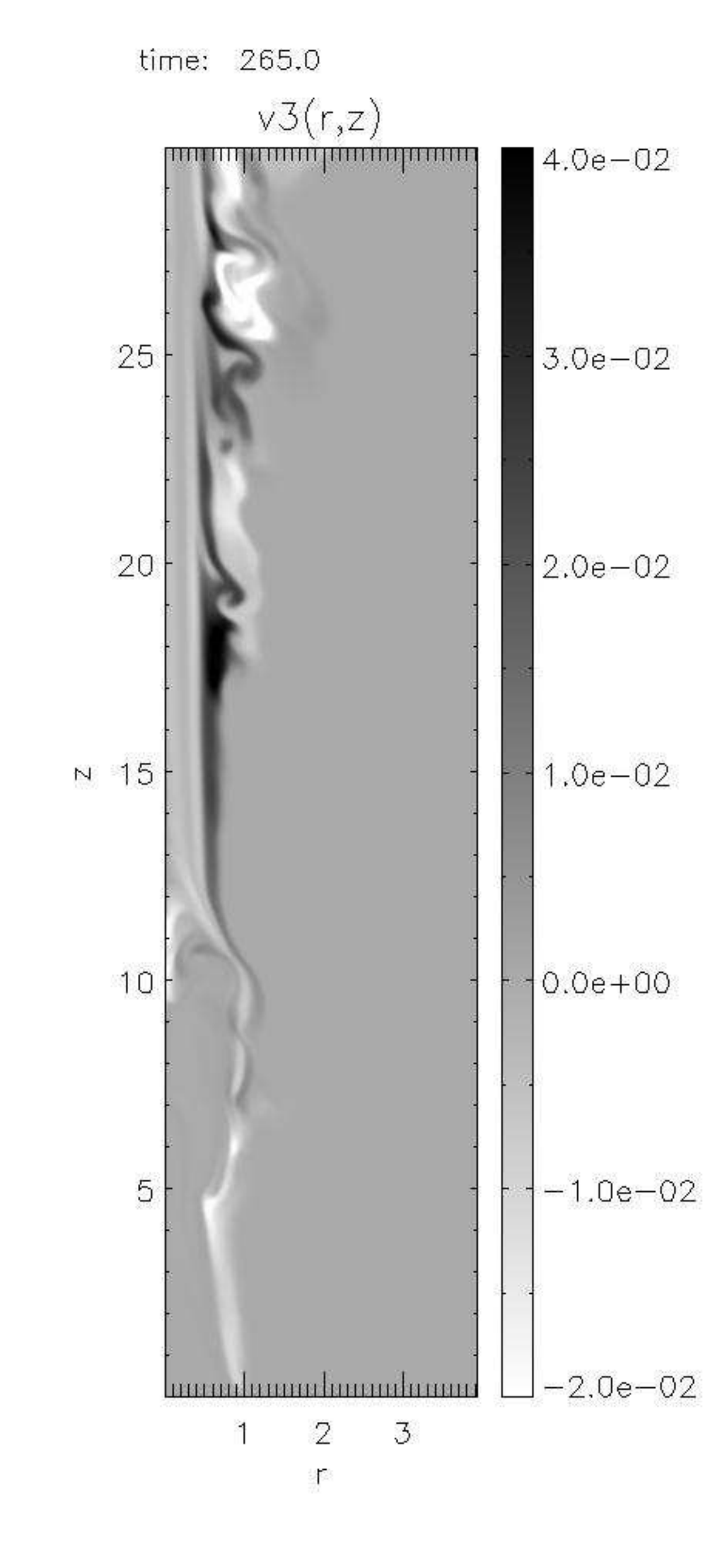}   
 \includegraphics[width=4.0cm,height=6cm]{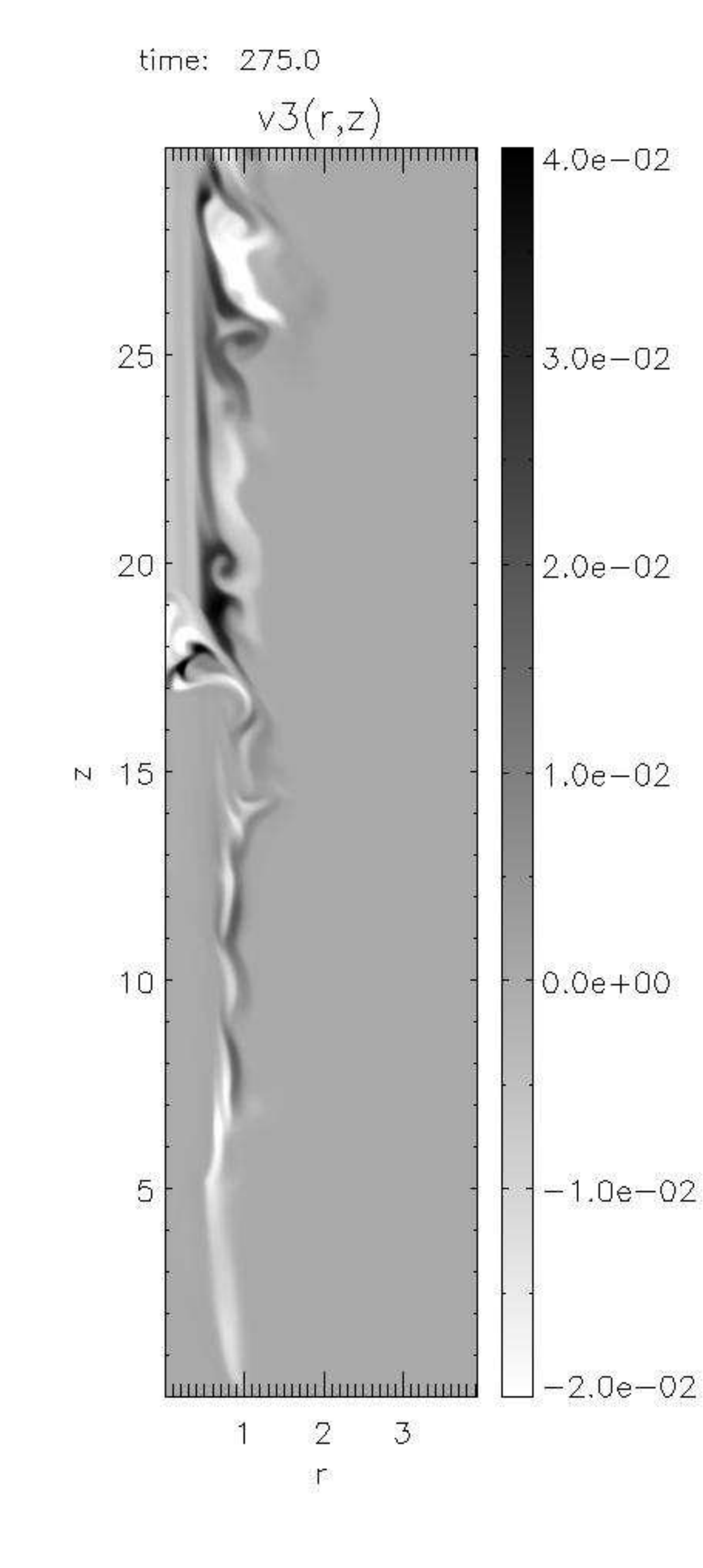}   
 \includegraphics[width=4.0cm,height=6cm]{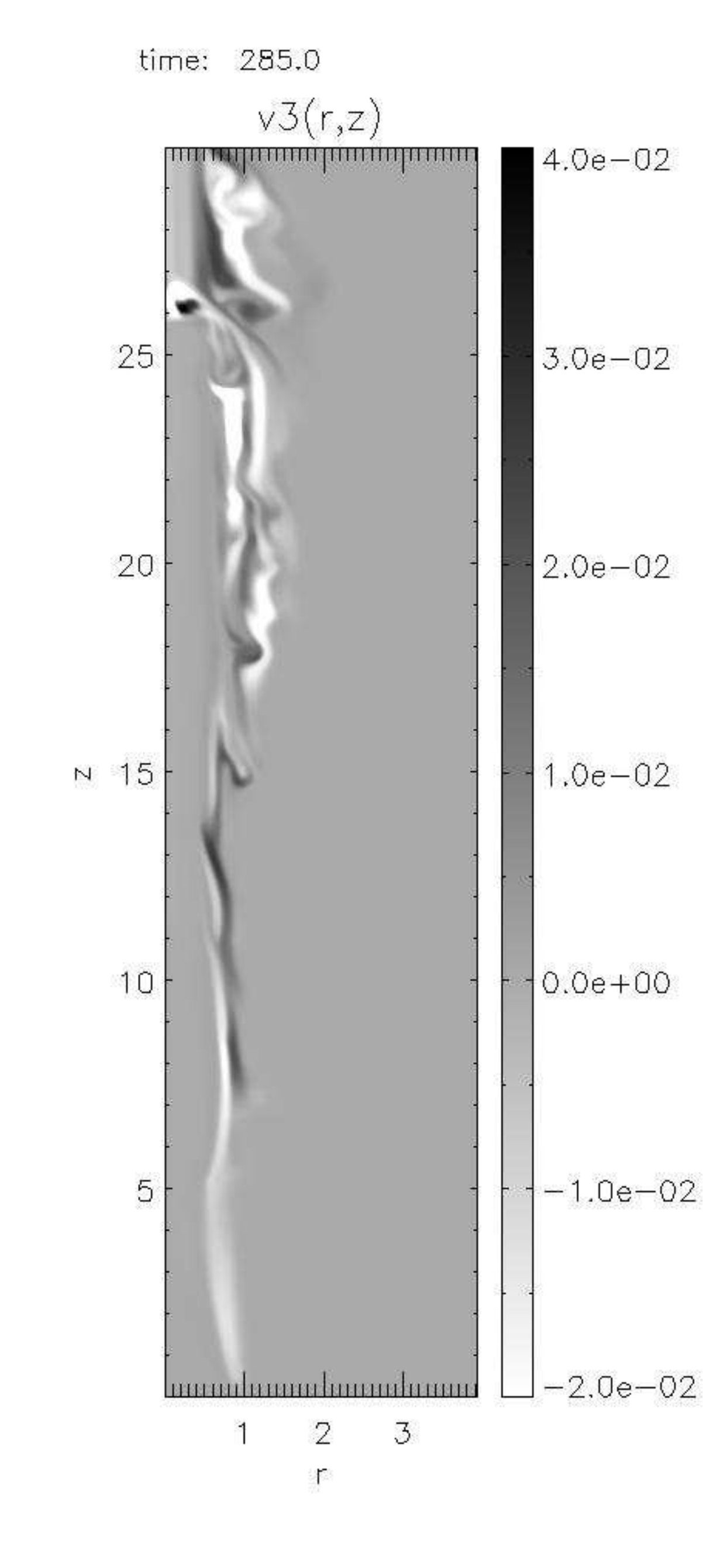}   
 \includegraphics[width=4.0cm,height=6cm]{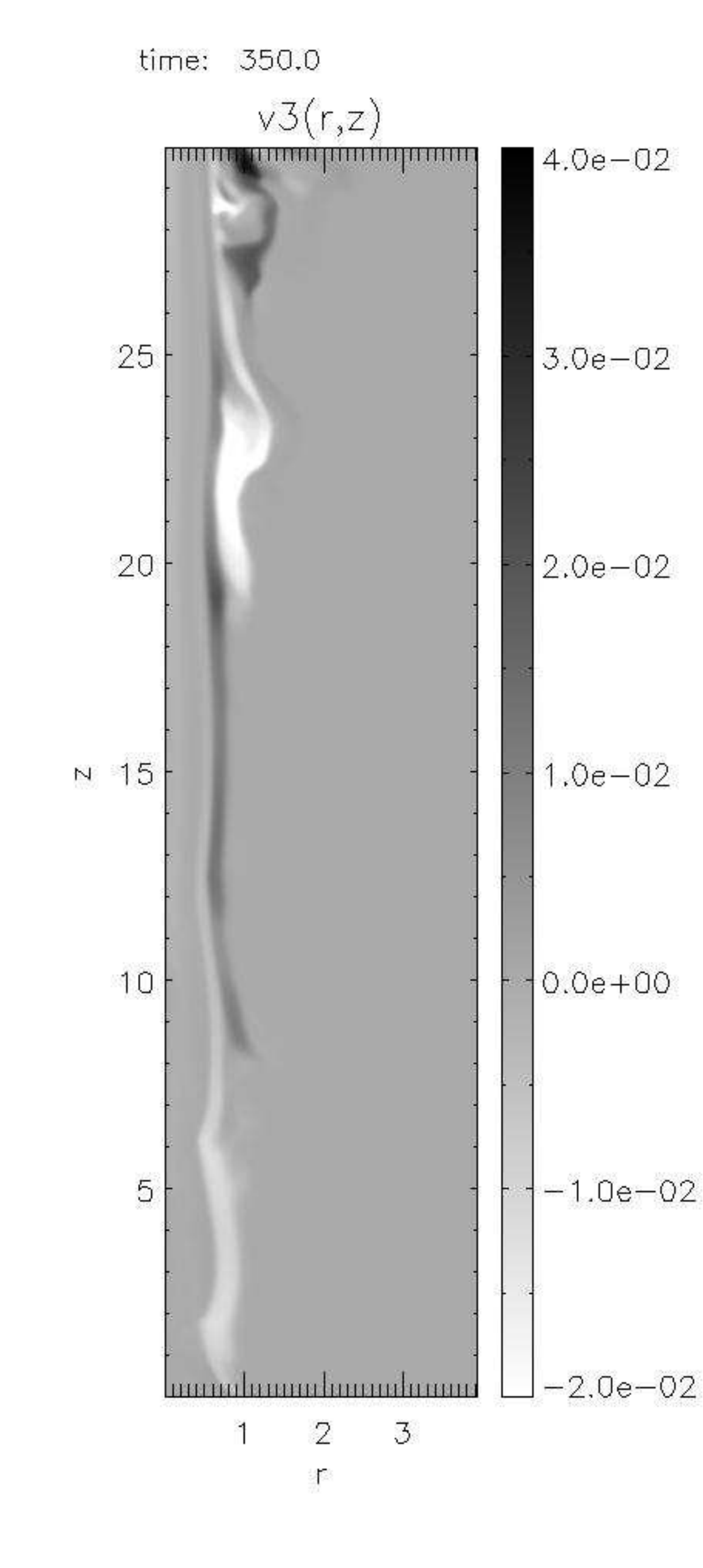}\\ 
 \includegraphics[width=4.0cm,height=6cm]{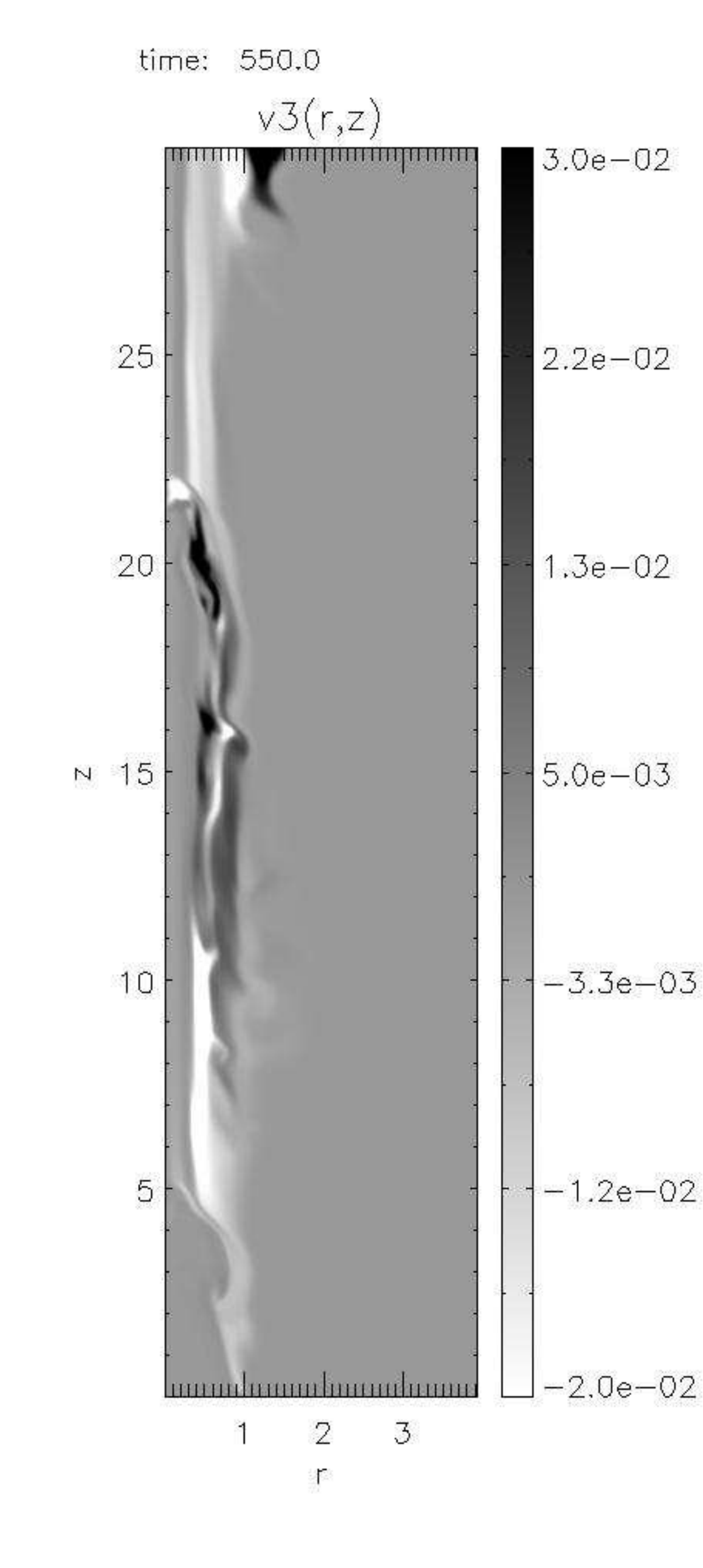}   
 \includegraphics[width=4.0cm,height=6cm]{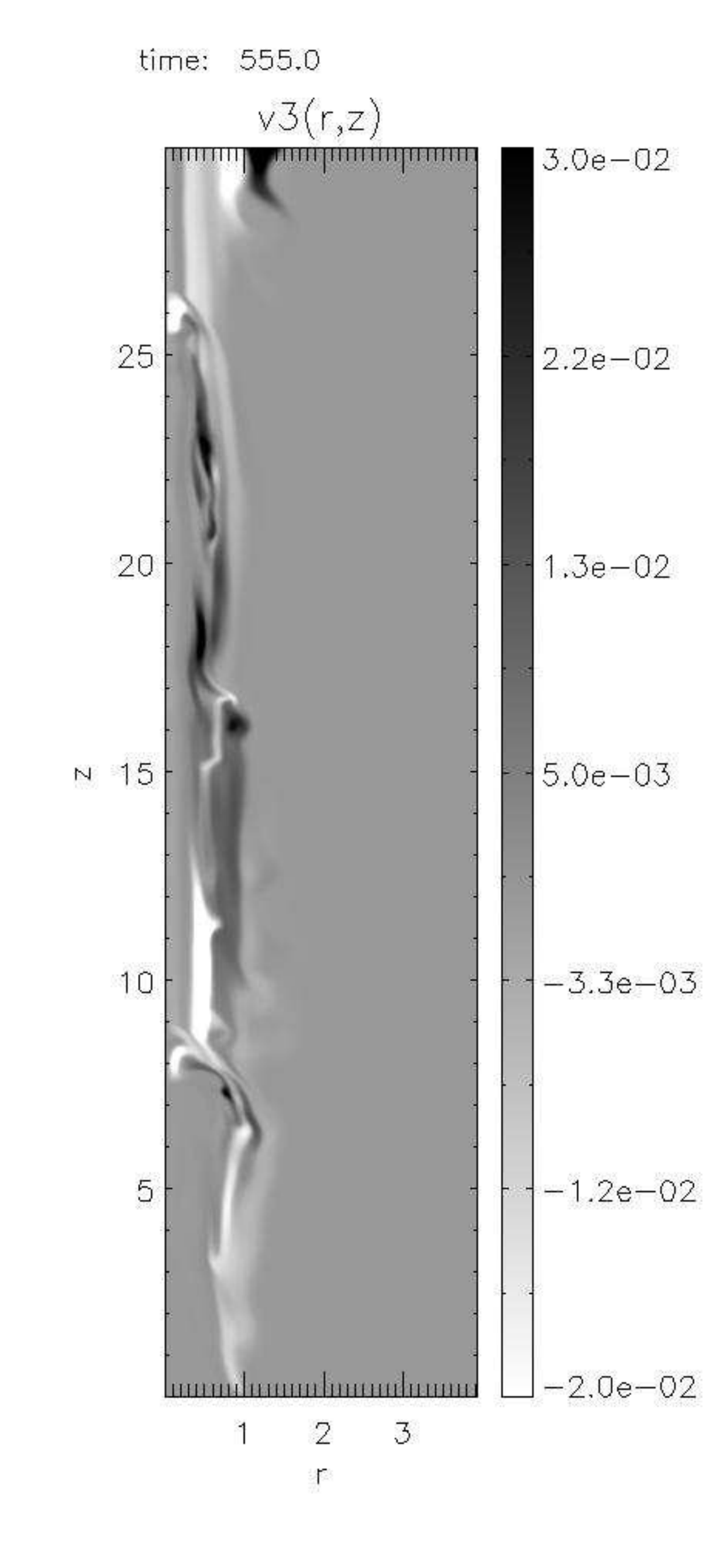}   
 \includegraphics[width=4.0cm,height=6cm]{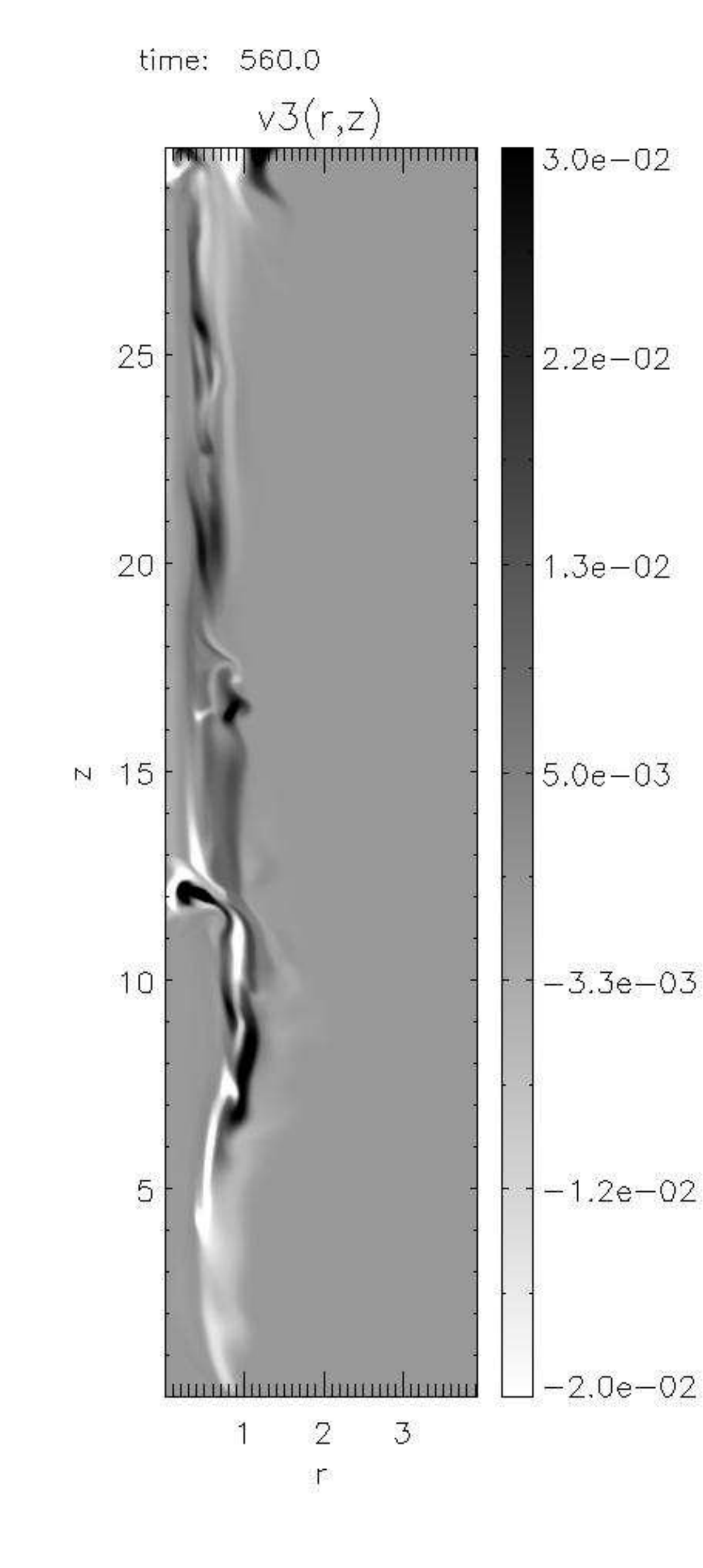}   
 \includegraphics[width=4.0cm,height=6cm]{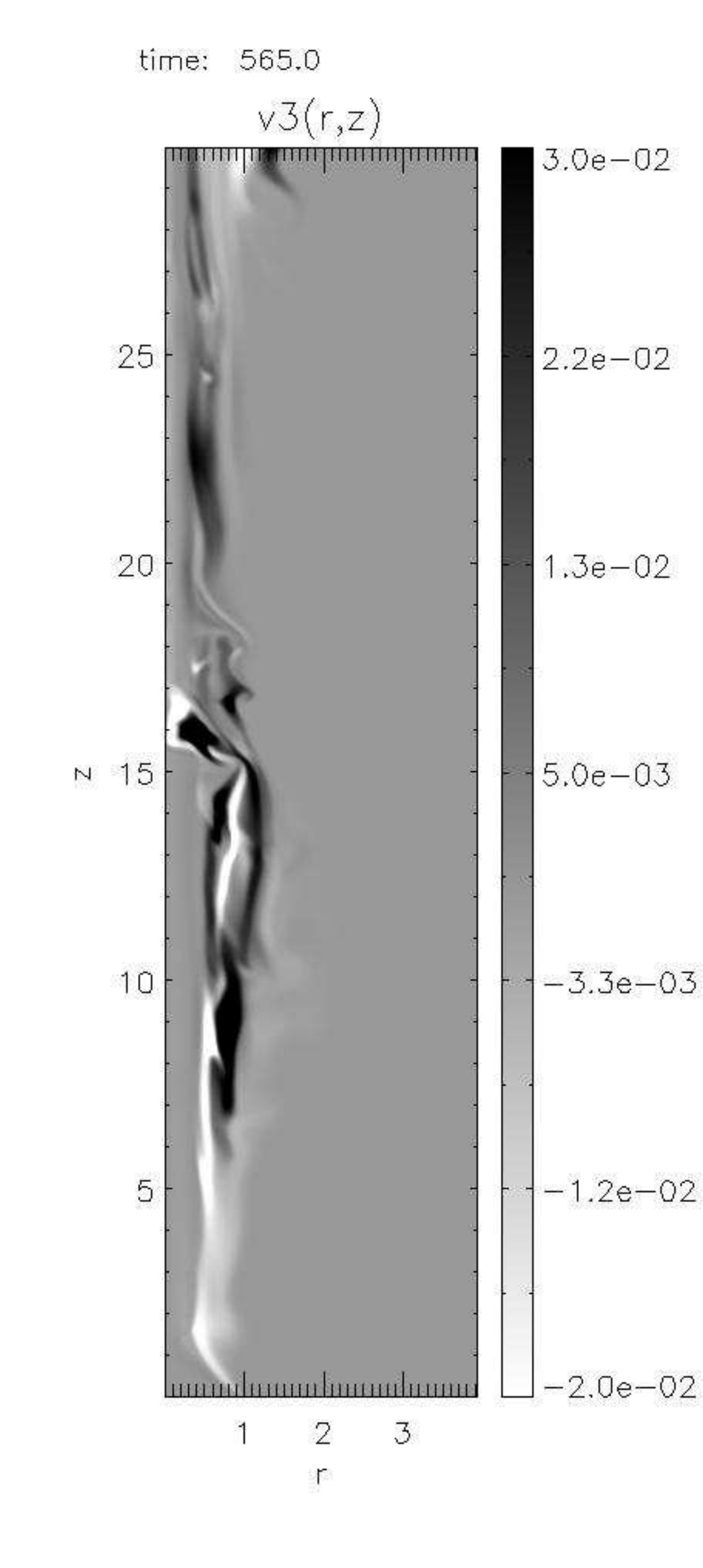}\\ 
 \includegraphics[width=4.0cm,height=6cm]{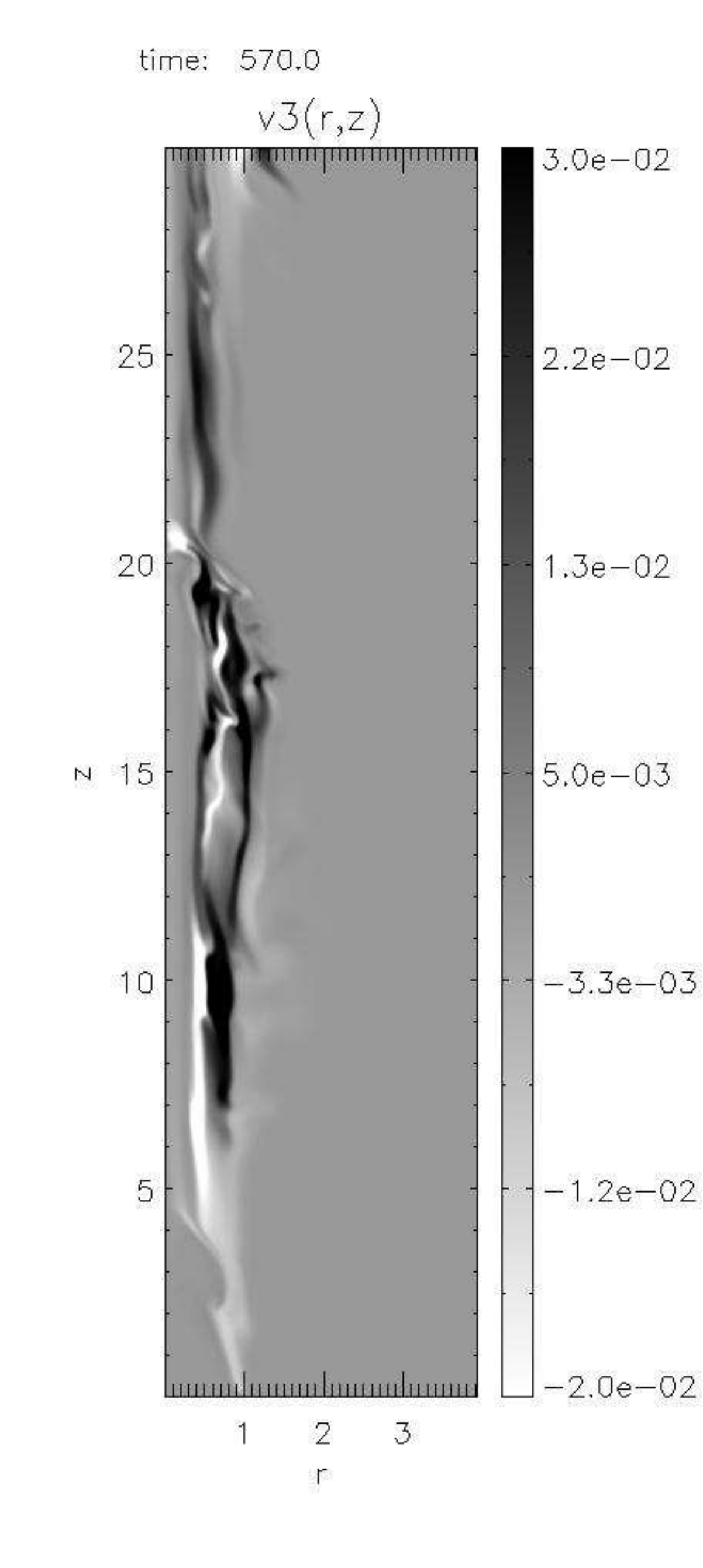}   
 \includegraphics[width=4.0cm,height=6cm]{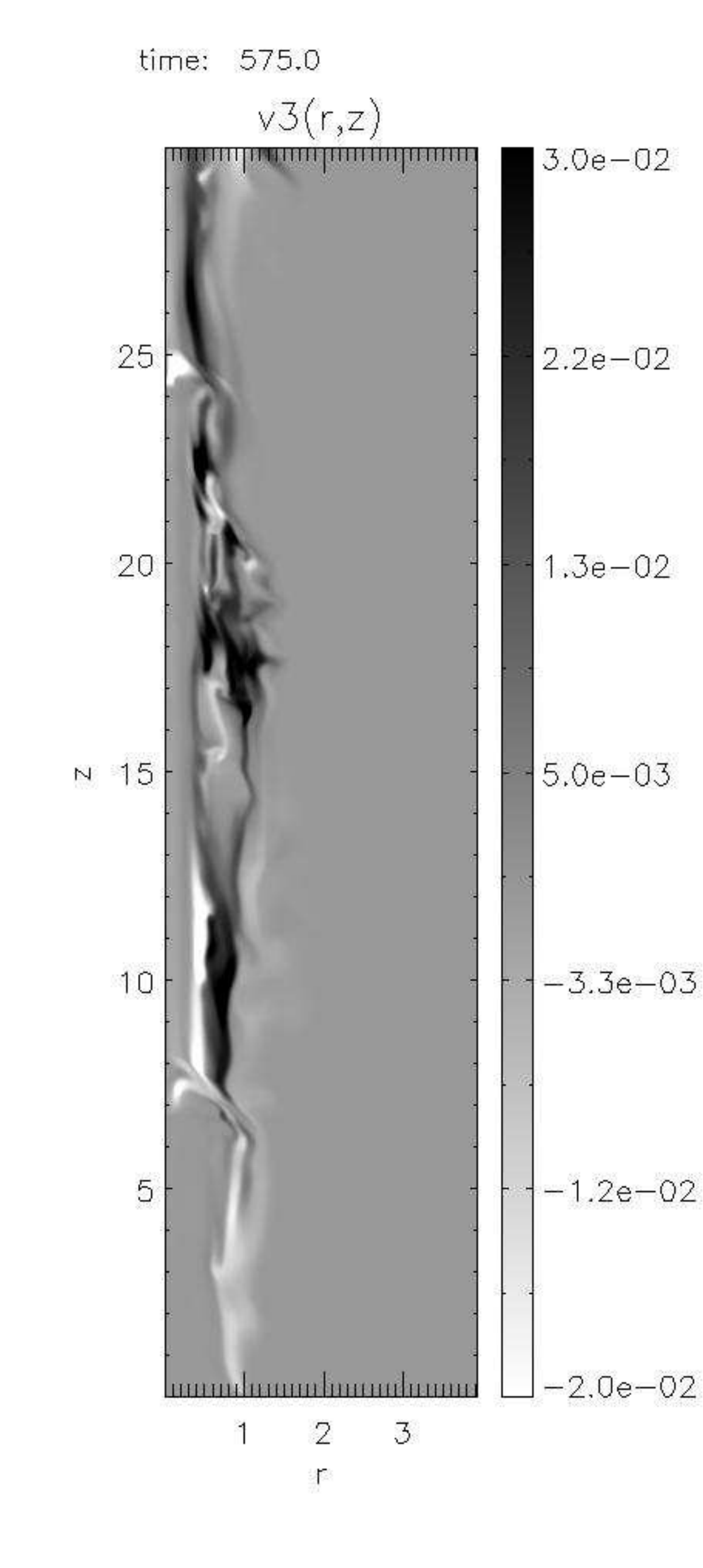}   
 \includegraphics[width=4.0cm,height=6cm]{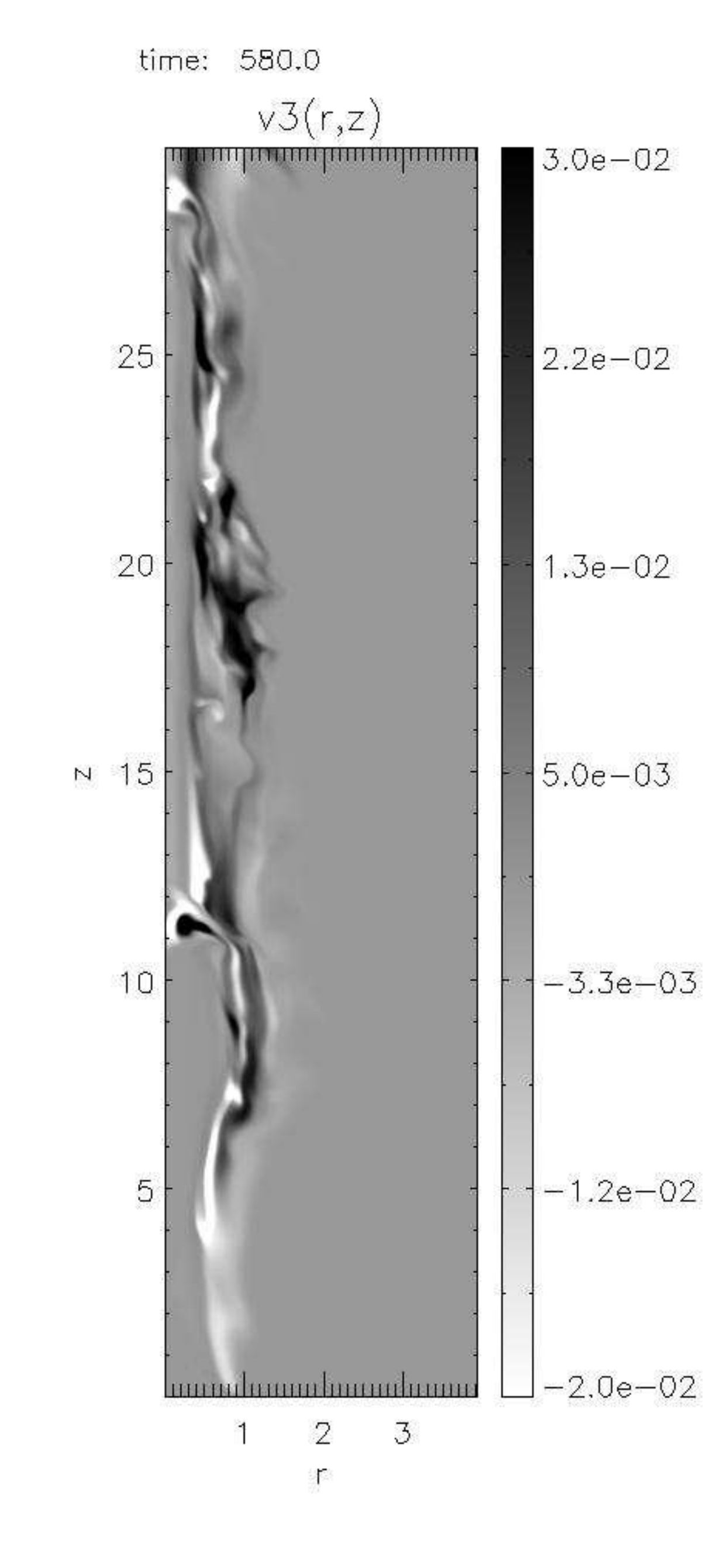}   
 \includegraphics[width=4.0cm,height=6cm]{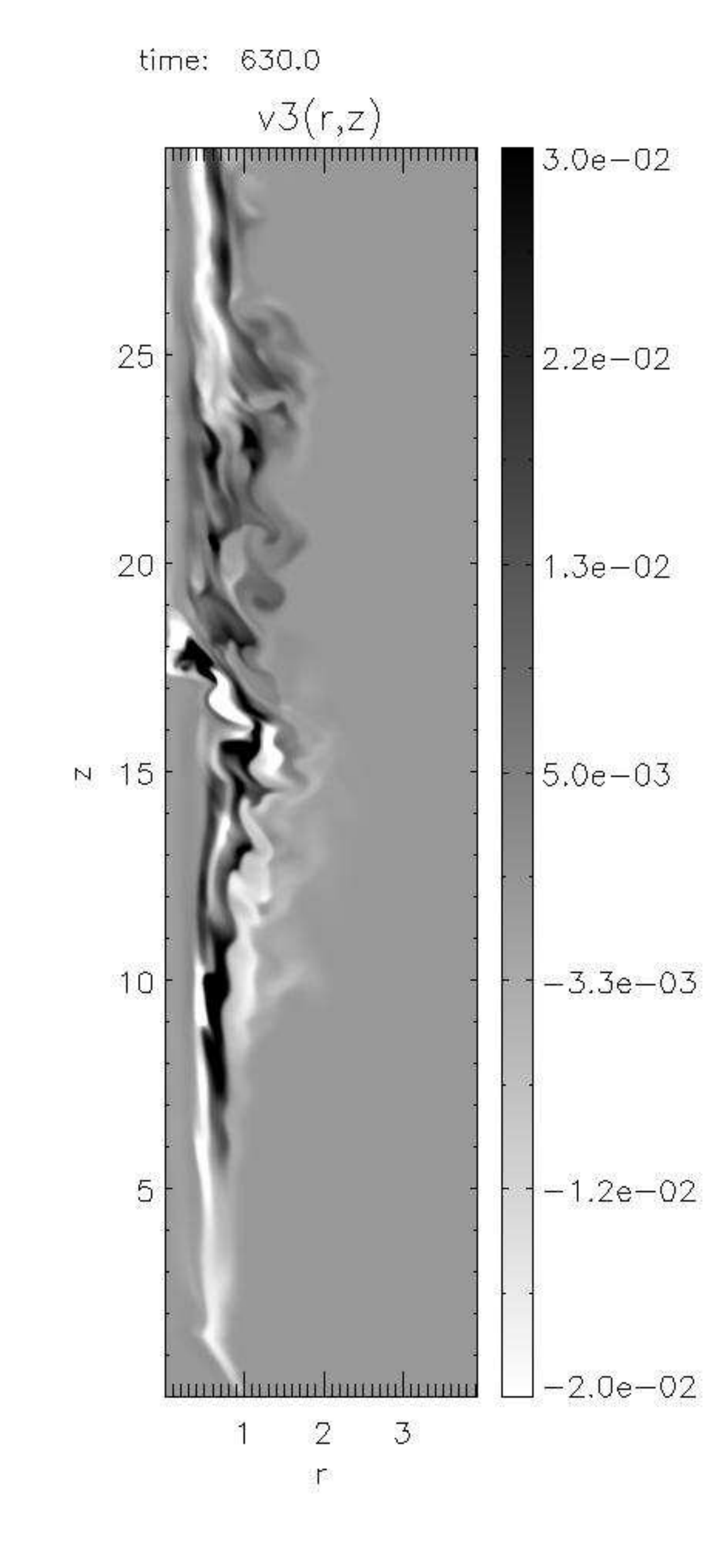}   
\caption{Axisymmetric simulation R16 continued with dynamical time steps
  $t = 265, 275, 285, 250, 550, 555, 560, 565, 570, 575, 580, 630$.
\label{fig:sim_run16b}
}
\end{figure*}

Simulation R16 runs with the same parameter setup as R12, but in 
difference applies a time-dependent injection speed,
$v_{\rm jet}(t) =  v_{\rm jet,0}   (1.0 + 0.5 \sin(0.3 t))$.
This leads to subsequent internal shocks along the jet, which
due to $B_{\phi}$-compression also force the jet material in rotation
(see Fig.~\ref{fig:sim_run16a}, \ref{fig:sim_run16b}).
In addition, each of the subsequent shock fronts further
entrains ambient material. 
Accompanying shocks set this material into rotation as well.
For this setup, it might be interesting to estimate the astrophysical
parameters.
Assuming a jet radius of 20\,AU, the separation between the generated
knots would correspond to about 600\,AU (one grid length).
The knot pattern separation would then correspond to a time scale of 
15\,years,
estimated from the 50 time units the knot pattern needs to traverse 
the 600\,AU grid (one time unit $\Delta t =1$ corresponds to about 
$10^7$s  assuming a jet mean velocity of 300\,km/s).
This time scale is of course put in by hand, but is not very different
from observationally derived numbers.

\begin{figure}
\centering
\includegraphics[width=7cm,height=11cm]{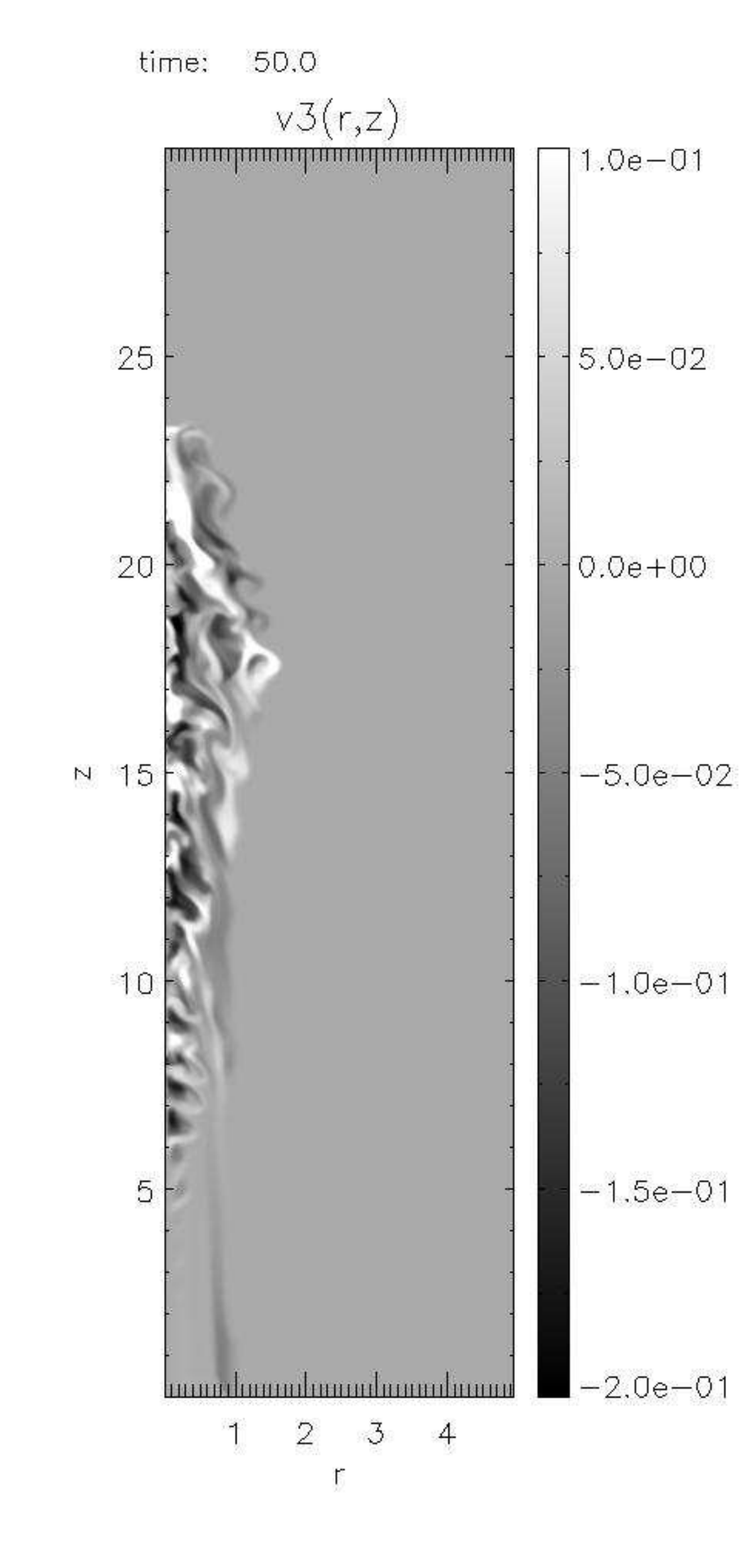}   
\caption{Toroidal velocity distribution of axisymmetric simulation run R19.
\label{fig:sim_run19}
}
\end{figure}

Simulation run R19 with lower Mach / Alfv\'en Mach numbers compared to R16 
has a slightly higher frequency in the injection speed variation.
Together this results in more internal substructure, seen e.g. in the toroidal 
velocity distribution (see Fig.~\ref{fig:sim_run19}), but also in the other
dynamical variables (not shown).
(and density,
\ref{fig:sim_run01}
not shown).

\subsection{Oblique reflection shocks}
\begin{figure*}
\centering
 \includegraphics[width=5.0cm,height=9cm]{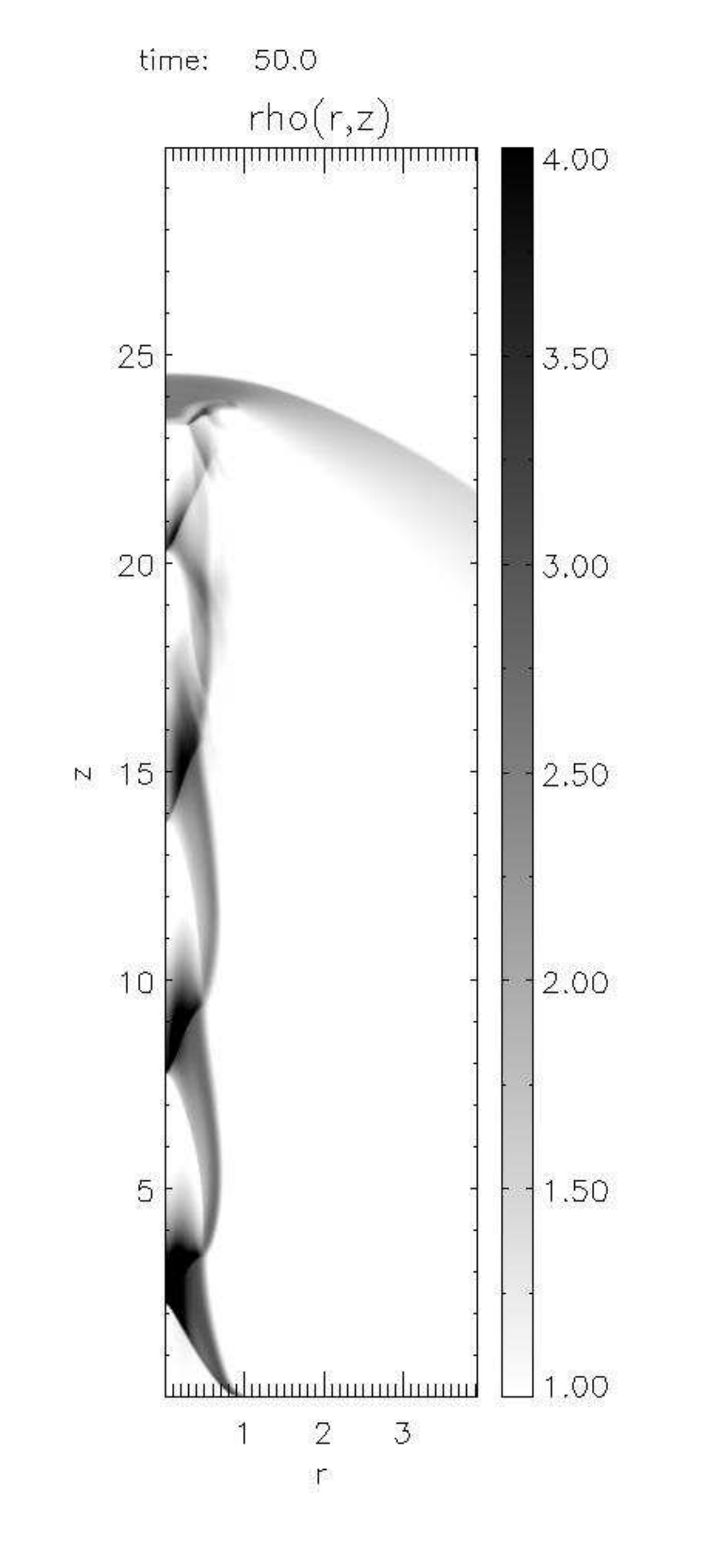}   
 \includegraphics[width=5.0cm,height=9cm]{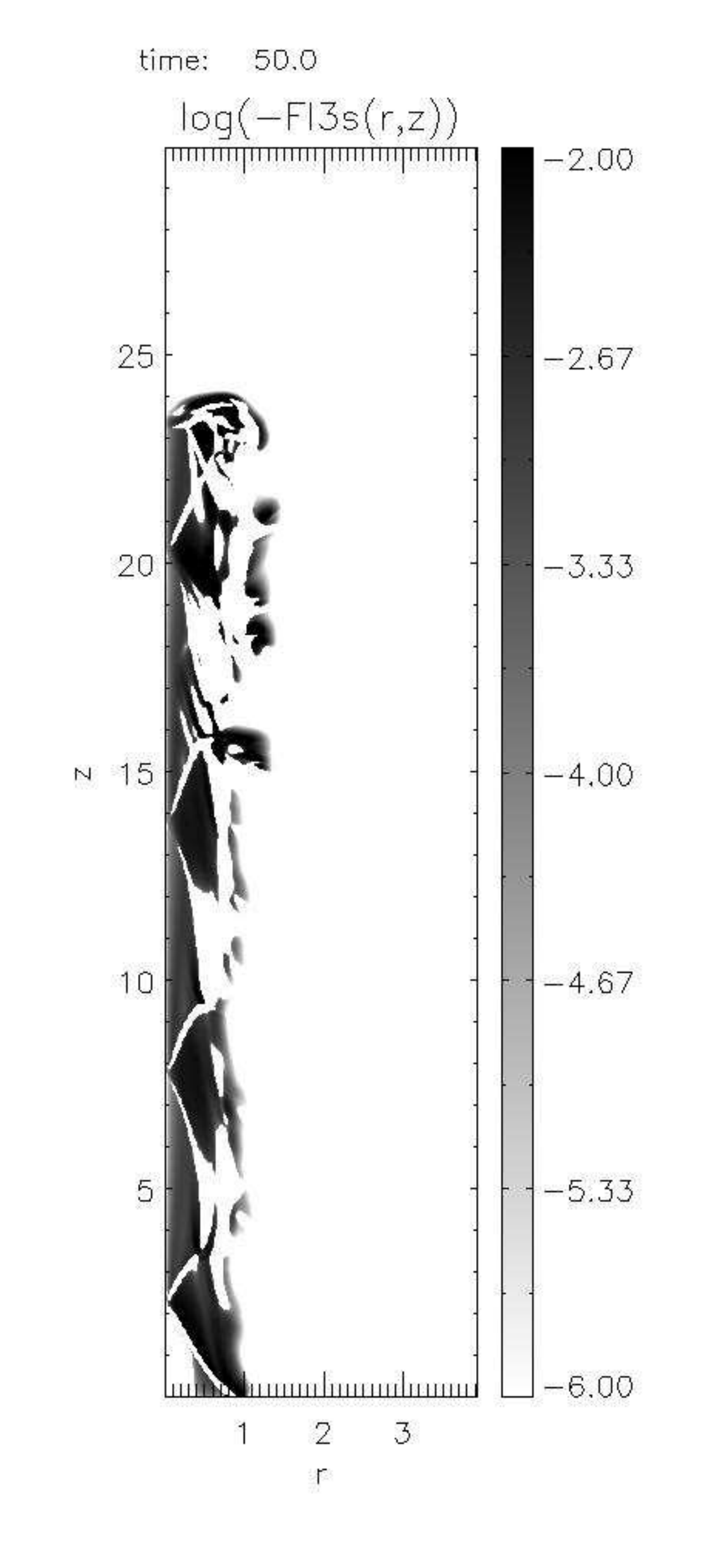}   
 \includegraphics[width=5.0cm,height=9cm]{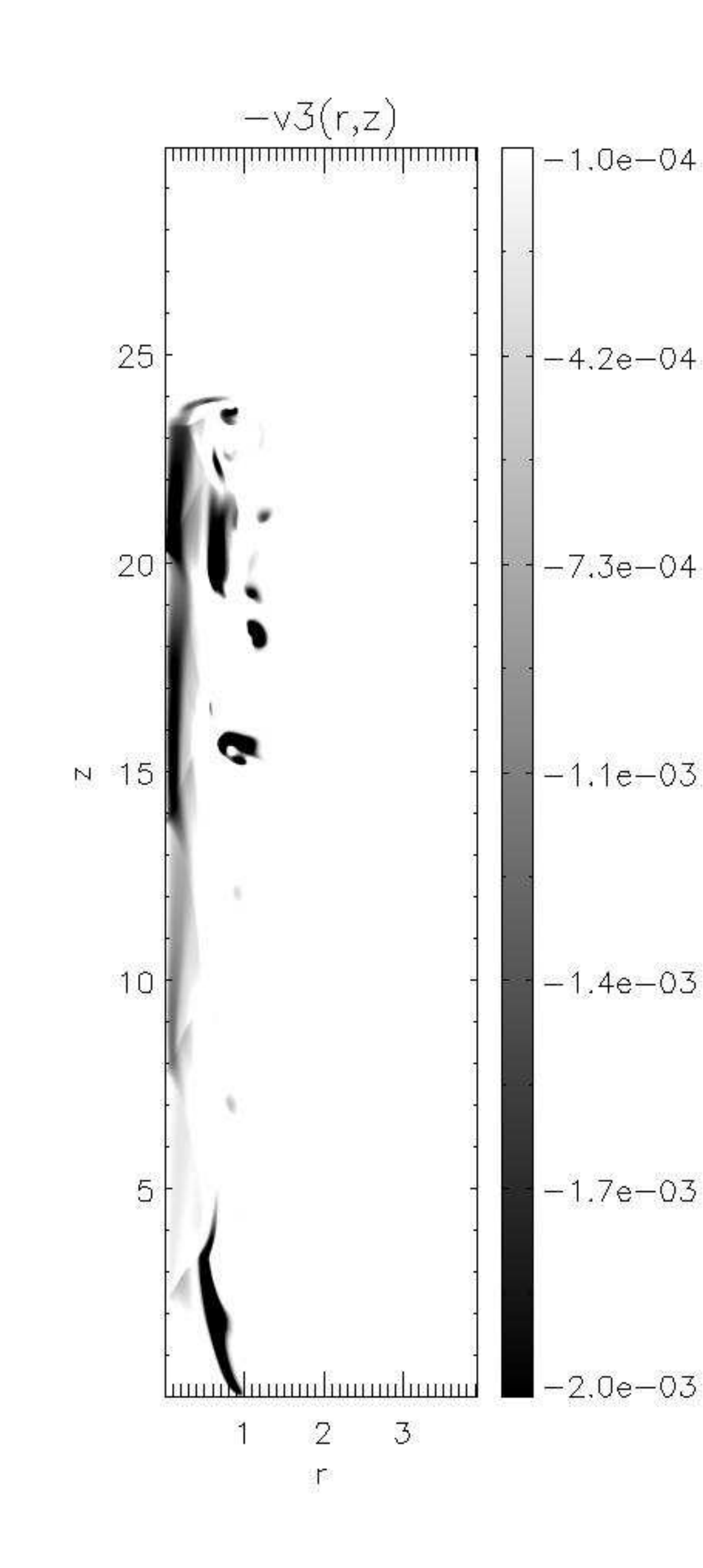}\\ 
 \includegraphics[width=5.0cm,height=9cm]{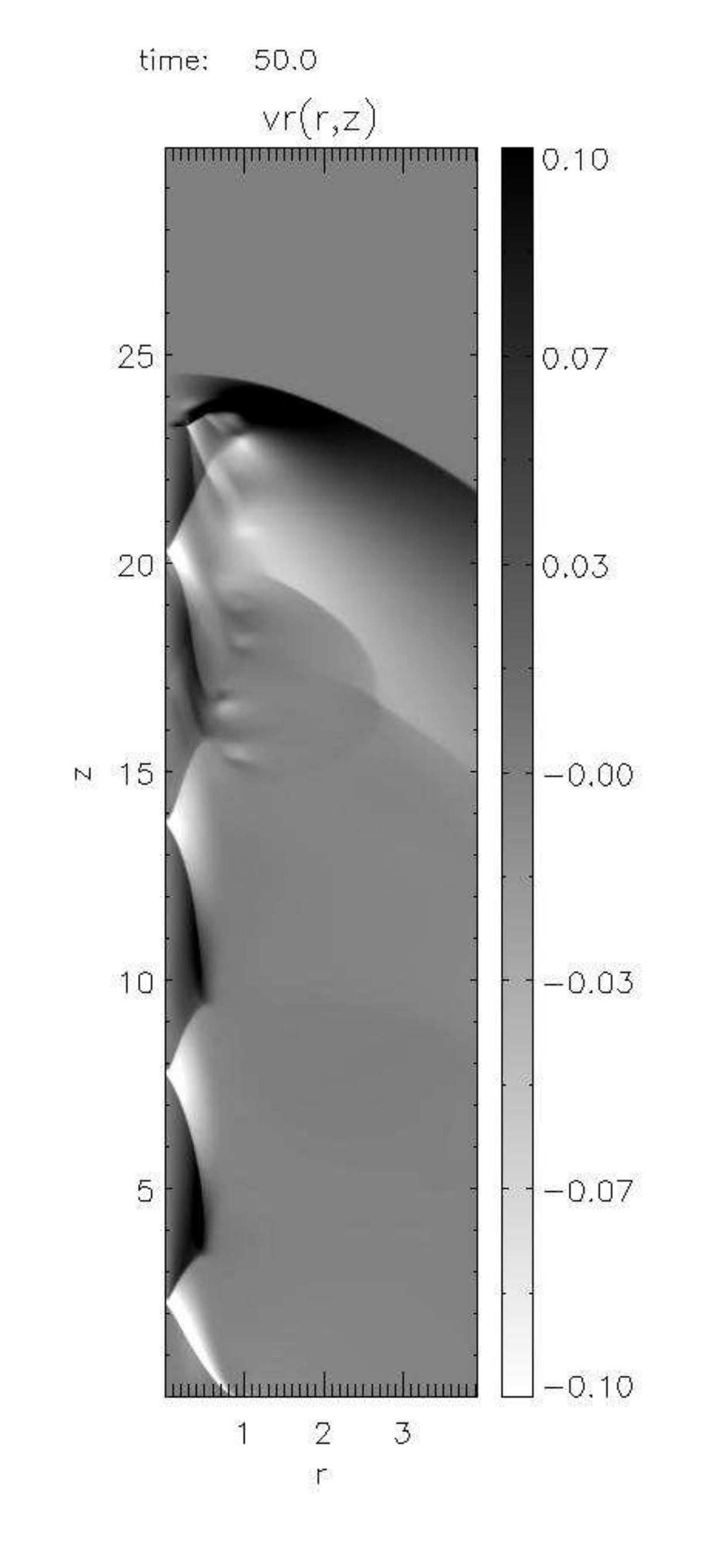}   
 \includegraphics[width=5.0cm,height=9cm]{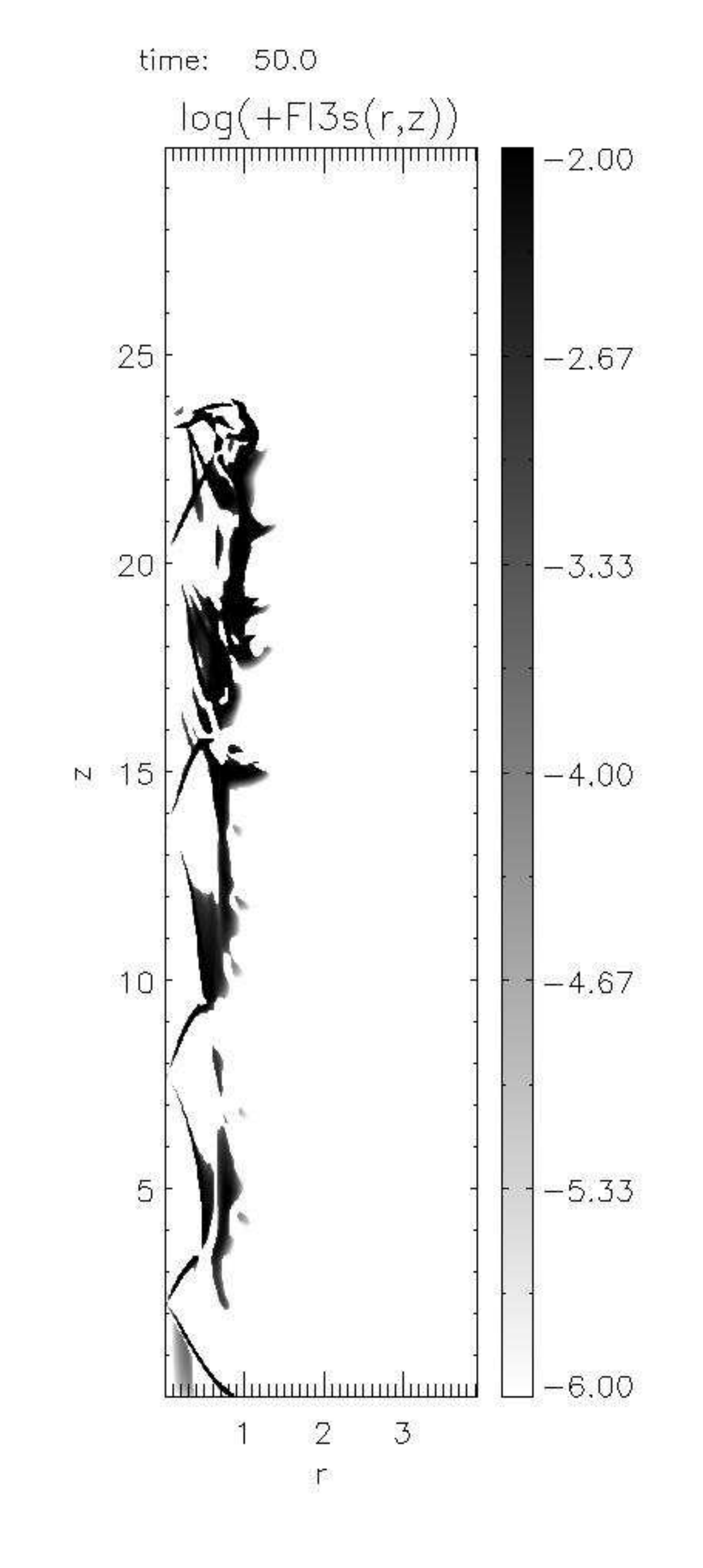}   
 \includegraphics[width=5.0cm,height=9cm]{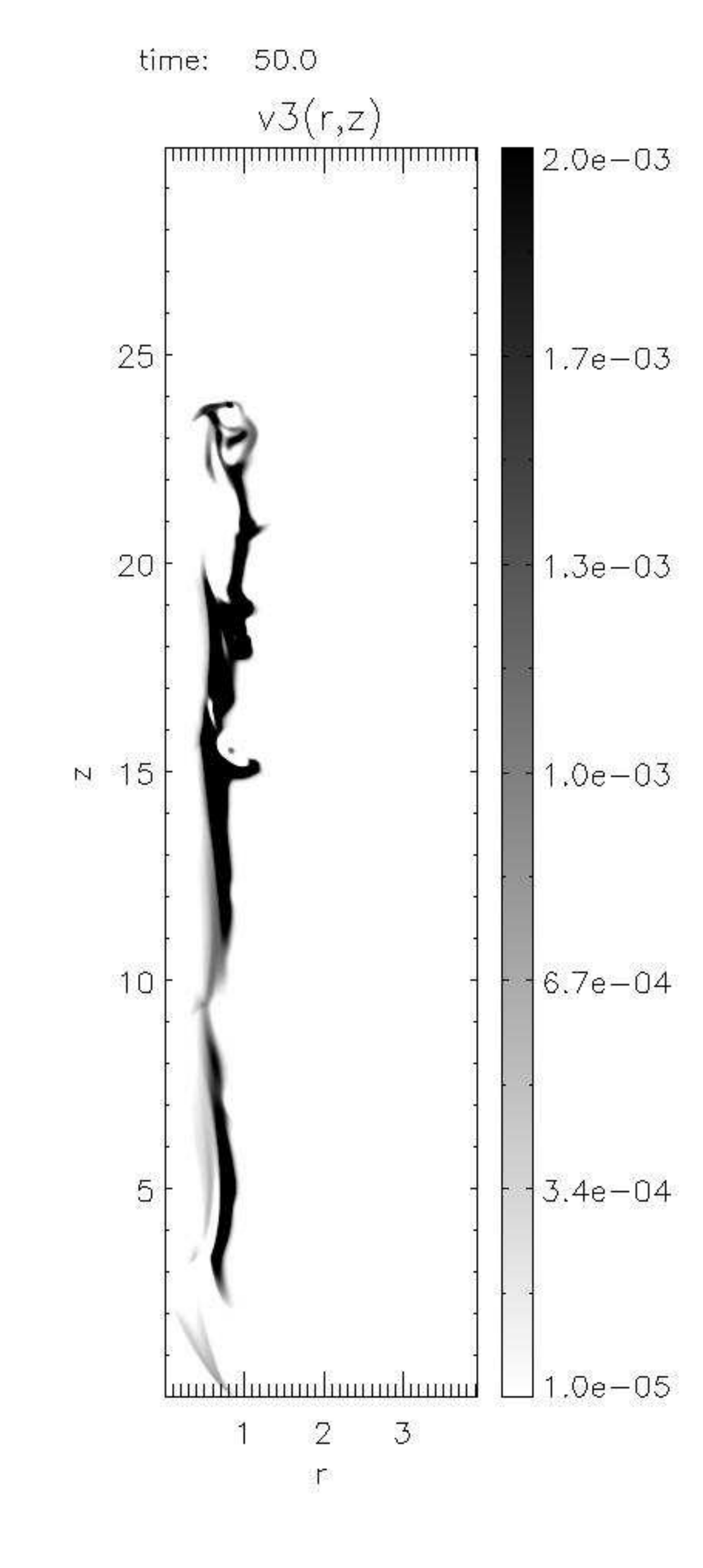}   
\caption{Axisymmetric simulation run R12.
 Shown are the density distribution ({\it top left}), 
 the radial velocity distribution  ({\it lower left}),
 the negative ({\it top}) and positive ({\it bottom}) toroidal specific Lorentz 
 force components ({\it middle}),
 and negative ({\it top}) and positive ({\it bottom}) toroidal velocity
 components ({\it right}) at dynamical time $t=50$.
\label{fig:sim_run12b}
}
\end{figure*}

Jets with low (internal) Mach numbers develop standing oblique relection 
shocks internal to the jet flow which do not further move as jet and bow 
shock propagate.
As an example Fig.~\ref{fig:sim_run12b} shows the dynamical state of 
simulation run R12 at an intermediate time step $t=50$.
The reflection shocks are most clearly seen in the density distribution
and the radial velocity distribution. The change in propagation speed
along the jet axis is about 15\% across each shock region. 

Along the jet flow the toroidal Lorentz force is predominantly
negative leading to an according rotation of the jet material.
The shock structure consists of a thin layer with positively
directed toroidal Lorentz force followed (direction downstream) 
by a extended region of positively directed toroidal Lorentz force.
The latter force is dominating leading to an overall negative
rotation of the jet beam.

Comparing the toroidal Lorentz force and toroidal velocity distribution
in Fig.~\ref{fig:sim_run12b} gives interesting insight on the acceleration
process.
The positive forces are strongly localized across the relection shocks.
However, we don't see there effect in a positive jet rotation with in the
jet beam.
In comparison, the negative forces are distributed over a wide range
between the internal shocks. As a result the (negative) rotation velocity
of the jet beam increases along the propagation direction. 
The increase is step-wise with each of the inter-shock region adding
to the overall rotation (see the step-wise increase of dark color along the
jet flow in the upper right Fig.~\ref{fig:sim_run12b}). 

Along the jet beam the negative Lorentz force wins and the overall jet
rotation is negative.
The situation is somewhat different for the material which is entrained 
along the jet flow. Here the negative Lorentz force is more strongly 
localized and the positive force wins resulting in an overall positive
rotation of the ambient (entrained) flow.

This picture of a gradual increase of rotation by cumulative action of 
Lorentz forces along the path of the material is also supported by the
small rotation feature visible near the jet head.
A closer look shows that the respective Lorentz force and velocity patterns
are slightly shifted. 
This tells us the jet $\phi$-acceleration is not, say in situ,
but that what we observe in rotation is the result of previously acting
Lorentz forces.
The same trend can be seen in the small scale toroidal velocity structures
around the jet in the positive velocities where observe a similar shift
between the positive Lorentz force and the positive toroidal velocity.

\subsection{Time evolution of angular momentum}
The magnetic field structure resulting from our rather basic and general
simulation setup evolves into a complex helical field distribution 
implying a similarly complex electric current system.
With the propagation of the outflow and the evolving jet-internal 
dynamics 
the toroidal and poloidal field components change direction resulting 
in a similar variation of the torques along and across the jet.
This is demonstrated exemplarily in Fig.~\ref{fig:sim_run01}, 
where we show the distribution of the specific toroidal Lorentz forces,
resp. the acceleration in toroidal direction.
Comparison of positive (right) and negative (left) Lorentz forces,
highlights the filamentary structure of the force distribution.
The structural details clearly depend on the specifics of the simulation 
setup and deserve a closer look by a future investigation.
Observational indications for such a filamentary structure may be obtained 
from radial velocity profiles across the jet if the spatial resolution will
be sufficient.

The time-evolution shown in Fig.~\ref{fig:sim_run01} is typical for the early
jet evolution - the jet is penetrating the ambient medium and thereby changes
the dynamical state compared to its initial injection profile. 
However, jet material and ambient gas are coupled by the magnetic field which 
is anchored into both components.
The differential rotation between jet and ambient gas results in shear and 
torques.
In later evolutionary stages when the jet has bored its funnel through the 
ambient gas, the differential rotation decreases, and the twist in the field 
structure is reduced.
Jet injections of different velocity will then lead to internal shocks
(see simulation R16, Fig.~\ref{fig:sim_run16a}, \ref{fig:sim_run16b}).

\begin{figure}
\centering
\includegraphics[width=1.0\columnwidth]{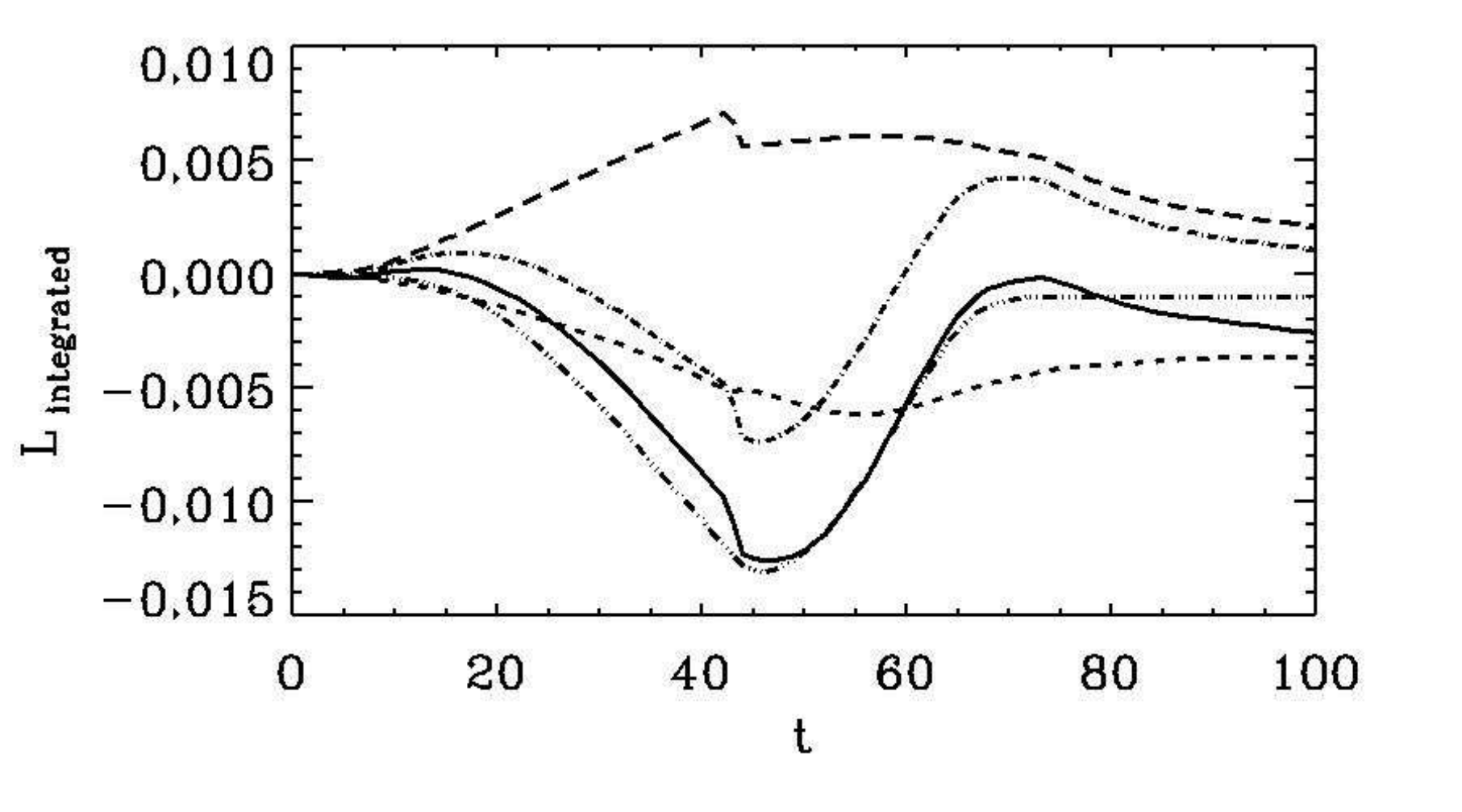}  
\caption{Time evolution of the domain-integrated angular momentum (a.m.)
contributions for example simulation R01. Shown is the
kinetic a.m. (positive contribution, {\it thin dashed}), 
kinetic a.m. (total contribution, {\it  thick dotted-dashed}), 
magnetic a.m. (negative contribution, {\it dashed}), 
kinetic a.m. (negative contribution, {\it thin dotted-dashed}), 
and the total a.m. ({\it solid line}).
Around time $t\simeq 45$ the jet terminal shock leaves the computational domain.
\label{fig:sim_am_evol_R01}
}
\end{figure}

Figure \ref{fig:sim_am_evol_R01} reveals the time sequence of the angular
momentum evolution as domain-integrated value for simulation R01.
Shown are the different contributions to the total angular momentum 
budget (i.e. not their specific value).
Since this is an density-weighted value, it is observationally more relevant
than the specific value (simply speaking, more gas will radiate more).
The observed jet rotation will not only depend on the rotation speed
alone, but also on how much material is in fact rotating with such speed.
The example of Fig.~\ref{fig:sim_am_evol_R01} shows at time $t=20$ the kinetic 
angular momentum on average more positive than negative.
This suggests that one may observe on average a positive rotation of the jet
although the negative velocity distribution is broader 
(see Fig.~\ref{fig:sim_run01}).


\begin{figure}
\centering
\includegraphics[width=1.0\columnwidth]{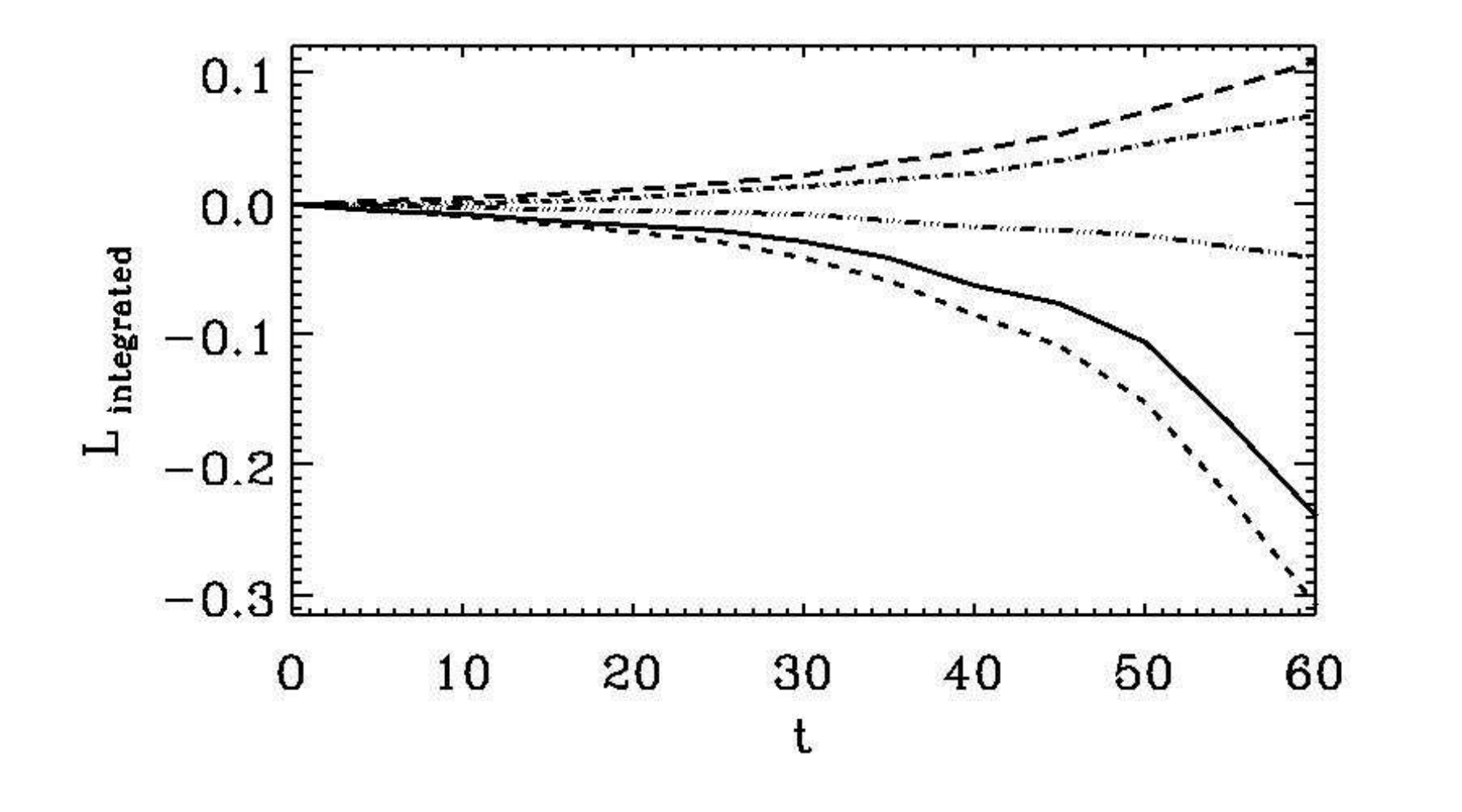}  
\caption{Time evolution of the domain-integrated angular momentum (a.m.)
contributions for example simulation R12. Shown is the
kinetic a.m. (positive contribution, {\it thin dashed}), 
kinetic a.m. (total contribution, {\it thick dotted-dashed}), 
magnetic a.m. (negative contribution, {\it dashed}), 
kinetic a.m. (negative contribution, {\it thin dotted-dashed}), 
and the total a.m. ({\it solid line}).
\label{fig:sim_am_evol_R12}
}
\end{figure}

At dynamical time $t=45$ the jet bow shock leaves the computational domain.
The jet material set into rotation by the processes we have discussed above 
will propagate out of the computational domain.
Since simulation R01 has a time-independent injection boundary condition,
no new shocks will be initiated which could enforce rotation
and the overall jet rotation will fade away - as indicated by 
Fig.~\ref{fig:sim_run01}.

As for simulation R01 we show in Fig.~\ref{fig:sim_am_evol_R12} 
the time evolution for the different contributions 
to the angular momentum budget for simulation R12.
In difference to R01 the amount of angular momentum is constantly growing.
This is due to the presence of the steady reflection shock which continue
to convert magnetic angular momentum to kinetic angular momentum, even when
the bow shock has left the computational domain.

\section{Conclusions}
%
We have performed axisymmetric MHD simulations of jet propagation
into an ambient gas applying the PLUTO code.
The goal of our study was to investigate how the gas flow in jets and 
outflows is set in rotation by Lorentz forces arising MHD shocks.

The process we discuss relies on a helical magnetic field structure 
which carries magnetic angular momentum from the jet base to the
asymptotic jet.
This could be converted into kinetic angular momentum by action of a
toroidal torque which is induced by shock compression of the toroidal 
field.
To demonstrate the applicability of the principal mechanism, for
simplicity we have injected a {\em non-rotating} jet material into 
the ambient gas for most cases.

Our simulations show that the proposed mechanism is indeed feasible, 
and accelerates the jet in toroidal direction by Lorentz forces which 
are induced by a shock compressed toroidal magnetic field.
Depending on the magnetohydrodynamic parameters of the jet flow,
the resulting rotational velocities are in the range of $0.1\% - 1\%$
of the jet propagation speed.

Jets with a high external Mach number gain the highest rotation speed.
This is because the mechanism works most efficient in the termination
shock.

The resulting velocity field is highly complex resulting in a filamentary
structure in the toroidal torque and toroidal velocity distribution.
The material which is set in rotation across the terminal shock, 
flows back respective to the propagating jet and forms a cocoon with a 
highly tangled rotation pattern.

Interestingly, 
of jets with the same initial magnetic angular momentum budget
(i.e. same $\sim B_{\phi}B_z$), those with a low poloidal field
(low magnetic flux) experience a stronger toroidal torque, and, 
thus gain higher rotational speed.
This is a consequence of the higher compression rates due to the
weaker poloidal field (more J-type than C-type shock).

For the jets with steady injection, the toroidal velocity field
decays as soon as the bow shock has left the computational domain.
On the long term the injected gas flows along the channel bored
by the jet during previous time steps and only weakly interacts 
with the ambient medium.
For jets with time-dependent injection, further shocks are generated
at a constant rate maintaining the jet rotation.

Jets with low internal Mach number develop standing oblique reflection 
shocks which constantly (but stepwise) accelerate the jet material
in toroidal direction.

In summary, by demonstrating the feasibility of toroidal jet acceleration 
by MHD shocks in a helical magnetic field, we propose the following 
implications.

i) Jets of young stars are typically seen in radiation of shocked 
   gas. Thus we expect this material to be affected by toroidal
   torques along its path from the jet launching region close to the
   star up to asymptotic regime of a collimated jet.
   To extrapolate from observed angular velocities far from the source 
   to the angular momentum of the jet material close to its origin
   is therefore questionable.

ii) Jet rotation, as observationally indicated in jets from young stars
   is feasible. Jets emerge from rotating disks and are ejected with
   kinetic angular momentum. In addition to this, MHD shocks could
   additionally accelerate the jet material by further converting magnetic 
   to kinetic angular momentum.

iii) Similar processes can be expected from relativistic stellar jets
   and extragalactic jets, leading to a rapid rotation in these jets.

The results presented in this paper are derived from kinetic / dynamical 
modeling.
Preliminary simulations considering cooling indicate not major
effect, although we see that cooling results in a somewhat 
stronger shock compression and therefore slightly higher
maximum rotation speeds.
Detailed comparison with observations would require also modeling
of radiation processes (radiative MHD), or post-processing
of emission maps, which is beyond the scope of this paper.

\begin{acknowledgements}
I thank the Andrea Mignone and the PLUTO team for the possibility to use 
their code. 
I also acknowledge many interesting and fruitful discussions with Bhargav 
Vaidya and Oliver Porth.
My thanks go as well to an unknown referee who contributed a number of
suggestions which clearly have improved the paper.
\end{acknowledgements}

\appendix

\section{Comparison simulations}

\begin{figure*}
\centering
\includegraphics[width=3.6cm]{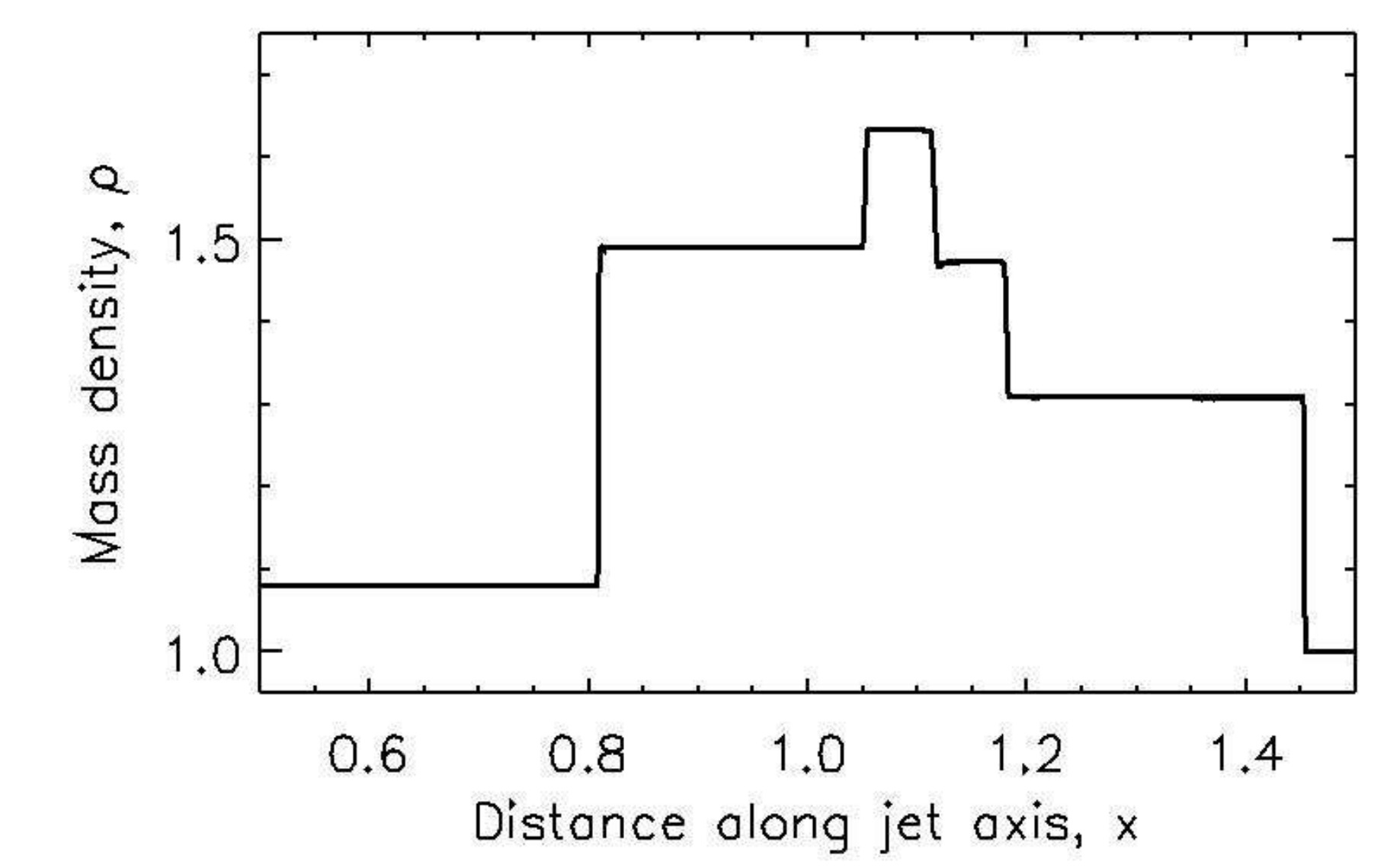}    
\includegraphics[width=3.6cm]{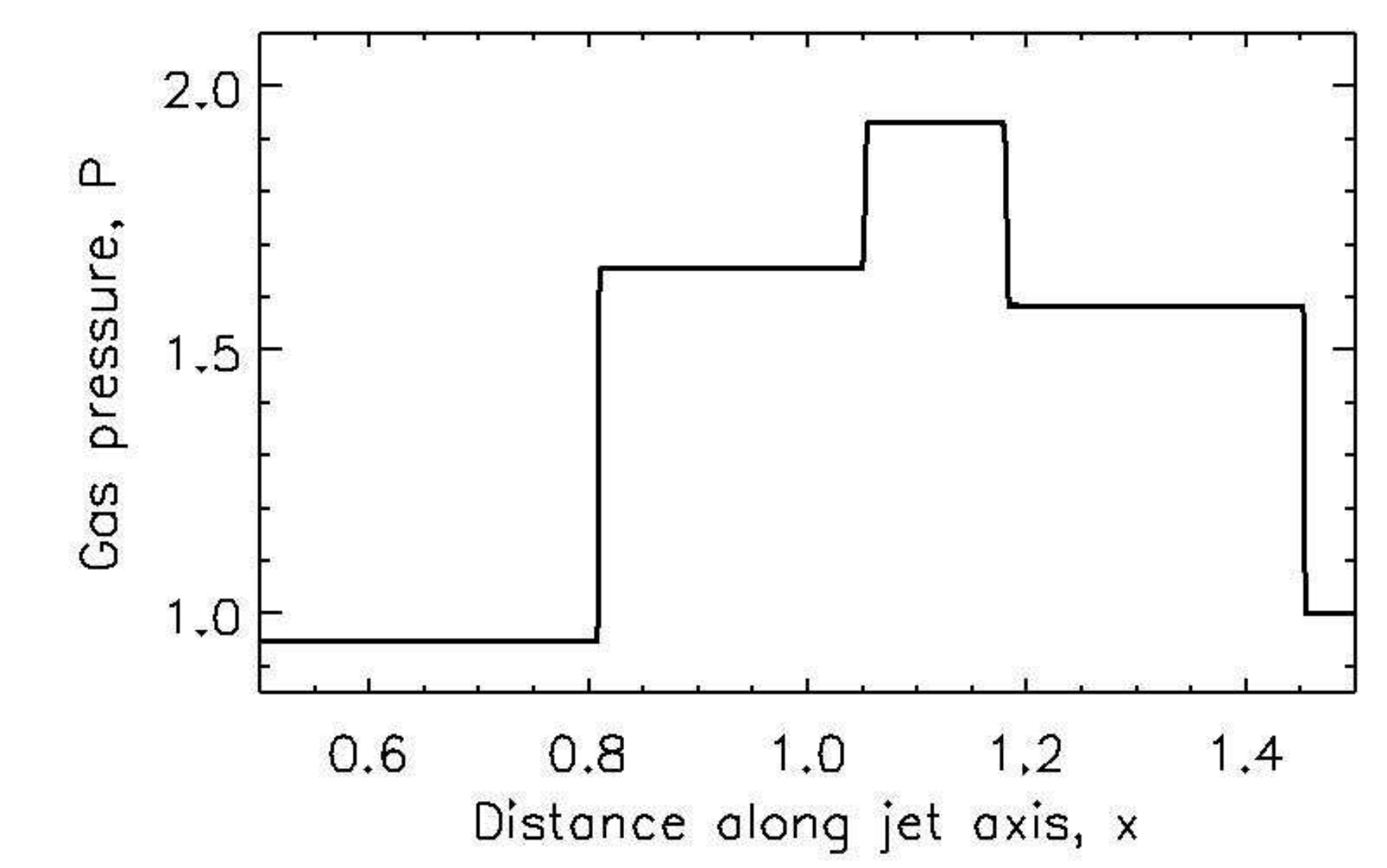}    
\includegraphics[width=3.6cm]{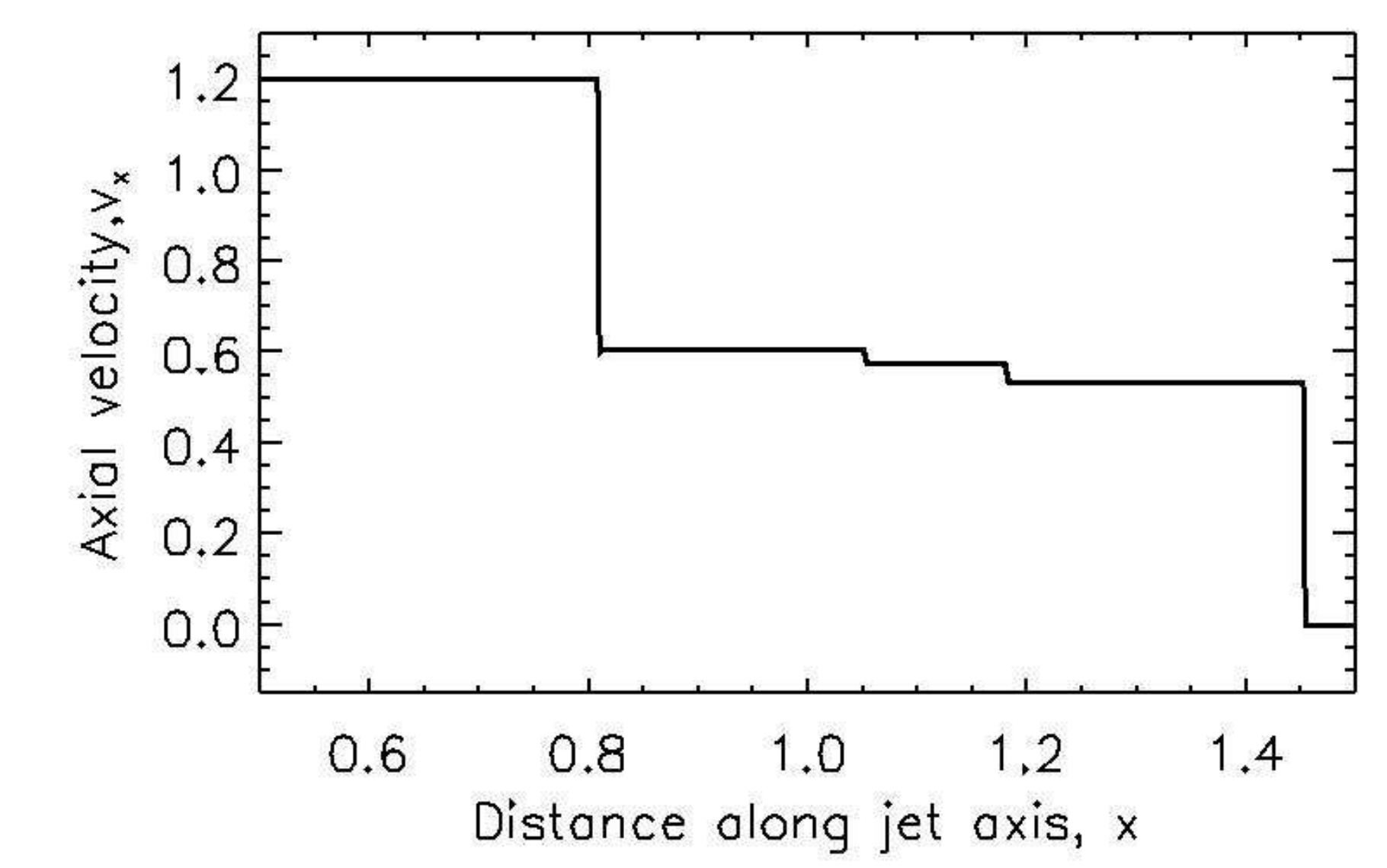}    
\includegraphics[width=3.6cm]{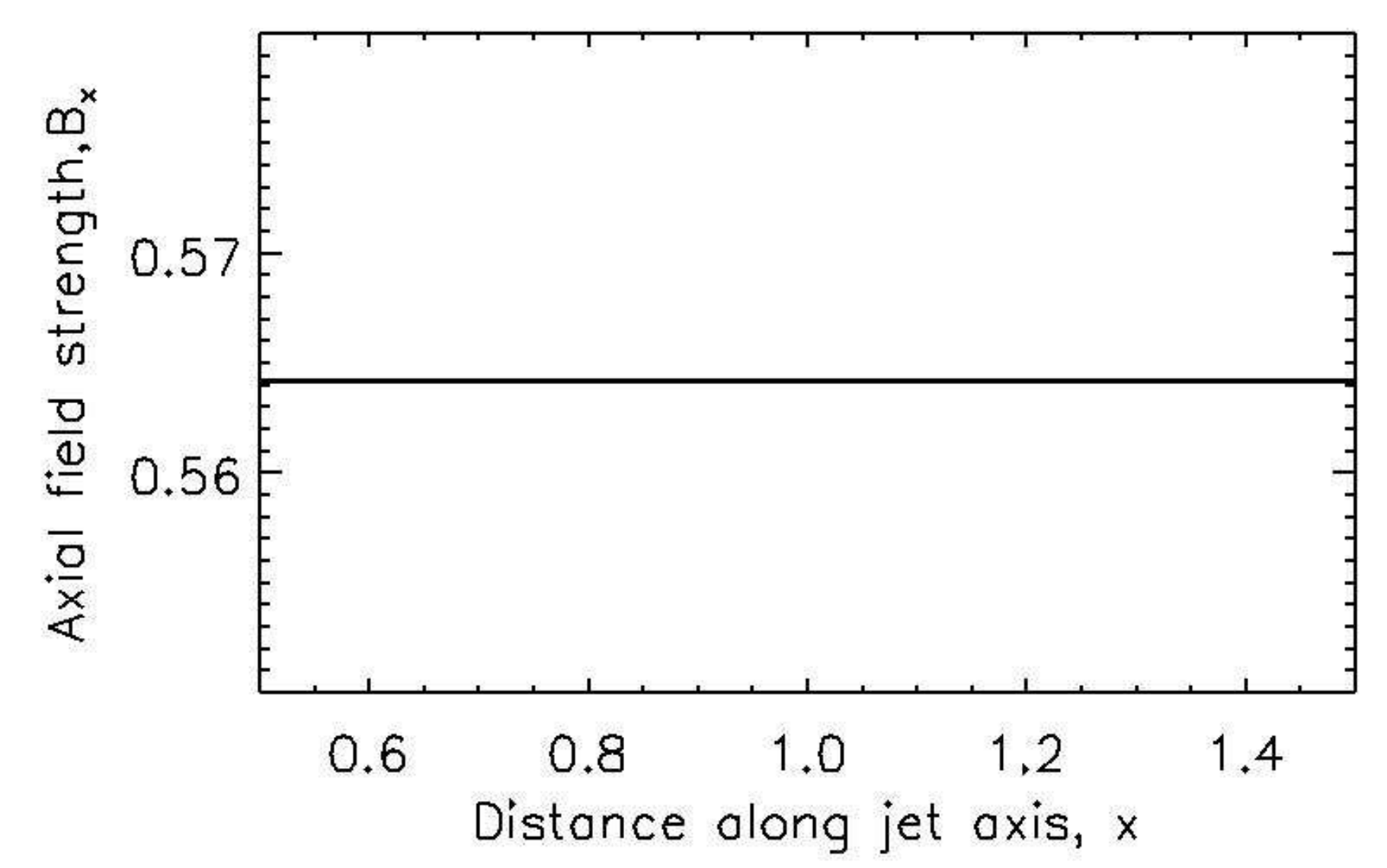}\\  
\includegraphics[width=3.6cm]{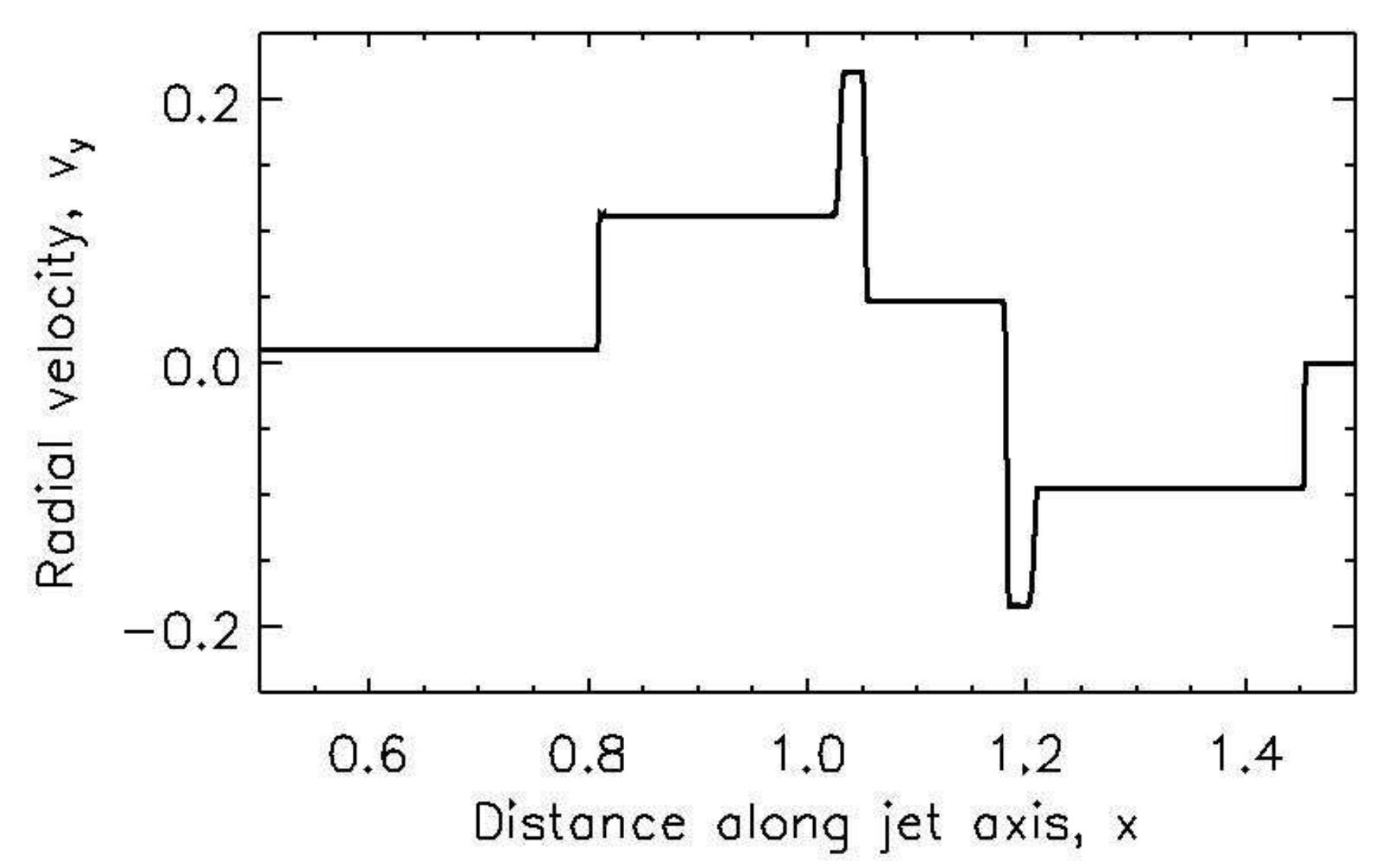}    
\includegraphics[width=3.6cm]{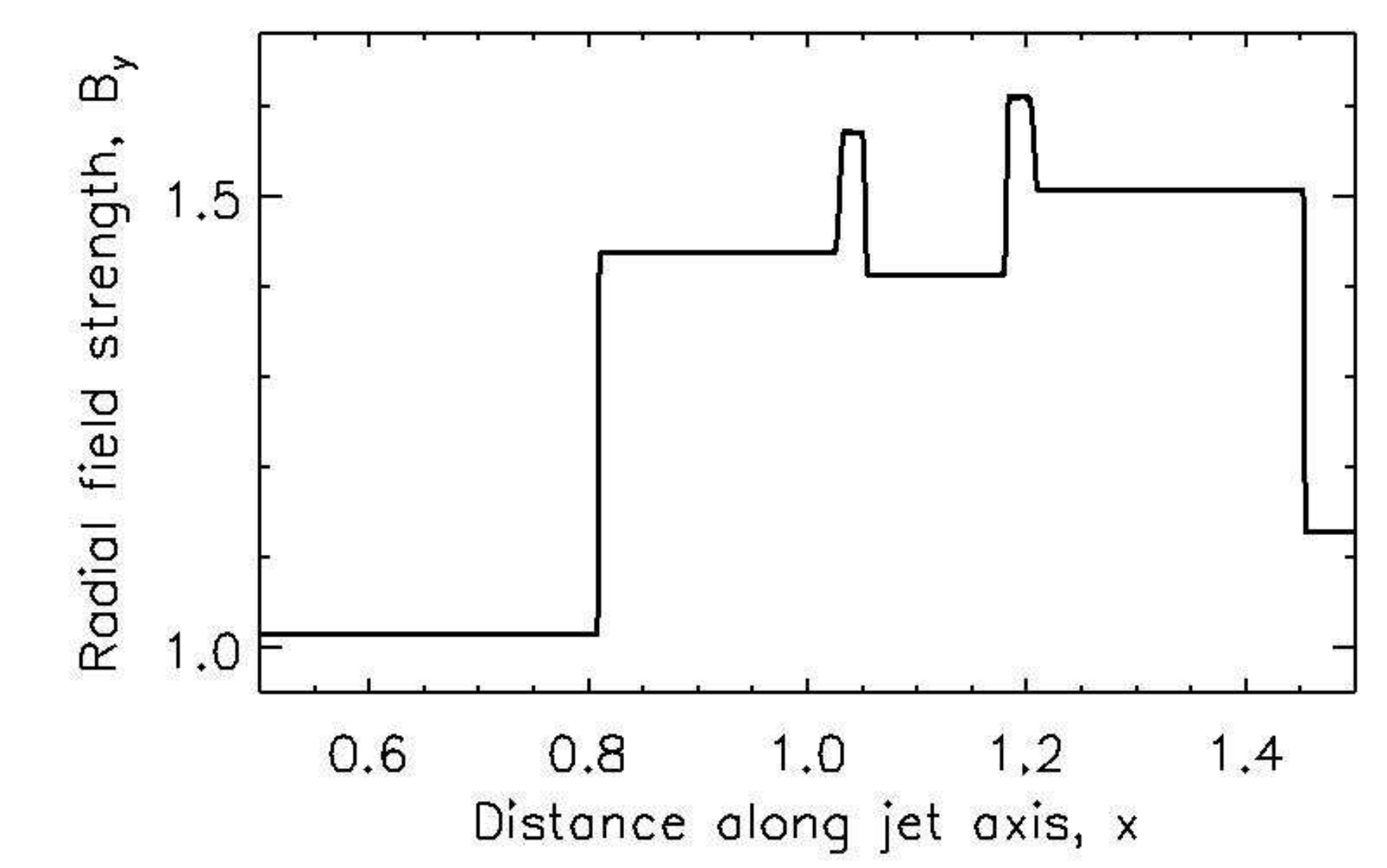}    
\includegraphics[width=3.6cm]{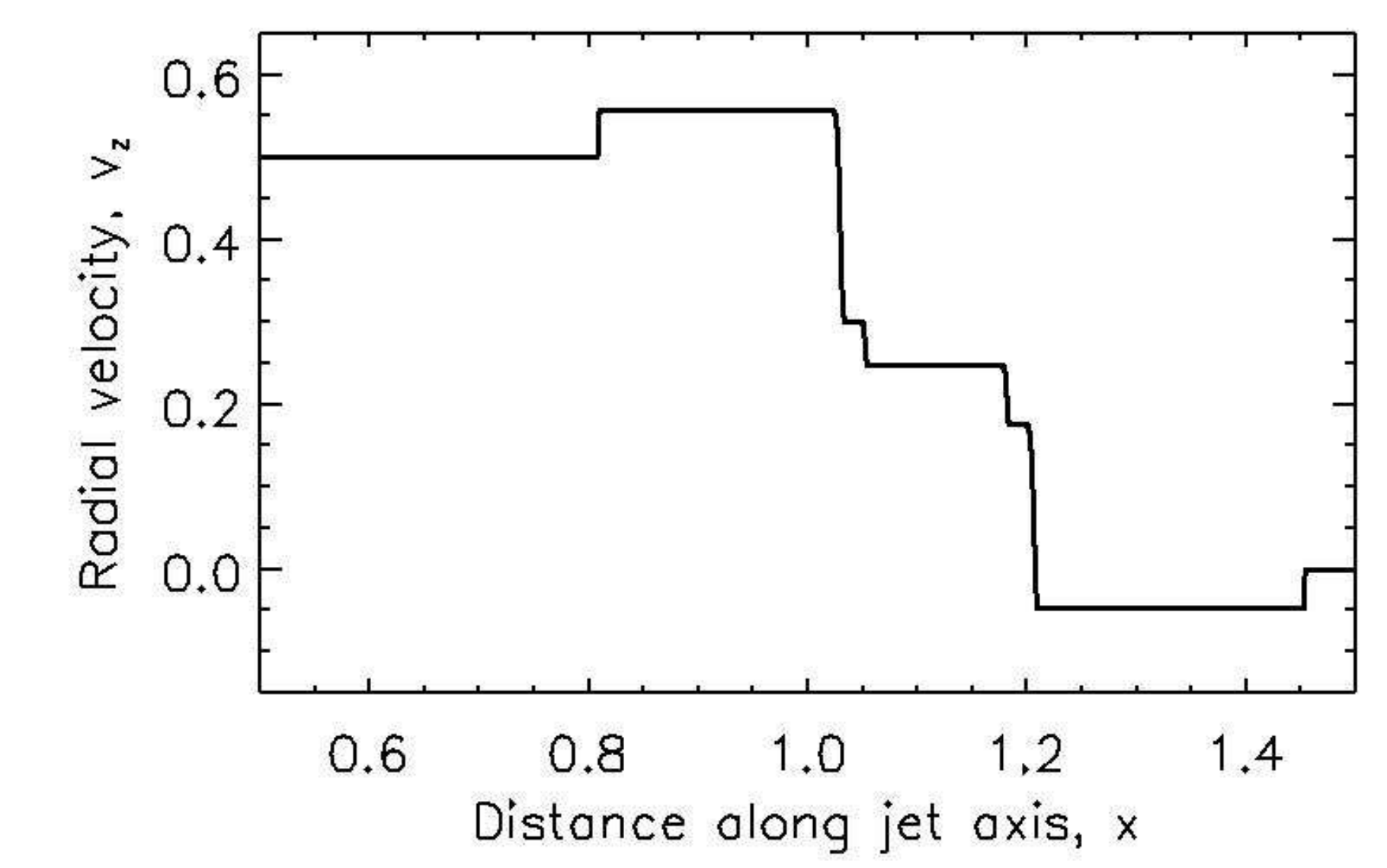}    
\includegraphics[width=3.6cm]{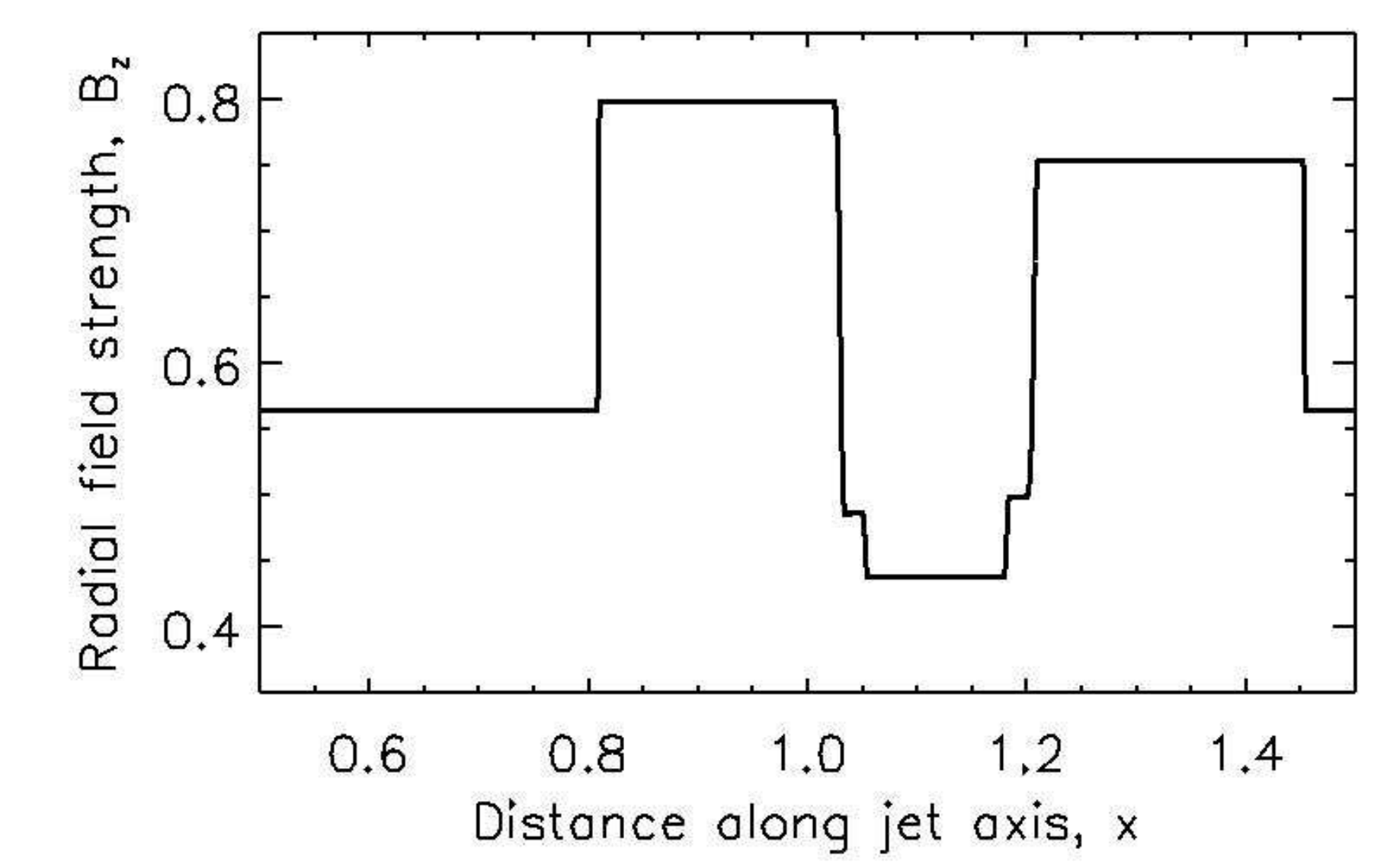}    
\caption{Simulation RJ11 in 1.5D Cartesian coordinates,
 equivalent to Fig.~2a of \citet{ryu95a}.
 Shown are:  density; gas pressure; axial and radial velocities;
 axial and radial magnetic field strength
 (from top left to bottom right) at dynamical time $t=10$
 (grid resolution 5000 cells per physical length 2.0).
}
\label{fig:ryu11}
\end{figure*}

\begin{figure*}
\centering
\includegraphics[width=3.6cm]{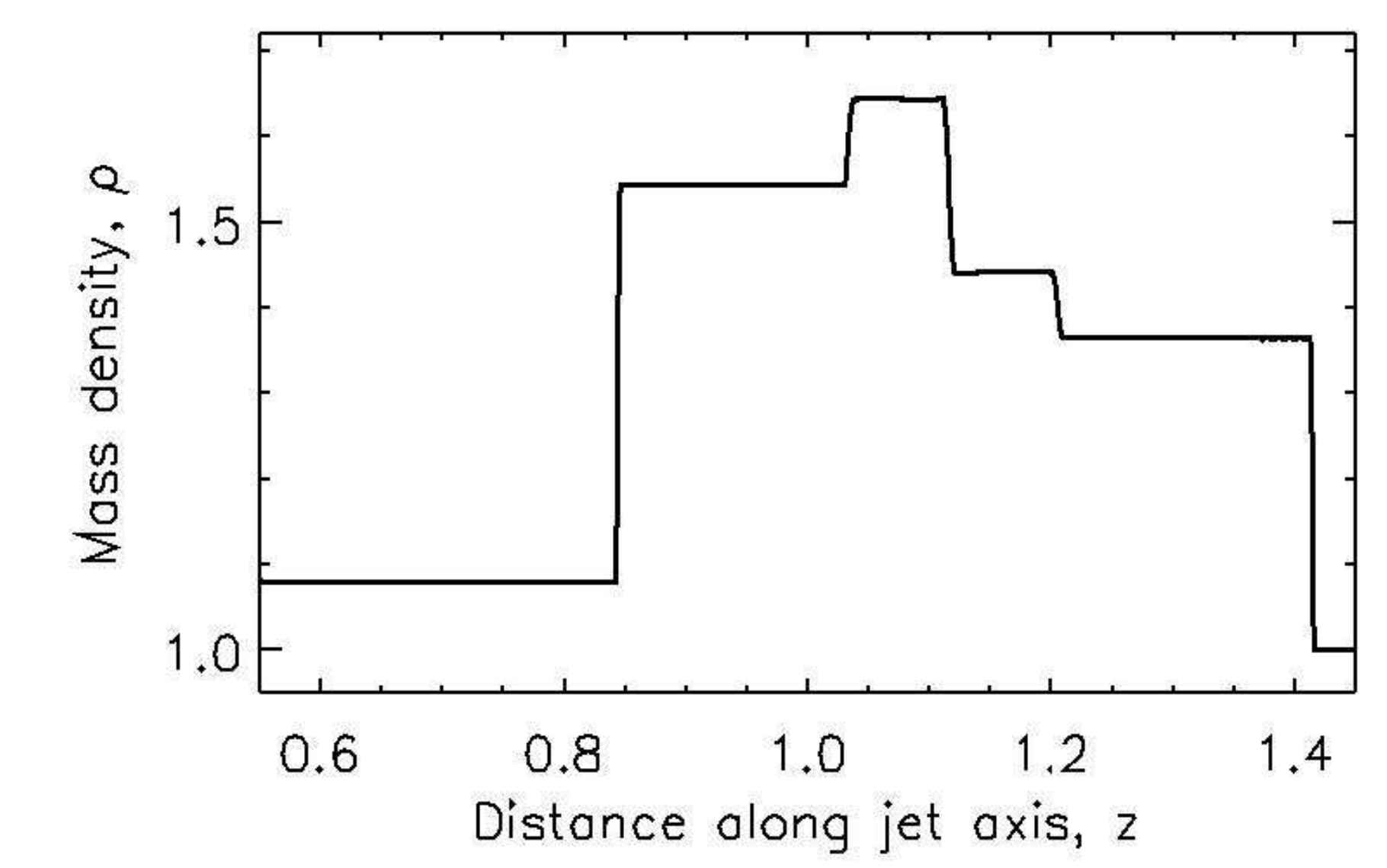}   
\includegraphics[width=3.6cm]{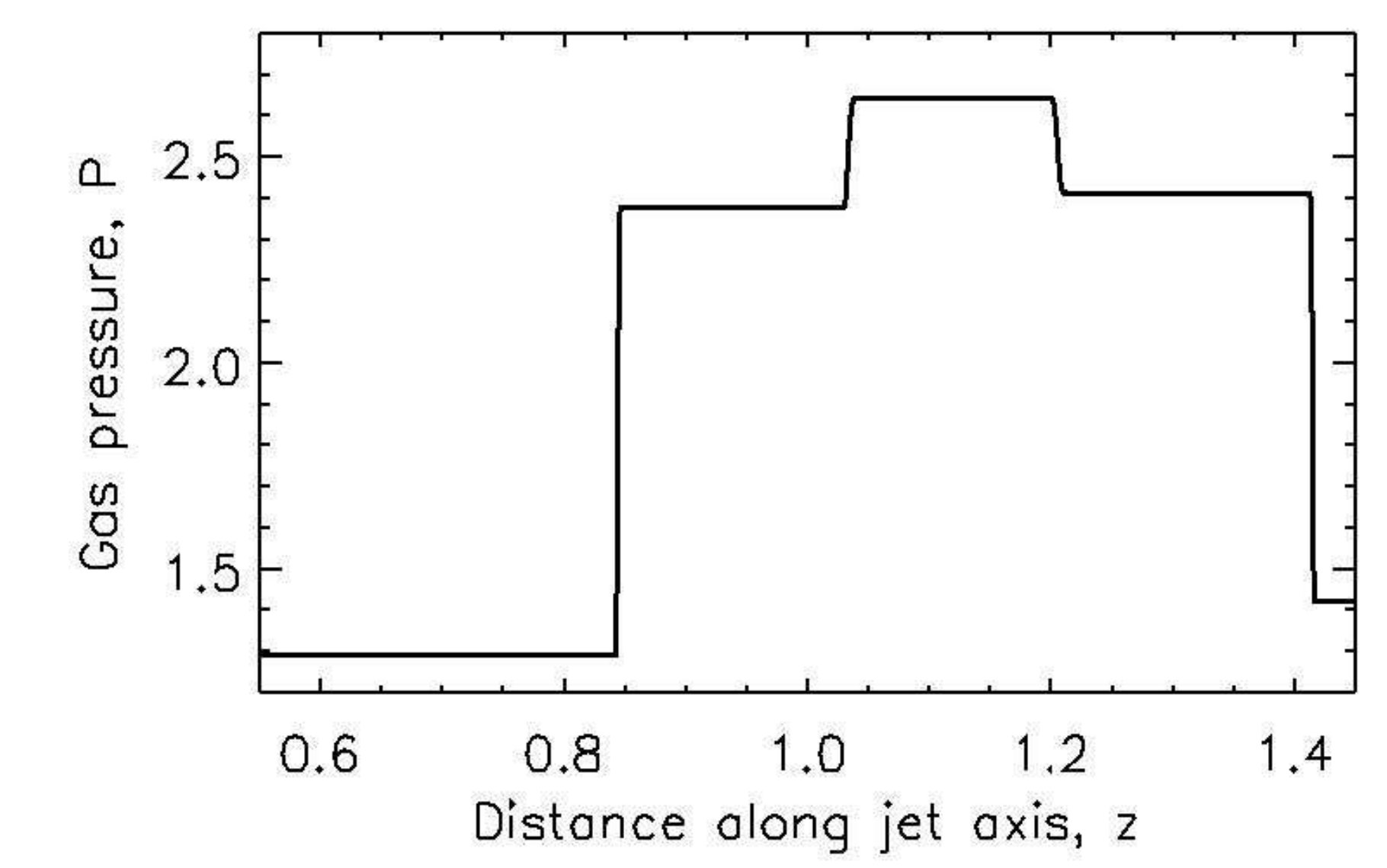}   
\includegraphics[width=3.6cm]{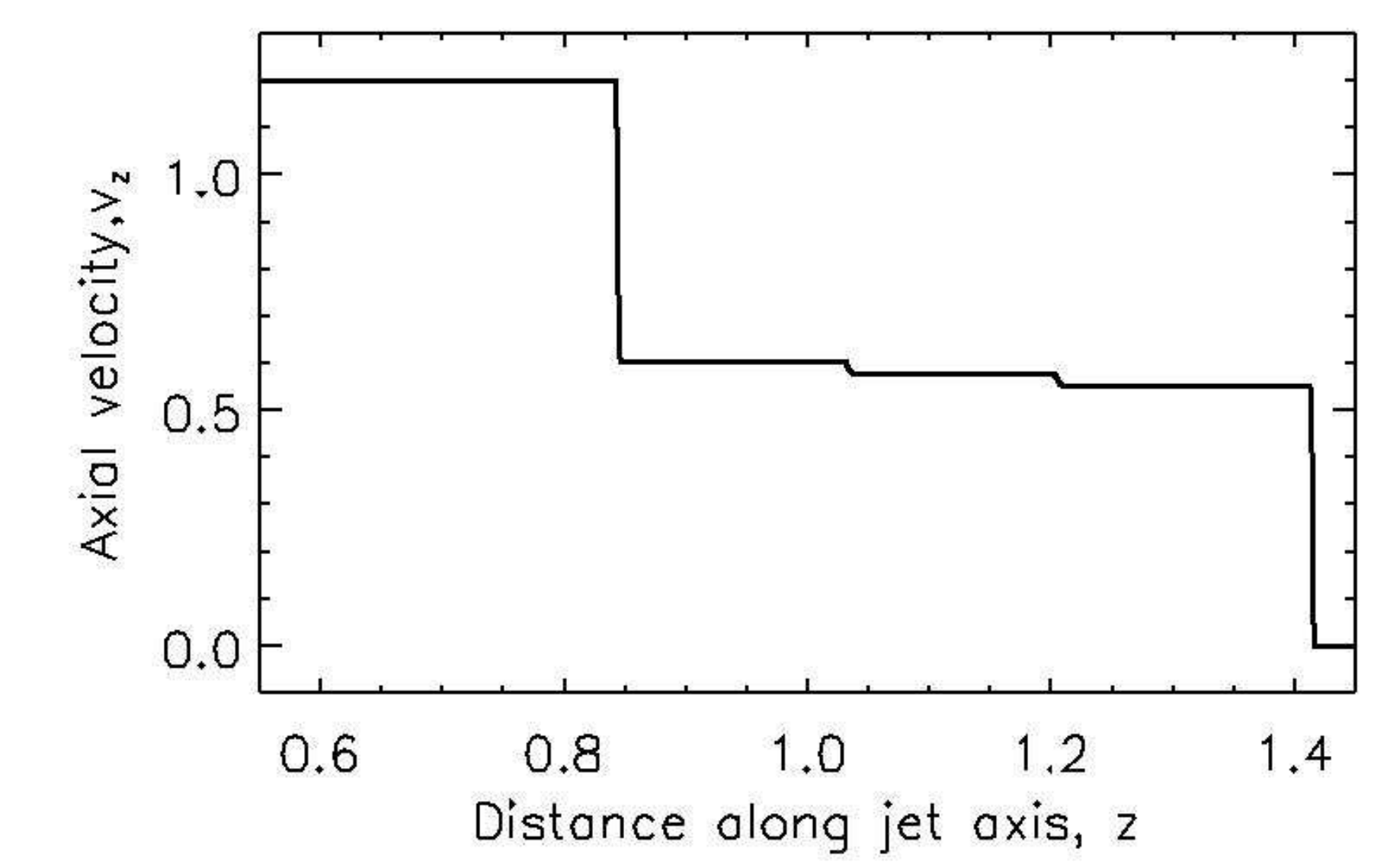}   
\includegraphics[width=3.6cm]{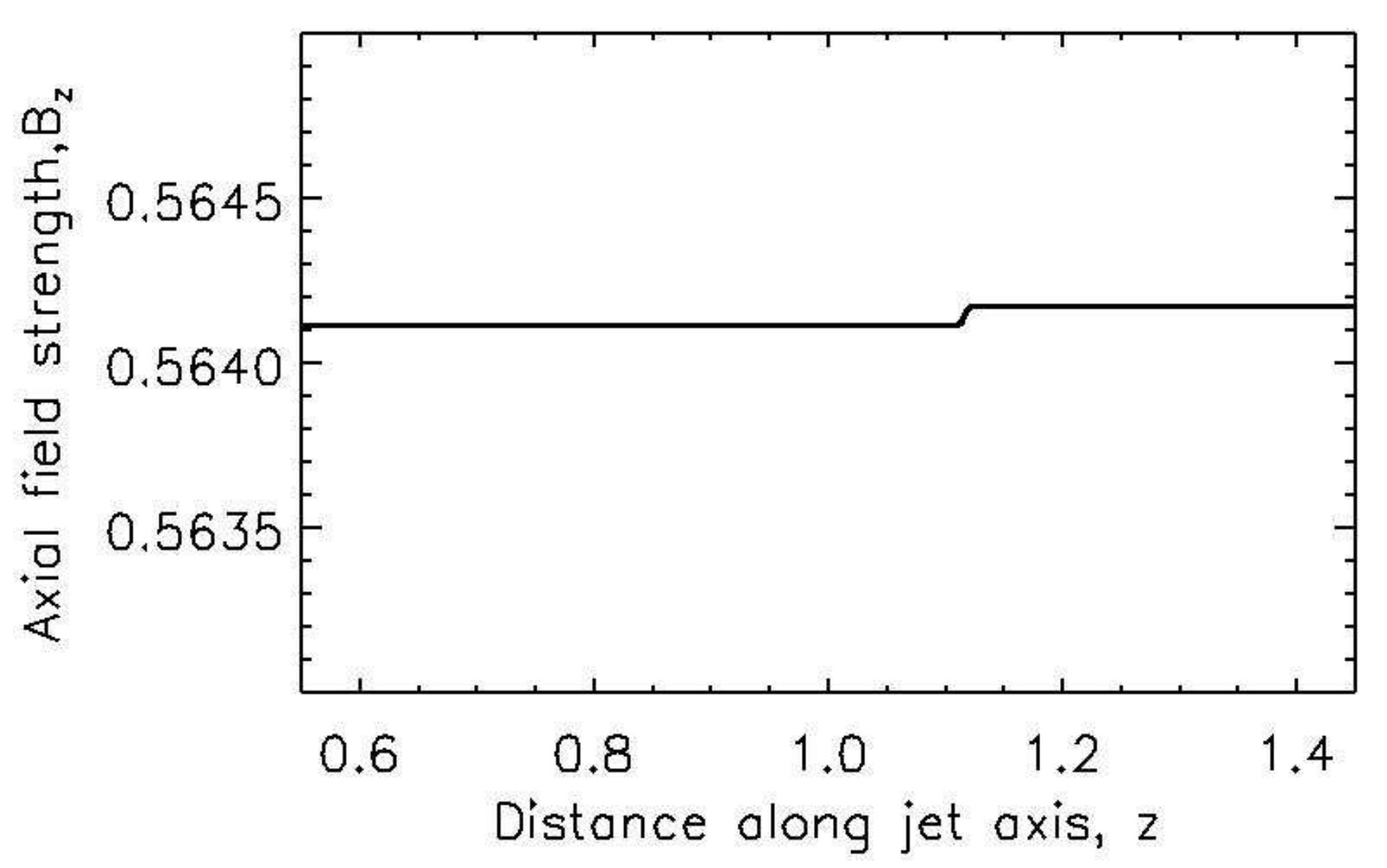}\\ 
\includegraphics[width=3.6cm]{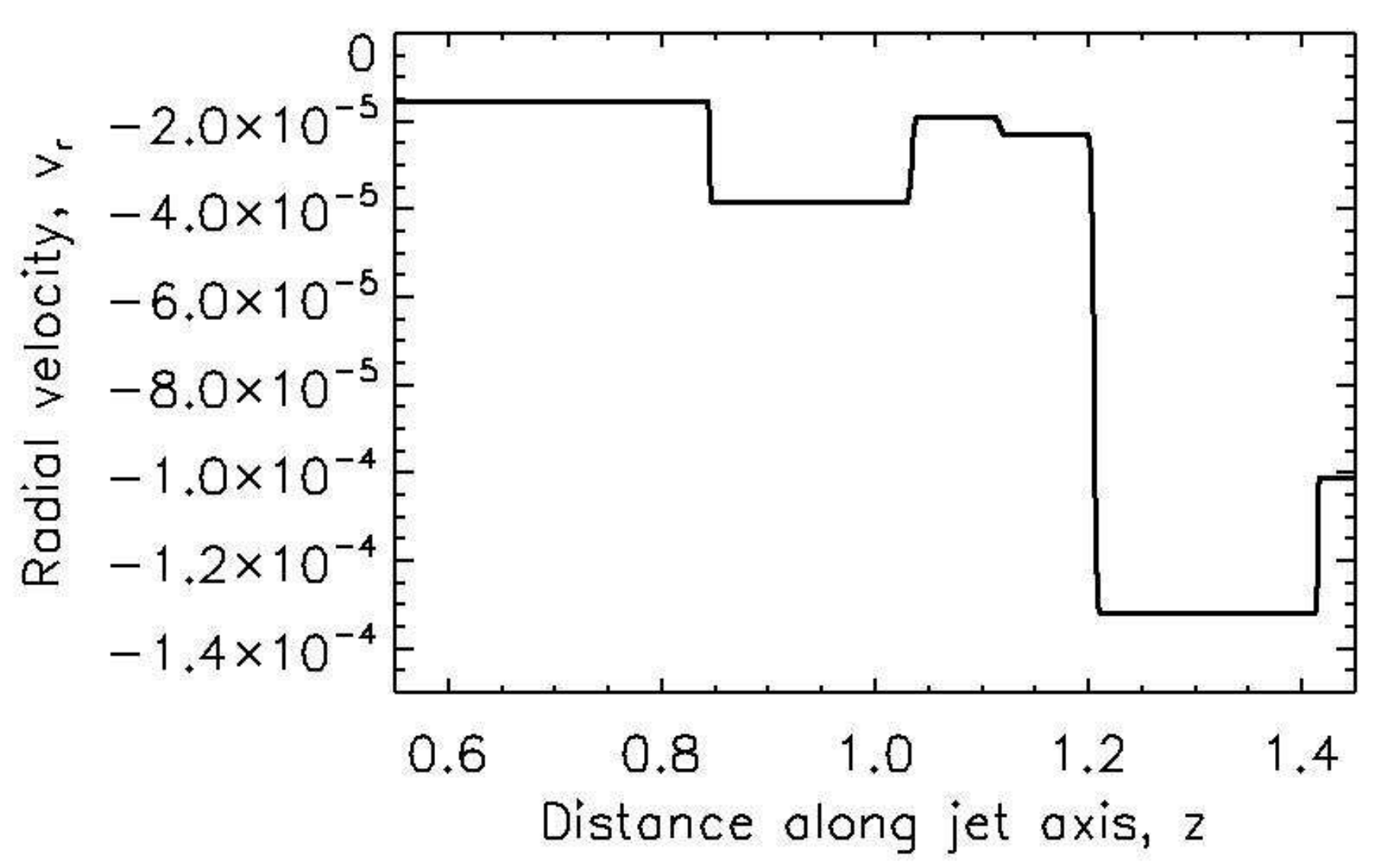}   
\includegraphics[width=3.6cm]{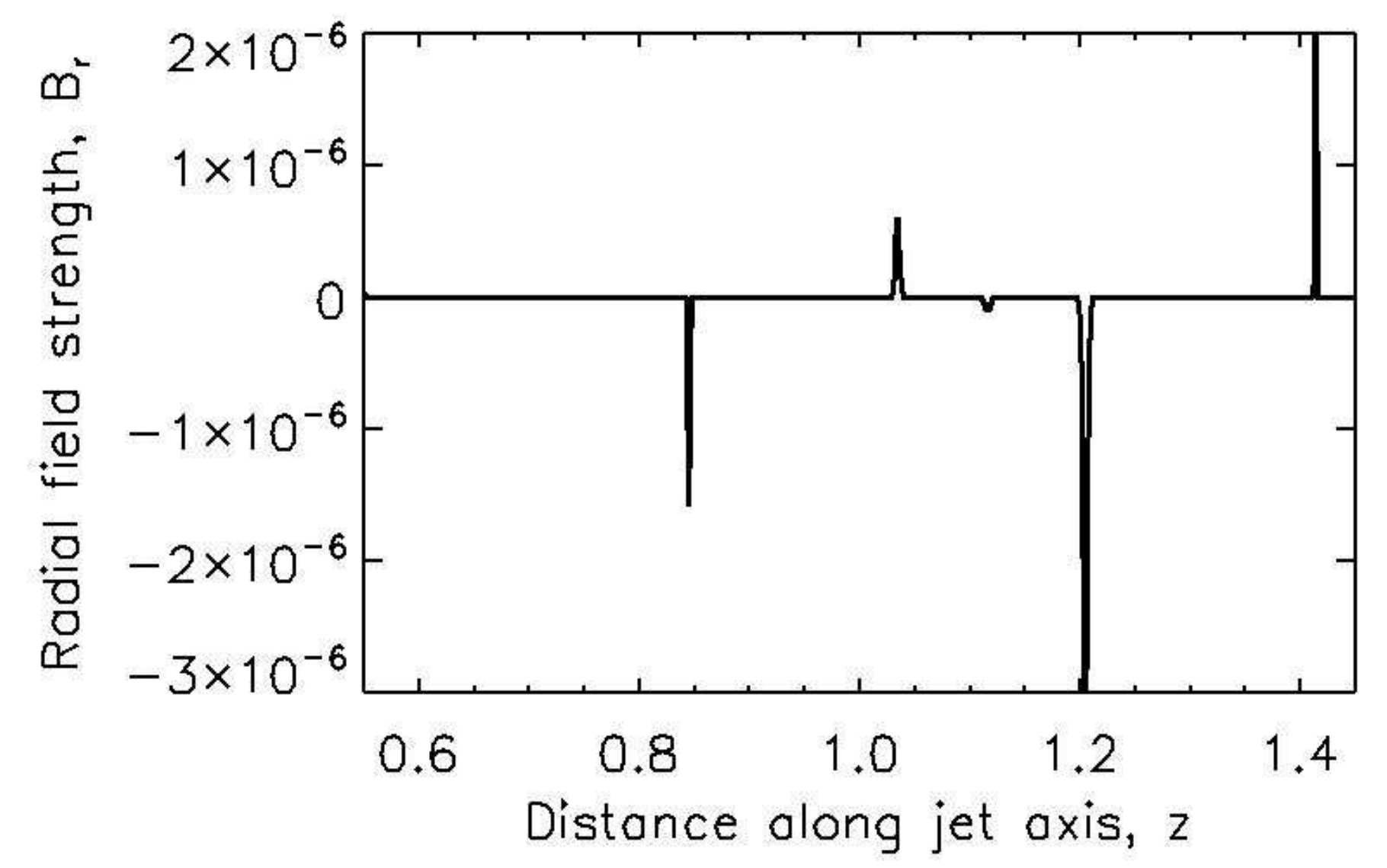}   
\includegraphics[width=3.6cm]{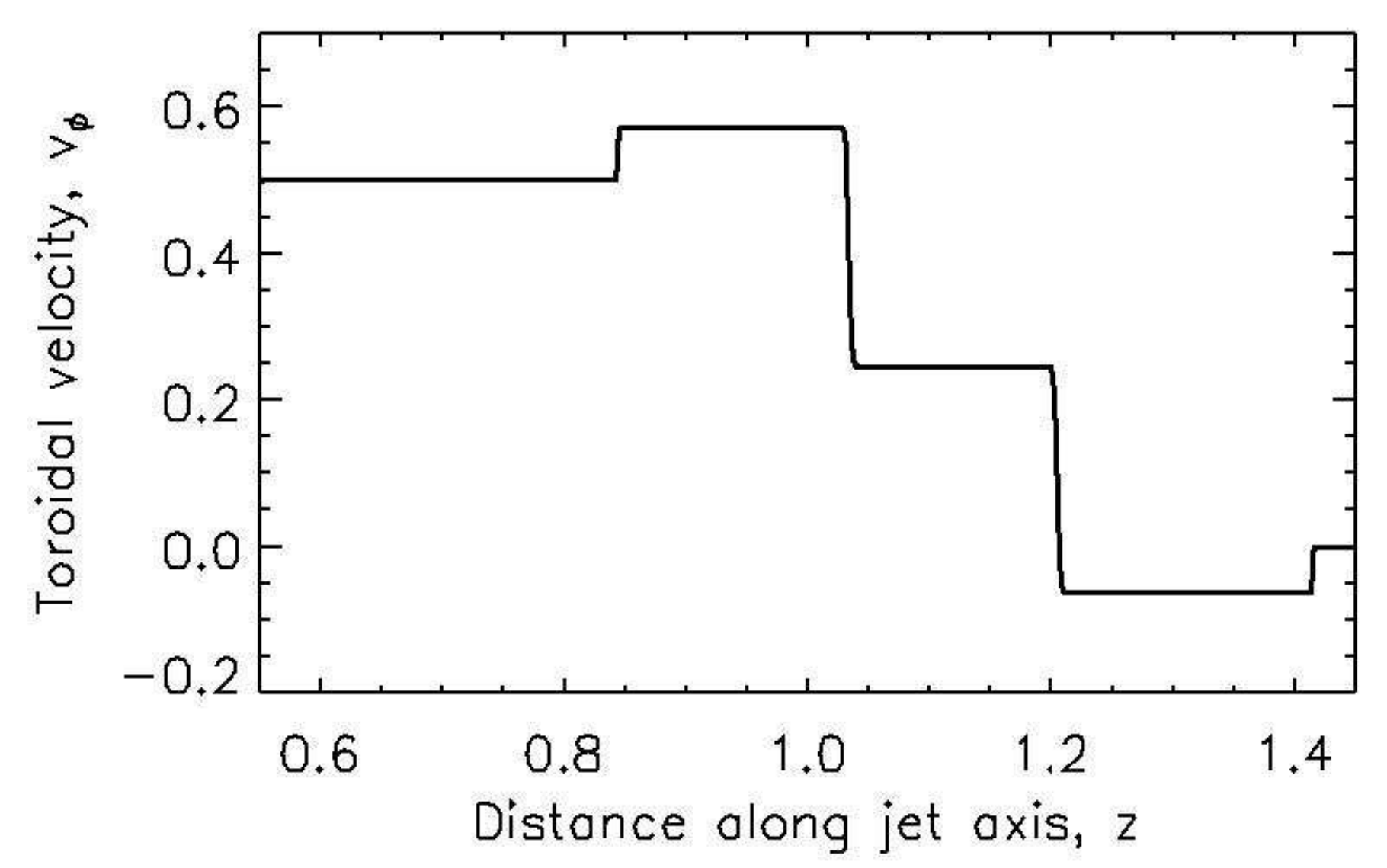}   
\includegraphics[width=3.6cm]{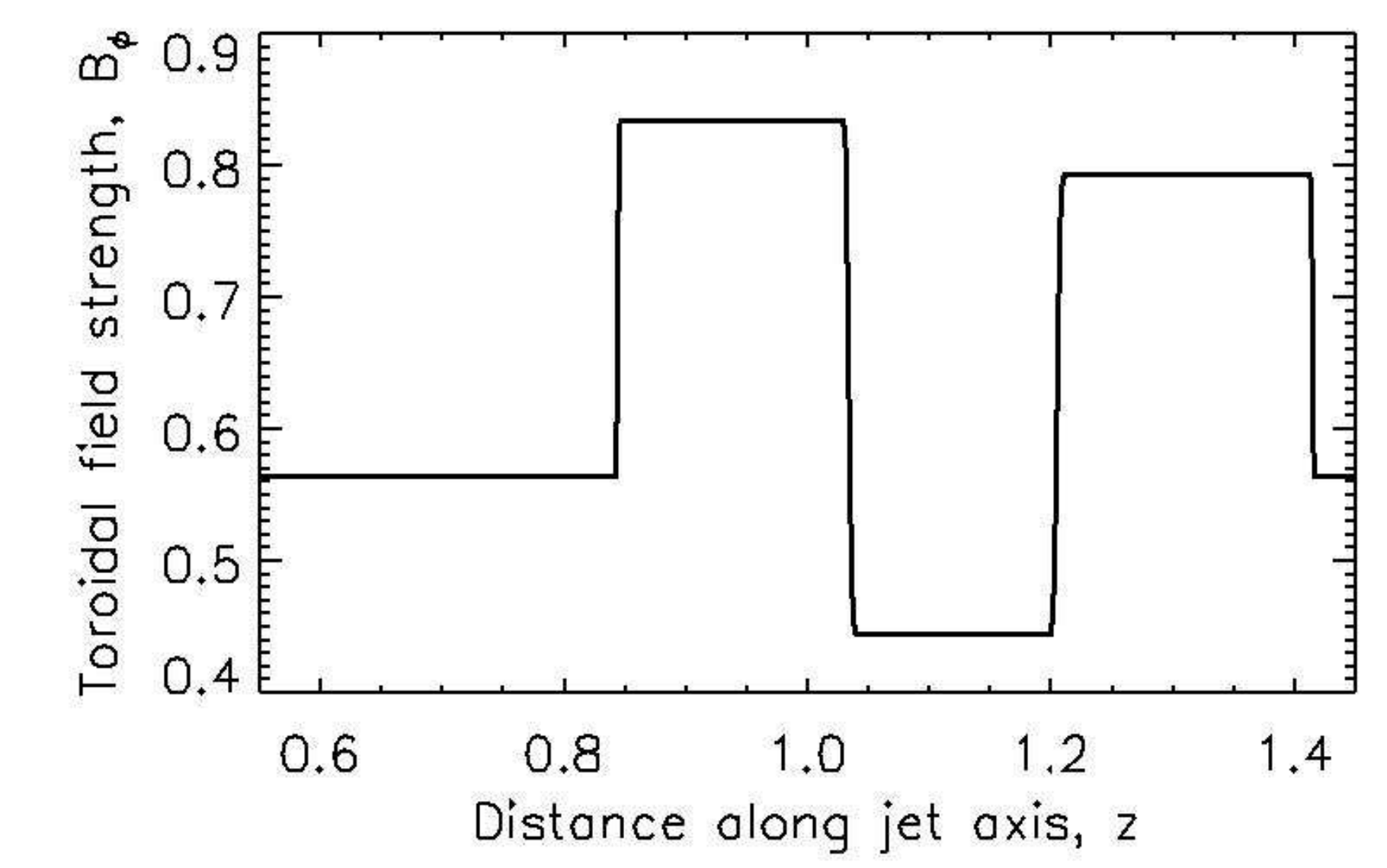}   
\caption{Simulation RJ09 in cylindrical coordinates and in axisymmetry, 
 similar to Fig.~2a of \citet{ryu95a}.
 Only one of the perpendicular vector components - the toroidal 
 component - is non-vanishing.
 Shown are:  density; gas pressure; axial, radial, and toroidal velocities;
 axial, radial and toroidal magnetic field strength
 (from top left to bottom right) at dynamical time $t=10$
 (grid resolution 3000 cells per physical length 2.0).
}
\label{fig:ryu09}
\end{figure*}

Here we show comparison simulations applying a similar set of parameters
(see Tab.~\ref{tab:para-1db}) as \citet{ryu95a}.
Figure \ref{fig:ryu11} shows simulation RJ11 which basically repeats the
simulation shown in \citet{ryu95a}, Fig.~2a, for which we have applied 
Cartesian coordinates as well.
Figure \ref{fig:ryu09} shows simulation RJ09 with similar dynamical 
parameters, but run in a cylindrical coordinate system, which can be
considered more appropriate to our general aim of investigating 
jet rotation.
In the latter case we have considered only one perpendicular vector 
component.
The initial setup with vanishing radial magnetic field and vanishing radial 
motion is very well conserved.

\begin{table*}
\scriptsize
\begin{center}
\caption{Parameter choice for the 1.5D simulations RJ09 (cylindrical
 coordinates) and RJ11 (Cartesian coordinates), which serve as 
 test cases to be compared to \citet{ryu95a}.
 The ambient density is ${\rho_{\rm ext}}=1.0$
 For comparison with Figs.~\ref{fig:ryu09},\ref{fig:ryu11} 
 note the magnetic field normalization $B \rightarrow B/\sqrt{4\pi}$.
\label{tab:para-1db}
}
\begin{tabular}{cccccccccccccc}
\tableline\tableline
\noalign{\smallskip} 
 ID     & 
 ${\rho_{\rm jet}}$      & 
 $v_{\rm p, jet}$        &
 $v_{\rm \phi, jet}$     &
 $P_{\rm jet}$           &
 $P_{\rm ext}$           &
 $B_x, B_z$              &
 $B_y, B_r$              &
 $B_z, B_{\phi}$ \\


\noalign{\smallskip}
\tableline
\noalign{\medskip}
\noalign{\medskip}
 RJ09 & 1.08 & 1.2 & 0.5 & 0.95 & 1.0 & 
        $B_{z,\rm jet} = B_{z,\rm ext} = 2.0$ & 
        $B_{r,\rm jet} = B_{r,\rm ext} = 0.0$ & 
        $B_{\phi,\rm jet} = B_{\phi,\rm ext} = 2.0$  \\
 RJ11 & 1.08 & 1.2 & 0.5 & 0.95 & 1.0 &
        $B_{x,\rm jet} = B_{x,\rm ext} = 2.0$ & 
        $B_{y,\rm jet} = 3.6, B_{y,\rm ext} = 4.0$ & 
        $B_{z,\rm jet} = B_{z, \rm ext} = 2.0$   \\
\noalign{\medskip}
\tableline
\end{tabular}
\end{center}
\end{table*}

\begin{thebibliography}{}

\bibitem[Anderson et al.(2003)]{ande03}
    Anderson, J.M., Li, Z.-Y., Krasnopolsky, R., Blandford, R.D. 2003, ApJ, 590, L107
\bibitem[Andre et~al.(1988)]{andr88}
    Andre, P., Montmerle, T., Feigelson, E.D., Stine, P.C., Klein, K. 1988, ApJ, 335, 940
\bibitem[Bacciotti et al.(2002)]{bacc02}
    Bacciotti, F., Ray, T.P., Mundt, R., Eisl\"offel, J., \& Solf, J. 2002, 576, 222
\bibitem[Blandford \&  Payne(1982)]{blan82}
    Blandford, R.D., \& Payne, D.G., 1982, MNRAS, 199, 883
\bibitem[Bouvier et~al.(2007)]{bouv07}
    Bouvier, J., Alencar, S.H.P., Boutelier, T., Dougados, C., Balog, Z., 
    Grankin, K., Hodgkin, S.T., Ibrahimov, M.A., Kun, M., Magakian, T.Y., Pinte, C. 
    2007, A\&A, 463, 1017
\bibitem[Cabrit(2007)]{cabr07}
    Cabrit, S., 
     The accretion-ejection connexion in T Tauri stars: jet models vs. observations,
     in: J. Bouvier \& I. Appenzeller (eds.), 
     Star-disk interaction in young stars, IAU Symposium 243, 2007, p.203
 \bibitem[Camenzind(1990)]{came90}
    Camenzind, M. 1990, Magnetized disk-winds and the origin
      of bipolar outflows, in: G. Klare (ed.) Rev. Mod. Astron. 3,
      Springer, Heidelberg, p.234
\bibitem[Carrasco-Gonz{\'a}lez et~al.(2010)]{carr10}
    Carrasco-Gonz{\'a}lez, C., Rodr{\'{\i}}guez, L.F., Anglada, G.,
    Mart{\'{\i}}, J., Torrelles, J.M.,  Osorio, M. , 2010, Science, 330, 1209
 \bibitem[Casse \& Keppens(2002)]{cass02}
    Casse, F. \& Keppens, R. 2002, ApJ, 581, 988
 \bibitem[Cecil et al.(1992)]{ceci92}
    Cecil, G., Wilson, A.S., Tully, R.B. 1992, ApJ 390, 365
 \bibitem[Clarke et al.(1986)]{clar86}
    Clarke, D.A., Norman, M.L., \& Burns, J.O. 1986, ApJ, 311, L63
 \bibitem[Choi et al.(2011)]{choi11}
    Choi, M., Kang, M., Tatematsu, K. 2011, ApJ, 728, L34
 \bibitem[Chrysostomou et al.(2008)]{chry08}
   Chrysostomou, A., Bacciotti, F., Nisini, B., Ray, T.P., Eisl\"offel, J., Davis, C.J.,
   Takami, M. 2008, A\&A 482, 575 
\bibitem[Coffey et al.(2004)]{coff04}
   Coffey, D., Bacciotti, F., Woitas, J., Ray, T.P., Eisl\"offel, J. 2004, 604, 758
\bibitem[Correia et al.(2009)]{corr09}
   Correia, S., Zinnecker, H., Ridgway, S.T., McCaughrean, M.J. 2009, A\&A 505, 673
 \bibitem[Davis et al.(2000)]{davi00}
    Davis, C.J., Berndsen, A., Smith, M.D., Chrysostomou, A., Hobson, J. 2000, MNRAS 314, 241
 \bibitem[Dempsey \& Rieger(2009)]{demp09}
    Dempsey, P. \& Rieger, F. 2009, IJMPD, 18, 1651
 \bibitem[Fendt et al.(1995)]{fend95}
    Fendt, C., Camenzind, M., \& Appl, S. 1995, A\&A, 300, 791
 \bibitem[Fendt \& Camenzind(1996)]{fend96}
    Fendt, C., \& Camenzind, M. 1996, A\&A, 313, 591
 \bibitem[Fendt \& Cemeljic(2002)]{fend02}
    Fendt, C., \& Cemeljic, M. 2002, A\&A, 395, 1045
 \bibitem[Fendt(2006)]{fend06}
    Fendt, C. 2006, ApJ, 651, 272
 \bibitem[Fendt(2009)]{fend09}
    Fendt, C. 2009, ApJ, 692, 346
 \bibitem[Ferreira et al.(2000)]{ferr00}
    Ferreira, J., Pelletier, G., Appl, S. 2000, MNRAS, 312, 387
 \bibitem[Frank et al.(2000)]{fran00}
    Frank, A., Lery, T., Gardiner, T.A., Jones, T.W., Ryu, D. 2000, ApJ, 540, 342
 \bibitem[Gabuzda et al.(2004)]{gabu04}
    Gabuzda, D.C., Murray, E., Cronin, P. 2004 MNRAS, L351, 89
 \bibitem[Herrnstein et al.(1997))]{herr97}
    Herrnstein, J., Moran, J.M., Greenhill, L.J., Diamond, P.J., Miyoshi, M., Nakai, N., 
    Inoue, M. 1997, ApJ, 475, L17
 \bibitem[Kato et al.(2002)]{kato02}
    Kato, S.X., Kudoh, T., \& Shibata, K. 2002, ApJ, 565, 1035
 \bibitem[K\"ossl et al.(1990)]{koes90}
    K\"ossl, D., M\"uller, E. \& Hillebrandt, W. 1990, A\&A 229, 378
 \bibitem[Krasnopolsky et al.(1999)]{kras99}
    Krasnopolsky, R., Li, Z.-Y., \& Blandford, R. 1999, ApJ, 526, 631
 \bibitem[Laing et al. (2006)]{lain06}
    Laing, R.A., Canvin, J.R., Bridle, A.H., Hardcastle, M.J. 2006, MNRAS, 372, 510
 \bibitem[Launhardt et al. (2009)]{laun09}
    Launhardt, R., Pavlyuchenkov, Y., Gueth, F., Chen, X., Dutrey, A., Guilloteau, S.,
    Henning, Th., Pi\'etu, V., Schreyer, K., Semenov, D., 2009, A\&A, 494, 147
 \bibitem[Lavalley-Fouguet et al.(2000)]{lava00}
    Lavalley-Fouquet, C., Sabrit, S., Dougadous, C. 2000, A\&A, 356, L41
 \bibitem[Lee et al.(2007)]{lee07}
    Lee, C.-F., Ho, P.T.P., Palau, A., Hirano, N., Bourke, T.L., N., Shang, H., 
    Zhang, Q. 2007, ApJ, 670, 1188
 \bibitem[Lee et al.(2008)]{lee08}
    Lee, C.-F., Ho, P.T.P., Bourke, T.L., Hirano, N., Shang, H., Zhang, Q. 
    2008, ApJ, 685, 1026
 \bibitem[Lee et al.(2009)]{lee09}
    Lee, C.-F., Hirano, N., Palau, A., Ho, P.T.P., Bourke, T.L., Zhang, Q., 
    Shang, H. 2009, ApJ, 699, 1584
 \bibitem[Lyutikov et al.(2005)]{lyut05}
    Lyutikov, M., Pariev, V.I., Gabuzda, D.C. 2005, MNRAS, 360, 869
 \bibitem[Matt et al. (2008)]{matt08}
    Matt, S.P., Pudritz, R.E. 2008, ApJ, 678, 1109
 \bibitem[Matt et al. (2010)]{matt10}
    Matt, S.P., Pinzon, G., de la Reza, R., Greene, T.P. 2010, ApJ, 714, 989
 \bibitem[Meliani \& Keppens(2007)]{meli07}
    Meliani, Z., \& Keppens, R. 2007, A\&A 475, 785
 \bibitem[Mignone et al.(2007)]{mign07}
    Mignone, A., Bodo, G., Massaglia, S., Matsakos, T., Tesileanu, O., 
    Zanni, C., Ferrari, A. 2007, ApJS, 170, 228
 \bibitem[Moll (2009)]{moll09}
    Moll, R. A\&A, 507, 1203
 \bibitem[Movsessian et al.(2007)]{movs07}
    Movsessian, T.A., Magakian, T.Y., Bally, J., Smith, M.D., Moiseev, A.V., 
    Dodonov, S.N. 2007, A\&A, 470, 605
 \bibitem[O'Neill et al.(2005)]{onei05}
    O'Neill, S.M., Tregillis, I.L., Jones, T.W., Ryu D., ApJ, 633, 717
 \bibitem[Ouyed \& Pudritz(1997)]{ouye97}
    Ouyed, R., \& Pudritz, R.E., 1997, ApJ, 482, 712
 \bibitem[Ouyed et al.(2003)]{ouye03}
    Ouyed, R., Clarke, D.A., Pudritz, R.E. 2003, ApJ, 582, 292
 \bibitem[Pelletier \& Pudritz(1992)]{pell92}
    Pelletier, G., \& Pudritz, R.E. 1992, ApJ, 384, 117
 \bibitem[Porth  \& Fendt(2010)]{port10}
    Porth, O., \& Fendt, C. 2010, ApJ, 709, 1100
 \bibitem[Pudritz \& Norman(1983)]{pudr83}
    Pudritz, R.E., \& Norman, C.A. 1983, ApJ, 274, 677
 \bibitem[Pudritz et al.(2007)]{pudr07}
    Pudritz, R.E., Ouyed, R., Fendt, Ch., Brandenburg, A. 2007,
    in: B.~Reipurth, D.~Jewitt, \& K.~Keil (eds.), Protostars \& Planets V,
    University of Arizona Press, Tucson, 2007, p.277
 \bibitem[Ramsey \& Clarke (2011)]{rams11}
    Ramsey, J.P., Clarke, D.A. 2011, ApJ, 728, L11
 \bibitem[Ray et~al.(1997)]{ray97}
    Ray, T.P., Muxlow, T.W.B., Axon, D.J., Brown, A., Corcoran, D.,
    Dyson, J., Mundt, R. 1997, Nature, 385, 415
 \bibitem[Rieger \& Mannheim(2002)]{rieg02}
    Rieger, F.M. \& Mannheim, K. 2002, A\&A, 396, 833
 \bibitem[Rieger \& Duffy(2004)]{rieg04}
    Rieger, F.M. \& Duffy, P. 2004, ApJ, 617, 155
 \bibitem[Rosen et al.(1999)]{rose99}
    Rosen, A., Hardee, P., Clarke, D.A., Johnson, A. 1999, ApJ, 510, 136
 \bibitem[Ryu \& Jones(1995)]{ryu95a}
    Ryu, D., Jones, T.W. 1995, ApJ, 442, 228
 \bibitem[Ryu et al.(1995b)]{ryu95b}
    Ryu, D., Jones, T.W., Frank, A. 1995, ApJ, 452, 785
 \bibitem[Soker (2005)]{soke05}
    Soker, N. 2005, A\&A, 435, 125
 \bibitem[Stone \& Hardee(2000)]{ston00}
    Stone, J., Hardee, P. 2000, ApJ, 540, 192
 \bibitem[Todo et al.(1992)]{todo92}
    Todo, Y., Uchida, Y., Sato, T., Rosner, R. 1992, PASJ, 44, 245
 \bibitem[Todo et al.(1993)]{todo93}
    Todo, Y., Uchida, Y., Sato, T., Rosner, R. 1993, ApJ, 403, 164 
 \bibitem[Uchida et al.(1992)]{uchi92}
    Uchida, Y., Todo, Y., Rosner, R., Shibata, K. 1992, PASJ, 44, 227
 \bibitem[Vitorino et al.(2003)]{vito03}
    Vitorino, B.F., Jatenco-Pereira, V., \& Opher, R. 2003, ApJ, 592, 332
 \bibitem[Vlemmings(2008)]{vlem08}
    Vlemmings, W.H.T. 2008, A\&A, 484, 773
\bibitem[Vlemmings et~al.(2010)]{vlem10}
    Vlemmings, W.H.T., Surcis, G., Torstensson, K.J.E., van Langevelde, H.J. 
    2010, MNRAS, 404, 134
 \bibitem[Wiseman et al.(2001)]{wise01}
    Wiseman, J., Wootten, A., Zinnecker, H., McCaughrean, M. 2001, ApJ, 550, L87
 \bibitem[Woitas et al.(2005)]{woit05}
    Woitas, J., Bacciotti, F., Ray, T.P., Marconi, A., Coffey, D.,
    Eisl\"offel, J. 2005, A\&A, 432, 149
 \bibitem[Zapata et al.(2009)]{zapa09}
    Zapata, L.A., Schmid-Burgk, J., Muders, D., Schilke, P., Menten, K., Guesten, R. 
    2009, A\&A, 510, 2
\end{thebibliography}
\end{document}